\documentclass[article,prc,epsfig,floatfix,nofootinbib,aps]{revtex4}
\usepackage{epsfig}
\usepackage{epsf}
\begin{document}

\title{Narrowing the uncertainty on the total charm cross section and its effect
on the $J/\psi$ cross section}

\author{R. Nelson$^{1,2}$, R. Vogt$^{1,2}$ and A. D. Frawley$^3$}

\affiliation{
$^1$Physics Division, Lawrence Livermore National Laboratory, 
Livermore, CA 94551, USA \break
$^2$Physics Department, University of California at Davis, Davis, CA 95616, 
USA \break
$^3$Physics Department, 
Florida State University, Tallahassee, FL 32306, USA \break
}

\begin{abstract}
We explore the available parameter space that gives reasonable fits to the
total charm cross section to make a better estimate of its true uncertainty.
We study the effect of the parameter choices on the energy dependence of the
$J/\psi$ cross section.
\end{abstract}

\maketitle

\section{Introduction}

Because the charm quark mass is finite, the total charm production cross
section can be calculated in perturbative QCD.  However, the charm
quark mass is relatively light so that there are large uncertainties due
to the choice of quark mass, factorization scale and renormalization scale
\cite{RVjoszo}.
Typical lower limits of the factorization and
renormalization scales are half the chosen charm quark mass \cite{RVjoszo,CNV}.
Here, the parton densities are subject to backward evolution since the
factorization scale is below the minimum scale of the parton densities.
In addition, for renormalization scales below 1 GeV, the strong coupling
constant $\alpha_s$ becomes large and the perturbative expansion is unlikely 
to converge.  Thus is it worth evaluating our assumptions concerning these
parameters to determine whether we can find a set of physically defensible
mass and scale parameters that reduce the cross section uncertainty.

In the color evaporation model of $J/\psi$ production, the $J/\psi$ cross
sections are calculated with the same set of mass and scale parameters as
open charm production \cite{Gavai:1994in}.  As we will show, the parameters 
used to calculate
the uncertainty on the charm quark cross sections in Refs.~\cite{RVjoszo,CNV}
do not place stringent bounds on the $J/\psi$ production cross section.  We
therefore seek to place limits on the $J/\psi$ cross section calculated in the 
color evaporation model for the first time.

In this paper we will explore the charm quark mass and scale
parameter space to reduce the uncertainty on the charm total cross section.
In Sec.~\ref{open}, we use existing data to set limits on the 
factorization and renormalization
scales.  We will then calculate the lepton distributions from heavy flavor decays
with our parameter limits and compare to data from RHIC and the LHC as a
reality check. Section~\ref{closed} describes how we use our open charm results 
in color evaporation
model calculations of $J/\psi$ production to determine the uncertainties on
quarkonium production, both as a function of incident energy and as a function
of the kinematic variables.  In Sec.~\ref{end}, we summarize our results.

\section{Setting Limits on the Total Charm Cross Section}
\label{open}

In our previous efforts to place uncertainties on the total charm cross section
using the same fiducial parameter set as the FONLL calculation, we found
a wide uncertainty band that grew larger at high center of mass energies
\cite{RVjoszo,RVhp08proc}.  At high energies, the lower limit was determined by 
a factorization scale of half the central value of the charm quark mass, assumed
to be 1.5 GeV.  In this region, the behavior of the gluon density at low
momentum fraction, $x$, when the parton densities are backwards evolved caused
the cross section to grow unphysically slowly with energy. The upper limit of
the uncertainty is obtained when the renormalization scale is equal to half
the charm quark mass.  Here the two-loop evaluation of the strong coupling
constant gives $\alpha_s > 0.5$, too large for convergence of the perturbative
expansion. Here we will use the total charm cross section data to obtain a more
physically motivated set of parameters for charm production.

We calculate the total hadronic charm production cross section in a $pp$
collision directly using the next-to-leading order (NLO) matrix elements 
\cite{NDE} for the total partonic cross section, $\hat \sigma$, 
\begin{eqnarray}
\sigma_{AB}(\sqrt{s},m^2) =  \sum_{i,j = q, \overline q, g}
\int dx_1 \, dx_2 \,  f_i^p (x_1,\mu_F^2) \,
f_j^p(x_2,\mu_F^2) \, \widehat{\sigma}_{ij}(\hat{s},m^2,\mu_F^2,\mu_R^2) \, \, ,
\label{nlo}
\end{eqnarray}
where $m$ is the charm quark mass, $\mu_F$ is the factorization scale, $\mu_R$
is the renormalization scale, $\sqrt{s}$ is the partonic center of mass energy,
$x_1$ and $x_2$ are the fractions of the parent proton's momenta carried 
by the colliding partons and $f_i^p$ are the proton parton densities.
The NLO calculation remains the state of the art for the
total cross section; there is still no complete NNLO evaluation of the total
cross section, especially at energies where $\sqrt{s} \gg m$.  We use the
central CT10 parton density set \cite{CT10PDFs}
but will also show the variation in the cross section based on all 52 variants
of the Hessian uncertainty matrix.

Since 
Eq.~(\ref{nlo}) is independent of the heavy quark kinematics, it is typical
to take $\mu_{R,F} = m$ as the central value and vary
the two scales independently within a `fiducial' region defined by $\mu_{R,F}/m$ 
with $0.5 \le \mu_{R,F}/m \le 2$ and $0.5 \le  \mu_R/\mu_F \le
2$. In earlier work, we used the following seven sets: $\{(\mu_F/m,\mu_R/m)\}$ =
\{(1,1),  (2,2), (0.5,0.5), (1,0.5), (2,1), (0.5,1), (1,2)\} 
\cite{RVjoszo,CNV}. 
The uncertainties from the mass variation and the combined scale variations 
listed above were then added in 
quadrature. The envelope containing the resulting curves,
\begin{eqnarray}
\sigma_{\rm max} & = & \sigma_{\rm cent}
+ \sqrt{(\sigma_{\mu ,{\rm max}} - \sigma_{\rm cent})^2
+ (\sigma_{m, {\rm max}} - \sigma_{\rm cent})^2} \, \, , \label{sigmax} \\
\sigma_{\rm min} & = & \sigma_{\rm cent}
- \sqrt{(\sigma_{\mu ,{\rm min}} - \sigma_{\rm cent})^2
+ (\sigma_{m, {\rm min}} - \sigma_{\rm cent})^2} \, \, , \label{sigmin}
\end{eqnarray}
defines the uncertainty on the total cross section as a function of center of
mass energy.  Here $\sigma_{\rm cent}$ is the 
cross section calculated with the central set,
$(\mu_F/m, \mu_R/m) = (1,1)$ and $m =1.5$ GeV, while $\sigma_{i,
{\rm max}}$ and $\sigma_{i, {\rm min}}$ are the maximum and minimum values of 
the cross section for a given mass ($i=m)$ or $(\mu_F/m,\mu_R/m)$ set 
in the fiducial region ($i = \mu$).  Although Eqs.~(\ref{sigmax})
and (\ref{sigmin}) have been written for the total cross section, the
corresponding maximum and minimum values of the differential distributions 
can be written similarly \cite{CNV}.

The charm quark mass we employ in our calculations is the Particle Data Group 
(PDG) value based on lattice
determinations of the charm quark mass in the $\overline{\rm MS}$ scheme
at $\mu = m$: $m(m) = 1.27 \pm 0.09$~GeV
\cite{latticemass}.  The fiducial $c \overline c$ parameter sets used in
FONLL calculations \cite{CNV} employ a higher charm quark mass, 
$m = 1.5$~GeV.  None of these fiducial parameter sets
give a particularly good fit to the total charm data.  When $n_f = 3$ flavors 
are used, as is proper for charm production, there is a wide uncertainty band 
on $\sigma_{\rm tot}$, especially at the center-of-mass energies appropriate for
colliders, $\sqrt{s} > 200$~GeV, primarily due to unconstrained gluon densities
at low $x$ for $\mu_F/m = 0.5 \leq \mu_0/m$ where $\mu_0$ is the minimum scale
of the parton densities.  Previous calculations with lower charm quark
masses but higher scales \cite{vogtHPC} agree better with data while avoiding 
backward evolution of the gluon density at low $x$. This bias of lower masses 
with
higher scales allows us to reduce the uncertainty in the charm production
cross section.

In principle, fitting the data is somewhat problematic since we neglect unknown
next-order uncertainties.  This is particularly true for charm where the mass
is relatively small and ${\mathcal O}(\alpha_s^4)$ corrections could be large.  
Indeed, approximate NNLO calculations show that, while the
scale dependence is reduced, the $K$ factor between the approximate NNLO and the
NLO results is similar
to that between the NLO and LO calculations \cite{charmNNLO1,charmNNLO2}.  
Since a full NNLO calculation is not
yet available, we feel a fit that narrows the uncertainties at collider
energies is useful, keeping in mind that a full NNLO calculation might yield a
good fit to the data with higher masses and somewhat lower scales.

For a fixed charm quark mass, we fit the factorization and renormalization 
scale parameters to a subset of the total charm production
data.  We use part of the fixed-target data measured with incident protons at 
beam energies 
$E_{\rm beam} = 250$ \cite{e769},
360 \cite{na16}, 400 \cite{na27}, 450 \cite{na50OC}, 800 \cite{e743,e653}, 
and 920 GeV \cite{HERAB-final}.  We do not include incident pion data in the
analysis because there have been no new global analyses of the pion parton
densities since 1999 \cite{grvpi} and none of the past pion fits are compatible
with modern proton parton densities. 
The Lexan bubble chamber (LEBC) was used
in the measurements of the NA16 \cite{na16}, NA27 \cite{na27} and E743 
\cite{e743} Collaborations.  LEBC allowed direct observation of the charm
production and decay vertices.  The first two measurements were made at CERN
\cite{na16,na27} while the last was made at Fermilab \cite{e743}.  The 800
GeV E653
measurement at Fermilab used an emulsion target to measure the primary 
production vertex and at least one decay vertex contained within the emulsion
volume \cite{e653}.  
While none of these experiments had very high statistics, their results
were very clean.  The E769 Collaboration used silicon vertex detectors to 
reconstruct $D$ meson ($D^\pm$, $D^0/\overline D^0$ and $D_s^\pm$)
decays \cite{e769}.  The NA50 data at $E_{\rm beam} = 450$ GeV
were obtained by studying the lepton pair invariant mass continuum over a 
range of nuclear targets.  The continuum was assumed to be a superposition of
dimuons from the Drell-Yan process and semileptonic decays of open charm.
Since the $A$ dependence of open charm and Drell-Yan production is compatible
with a linear growth, $\sigma_{pA} = \sigma_{pp}A$, the charm cross section was
obtained from a global fit to the four targets studied (Al, Cu, Ag and W) 
\cite{na50OC}.
The data from Refs.~\cite{e769,na16,na27,e743,e653} were evaluated in the 
review of Louren\c{c}o and W\"{o}hri and adjusted to the values we employ
in our fits
using the most up-to-date branching ratios for the measured decay channels
\cite{CarlosHermine}.

We also include total cross section data at $\sqrt{s} = 200$ GeV from RHIC.
There are data from both PHENIX \cite{PHENIX200} and STAR 
\cite{STAR04,STAR11,STAR11final}.
The PHENIX measurement is based on inclusive single electron $p_T$ distributions
in $pp$ collisions in the pseudorapidity interval $|\eta| < 0.35$.  The 
`non-photonic' electrons, assumed to come from heavy flavor decays, were 
extracted from the total electron spectrum by subtracting `photonic' 
(background) sources.  The shape of the resulting 
$p_T$ distribution is described by a superposition of charm and bottom 
contributions.  The charm contribution was extrapolated to $p_T = 0$ to obtain
the total charm cross section, $0.551 ^{+0.203}_{-0.231}$ mb \cite{PHENIX200}.
The first STAR data point was extracted from d+Au collisions by two independent
measurements \cite{STAR04}.  
They directly reconstructed $D^0 \rightarrow K^+ \pi^-$ decays
with $|y| < 1$ and $0.1 < p_T < 3$ GeV.  STAR also used inclusive non-photonic 
electrons to study semileptonic decays of charm. The initial result, 
$1.4 \pm 0.2 \pm 0.4$ mb \cite{STAR04},  was significantly higher than the 
PHENIX result but compatible within systematic uncertainties.  
After a reanalysis of the non-photonic electron data
and new $D$ meson measurements, the STAR cross section reported at Quark Matter
2011 came down to
$0.949 \pm 0.365$ mb \cite{STAR11} while the final result, obtained after our
analysis was finished, is reduced to $0.797 \pm 0.210 ^{+0.208} _{-0.262}$ mb 
\cite{STAR11final}.  This does not change the central value of the scale parameters,
only reduces the one standard deviation limits.
While the final STAR result is still higher than the PHENIX cross section, the two results are
now comparable within uncertainties.

We have made five different fits to combinations of the data just described: 
the fixed-target
data \cite{e769,na16,na27,e743,e653,HERAB-final} only; adding only the PHENIX 
data \cite{PHENIX200}; adding the
PHENIX and the 2004 STAR result \cite{STAR04};
and finally, including the 2011 STAR result \cite{STAR11} 
and, subsequently, checking how much the results changed when the final 2012 
STAR point \cite{STAR11final} was added to the 
PHENIX data.  The experimental uncertainties used in the fitting were obtained 
by adding the statistical and systematic uncertainties in quadrature.

In our analysis, the total charm cross sections were calculated for a range of 
charm quark masses between 1.18 and 1.54~GeV in steps of 0.03~GeV. At each mass,
we varied $\mu_F/m$ between 0.45 and 10.65 while simultaneously varying 
$\mu_R/m$ between 0.5 and 2.9.  The step size in $\mu_F/m$ and $\mu_R/m$ was 
0.05 in both cases. The $\chi^{2}$/dof for each parameter set was evaluated by 
comparing the calculated cross sections with each of the five subsets of the 
data considered. 

The best fit values of $\mu_F/m$ and $\mu_R/m$ are rather sensitive to the charm
mass.  In general
increasing the quark mass above 1.27 GeV decreases both $\mu_F/m$ and $\mu_R/m$.
It also tends to increase the $\chi^2$/dof for each fit.  If one
plots $\chi^2$/dof for a given value of $m$ as a function of either $\mu_F/m$ 
or $\mu_R/m$ while the other scale parameter is held fixed, typical parabolic
shapes with a minimum are found.  The parabolas grow narrower as the mass 
increases.  When $m \leq 1.2$ GeV, single variable parabolas of $\chi^2$/dof
are rather broad and prefer high $\mu_{R,F}/m$ values.
For $m \geq 1.5$ GeV, the fits give $\mu_{F,R}/m \leq 1$, close
to the minimum $\mu_F$ of the parton densities and in a region where
$\alpha_s(\mu_R^2)$ is rather large.

Because the charm quark mass was assigned the value of $m=1.27 \pm 0.09$~GeV 
by the PDG, we decided to add a penalty to the $\chi^{2}$ equal to
$(m-m_{\rm PDG})^{2} / {\Delta}m_{\rm PDG}^{2}$.  With this penalty for deviations 
from the PDG value of the charm quark mass, the minimum $\chi^{2}$/dof when
varying the charm quark mass within our chosen range was found for 
$m = 1.27$~GeV, the PDG mass, for all five subsets of the data.
 
The best fit results in all cases
are given in Table~\ref{chi2table}.  The $\chi^2$/dof for each fit is also 
shown. The largest $\chi^2$/dof is obtained when the 2004 STAR point is used 
since it is high relative to the $\sqrt{s}$ dependence of the other 
measurements. We note that the values of $\mu_F/m$ found with the later STAR
results is more in line with physical arguments than that obtained with the
2004 data.

\begin{table}
\begin{center}
\begin{tabular}{|c|c|c|c|} \hline
Fitted Data & $\mu_F/m$ & $\mu_R/m$ & $\chi^2$/DOF \\ \hline
fixed-target only & $1.1^{+1.00}_{-0.40}$ & $1.6^{+0.13}_{-0.08}$ & 1.03 \\
+ PHENIX       & $1.6^{+1.53}_{-0.56}$ & $1.6^{+0.09}_{-0.13}$ & 1.03 \\
+ STAR (2004)  & $2.8^{+2.73}_{-1.35}$ & $1.6^{+0.14}_{-0.10}$ & 1.53 \\
+ STAR (2011)  & $2.1^{+2.55}_{-0.85}$ & $1.6^{+0.11}_{-0.12}$ & 1.16 \\ 
+ STAR (2012)  & $2.1^{+2.21}_{-0.79}$ & $1.6^{+0.10}_{-0.11}$ & 1.06 \\ \hline
\end{tabular}
\end{center}
\caption[]{The factorization, $\mu_F/m$, and renormalization, $\mu_R/m$, 
scale uncertainties obtained by fitting subsets of the total charm cross section
data with $m = 1.27$ GeV.
}
\label{chi2table}
\end{table}

\begin{figure}[htbp]
\begin{center}
\begin{tabular}{cc}
\includegraphics[width=0.5\textwidth]{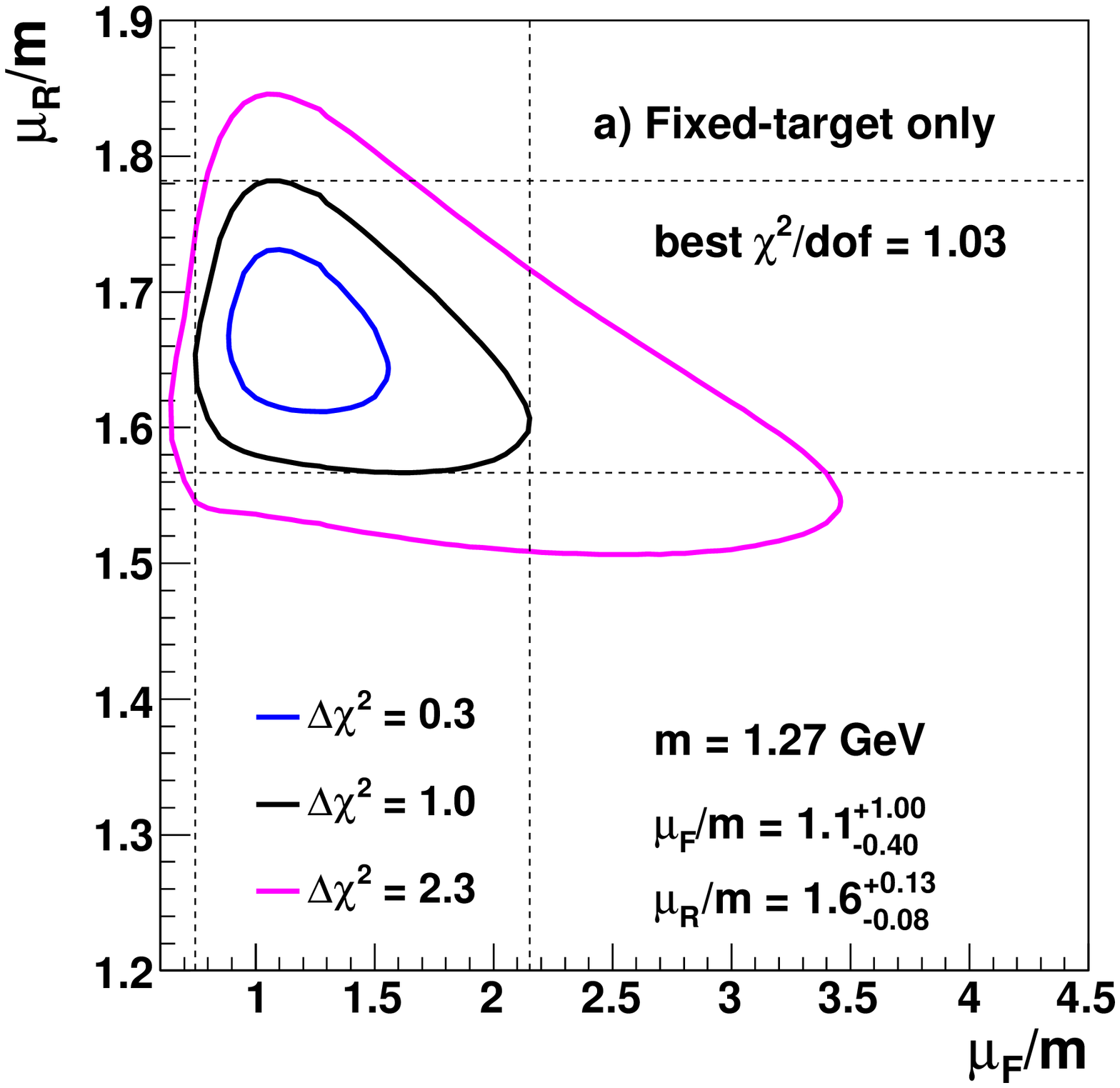} &
\includegraphics[width=0.5\textwidth]{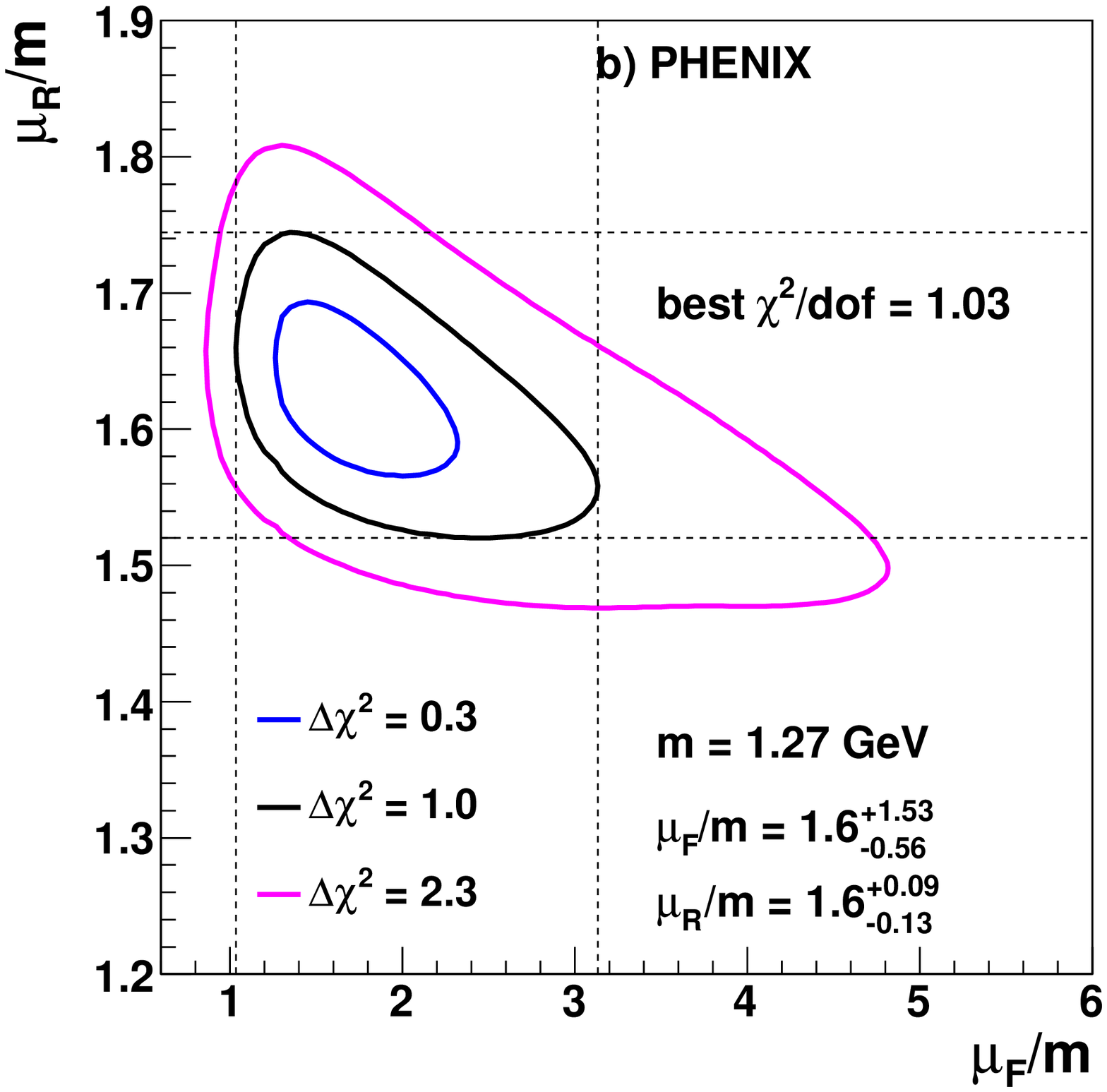} \\
\includegraphics[width=0.5\textwidth]{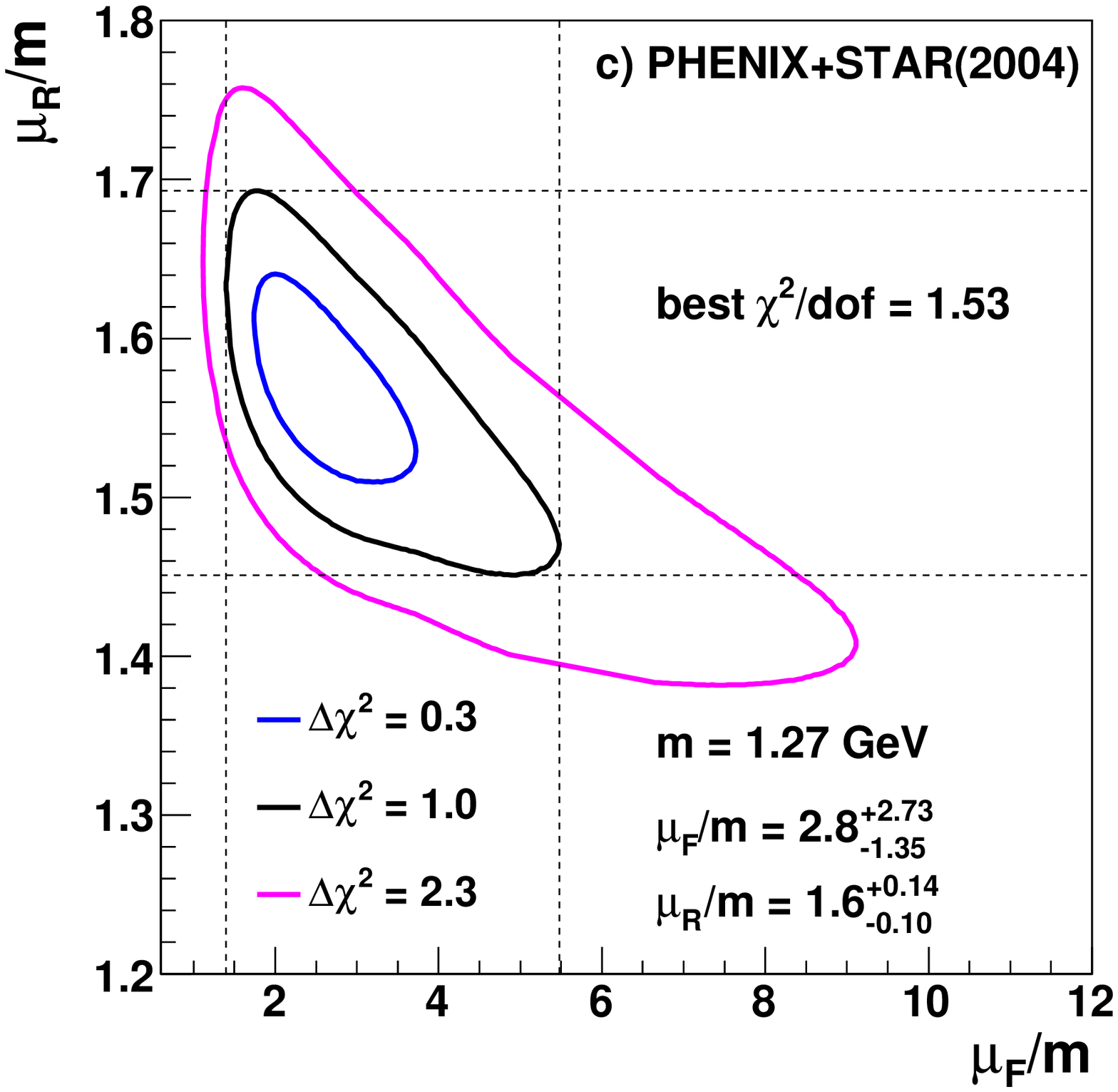} &
\includegraphics[width=0.5\textwidth]{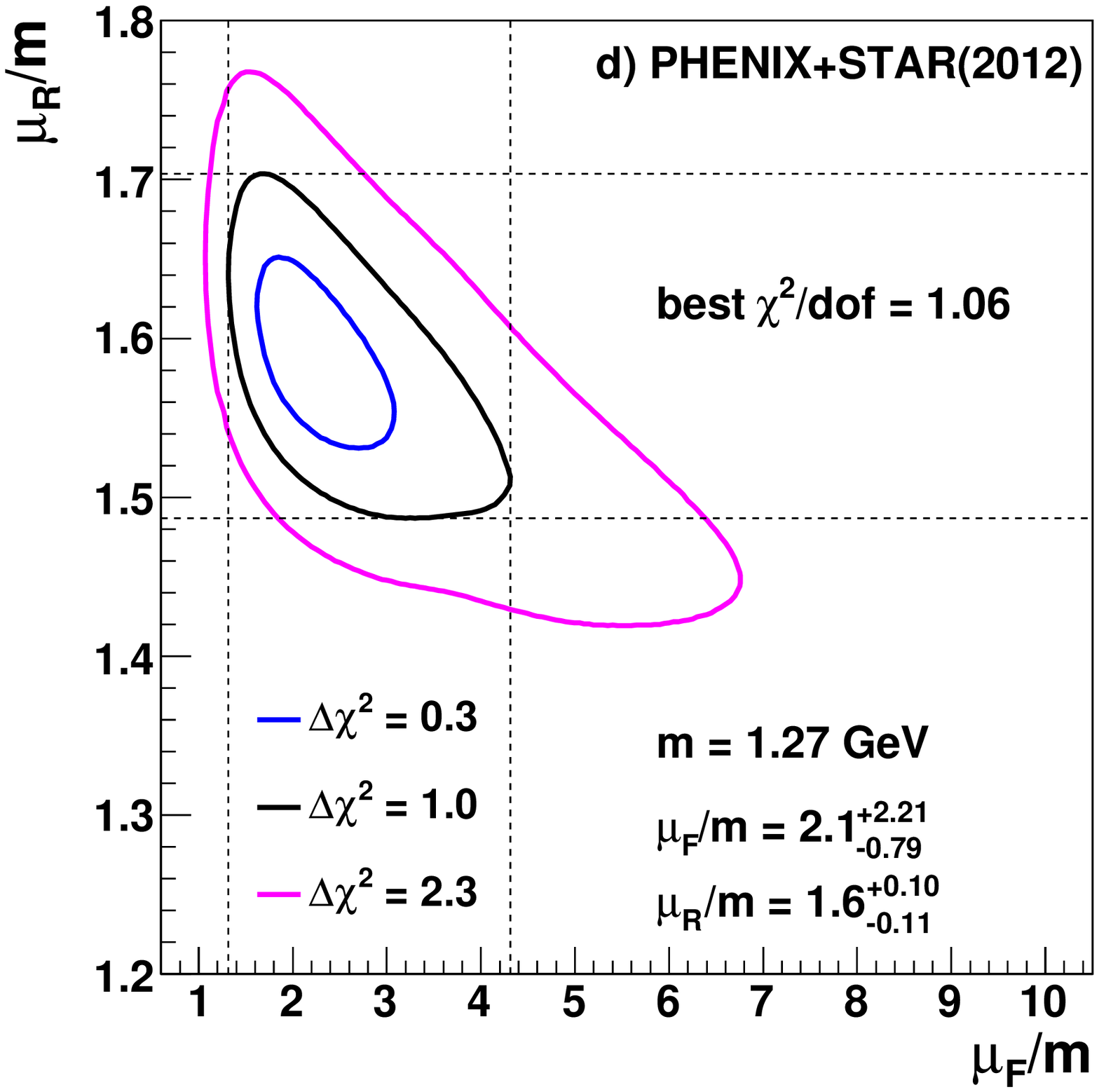}
\end{tabular}
\end{center}
\caption[]{ (Color online)
The $\chi^2$/dof contours for (a) fixed target data only, (b) 
including the PHENIX 200 GeV cross section, (c) including the STAR 
2004 cross section and (d) including the STAR 2012 cross section
but excluding the STAR 2004 cross section.  The best fit values are given for
the furthest extent of the $\Delta \chi^2 = 1$ contours.  Note that while the
$y$-axis range in the same in all four panels, the $x$-axis range varies
significantly.
}
\label{chi2fig}
\end{figure}

The uncertainties in the fitted parameters were evaluated from the $\chi^{2}$ 
distributions.
We show the $\chi^2$ fit contours in Fig.~\ref{chi2fig} for the four cases
represented in Table~\ref{chi2table}.  The $\chi^2$ contours in $\mu_F/m$ 
($x$-axis) and $\mu_R/m$ ($y$-axis) are depicted at $\Delta \chi^2 = 
0.3$, 1 and 2.3.  The one standard deviation uncertainty in the fitted value 
of $\mu_F/m$ ($\mu_R/m$) was taken as
the maximum extent of the $\Delta \chi^2 = 1$ contour along the 
$\mu_F/m$ ($\mu_R/m$) axis.  These uncertainties are included with the best fit 
parameter values in Table~\ref{chi2table}.  The one standard deviation 
uncertainty in the total cross section is the range of cross sections resulting
from all combinations of $\mu_F/m$ and $\mu_R/m$ contained within the 
$\Delta \chi^2 = 2.3$ contour.  The $\Delta \chi^2 = 0.3$ 
contour is included only to guide the eye.

Using the final STAR data point \cite{STAR11final} in the fitting results in
the same optimum parameter values for $\mu_F/m$ and $\mu_R/m$.  However, the
uncertainties on the parameter values are somewhat modified. The 
upper and lower limits on $\mu_F/m$ are reduced by 8\% and 4\%
respectively, while the limits on $\mu_R/m$ change by less than 1\%, see 
Table~\ref{chi2table}.  Since the analysis for this paper was 
completed before the latest STAR charm data release, we used the limits 
obtained with the preliminary point in our further analysis.  

Note the narrow range in $\mu_R/m$ relative to the much broader $\mu_F/m$ range,
even for fits to the fixed-target data only.  Indeed, the largest difference
in the fits to the various data sets is in the $\mu_F/m$ range.  
The $\mu_F/m$ range compatible with the data varies considerably for the
different fits, note the difference in $\mu_F/m$ ranges for the four panels
in Fig.~\ref{chi2fig}.  The 
fixed-target data probe a region of relatively large parton momentum fractions,
$x \sim 2m/\sqrt{s}$, equivalent to $0.06 < x < 0.12$ for $19.4 \leq \sqrt{s}
\leq 40$ GeV.  This range of $x$ is near the pivot point of the gluon
distribution, $xg(x,\mu_F^2)$, as a function of $x$ for a range of factorization
scales.  The fixed-target data are therefore rather insensitive to the evolution
of the gluon density as a function of $\mu_F$ so that the results skew toward
rather low values of $\mu_F/m$.

Including RHIC data in the fit introduces greater sensitivity to the low $x$
region although $x \sim 0.012$ at midrapidity is near the high $x$ edge of the 
low $x$ regime.  The PHENIX point, obtained earliest, has the lowest cross
section and thus requires a lower factorization
scale than when either of the two STAR points
are included.  The STAR cross section from 2004, more than a factor of two
larger than the PHENIX cross section, requires the largest factorization scale
of all the fits.  Note the high value, $\mu_F/m \sim 10$, required to close the
$\Delta \chi^2 = 2.3$ contour for this fit.  The newer STAR measurements 
\cite{STAR11,STAR11final}, based on reconstructed $D$ meson
decays in $pp$ collisions, rather than on d+Au collisions, gives a lower
best fit value of $\mu_F/m$ than the 2004 cross section but still higher
than either the fixed-target data only or with only the PHENIX measurement
at 200 GeV.

The value of $\mu_F/m$ is strongly dependent on the data sets used in the
fits. The uncertainty on $\mu_F/m$ is very asymmetric with a 100\% or greater
upper uncertainty relative to the best fit value.  The difference between the
lower limit of the uncertainty on $\mu_F/m$ and the best fit value is not as
large because there is
a much greater change in $xg(x,\mu_F^2)$ at lower factorization scales than when
$\mu_F \gg \mu_0$.  We finally note that the value of $\mu_F/m$ has the greatest
effect on the energy dependence of the total charm cross section.

The best fit value of $\mu_R/m$ is the same in all cases and the 
uncertainty is much smaller than for $\mu_F/m$.  These uncertainties are also
asymmetric but they typically differ by
less than 10\%, indicating that $\mu_R/m$ 
acts to fine tune the magnitude of the cross section.  Changing $\mu_R/m$
changes the total cross section by the same factor at all energies and does
not affect the energy dependence of the cross section.

\begin{figure}[htbp]
\begin{center}
\begin{tabular}{cc}
\includegraphics[width=0.5\textwidth]{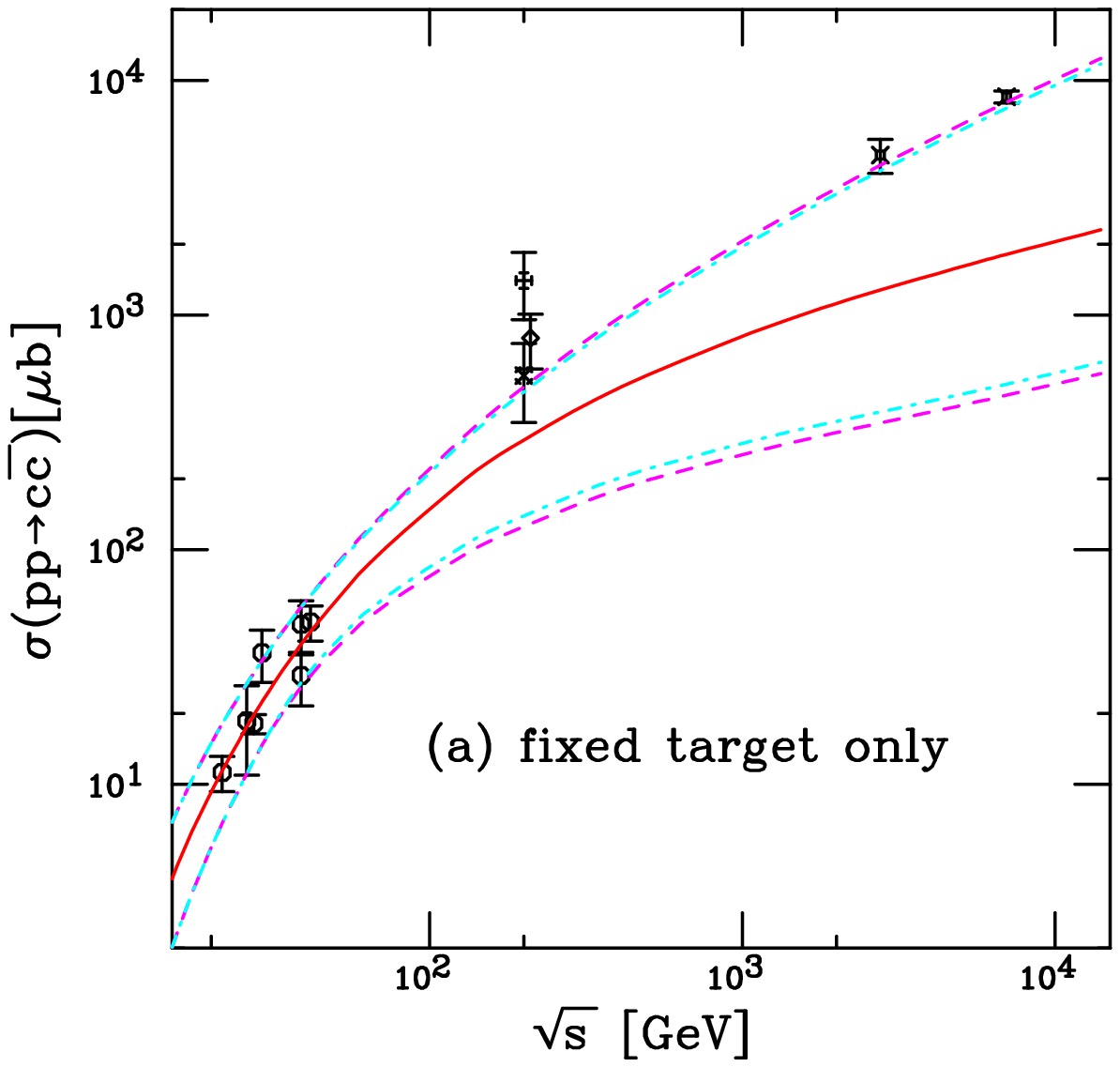} &
\includegraphics[width=0.5\textwidth]{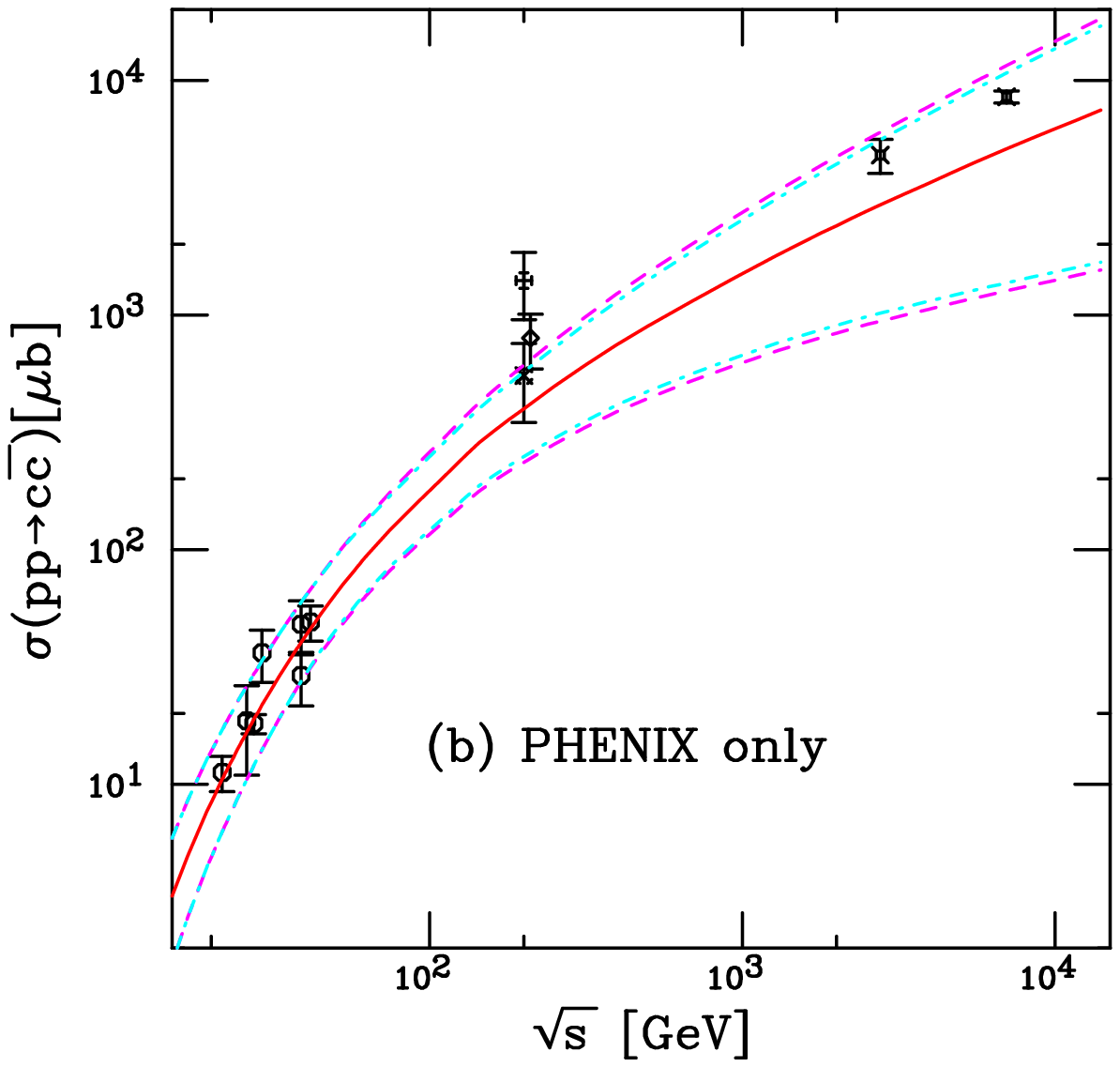} \\
\includegraphics[width=0.5\textwidth]{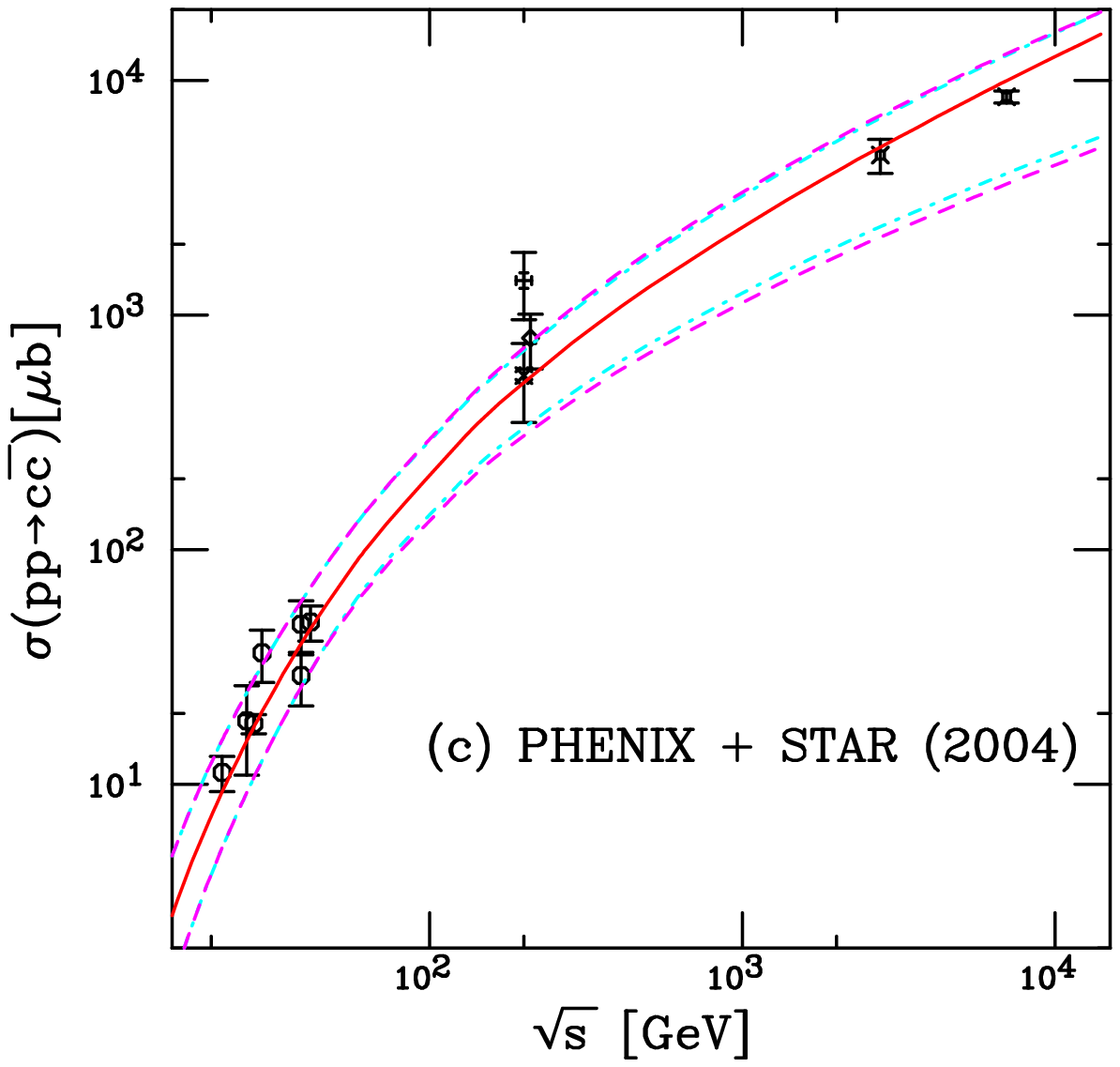} &
\includegraphics[width=0.5\textwidth]{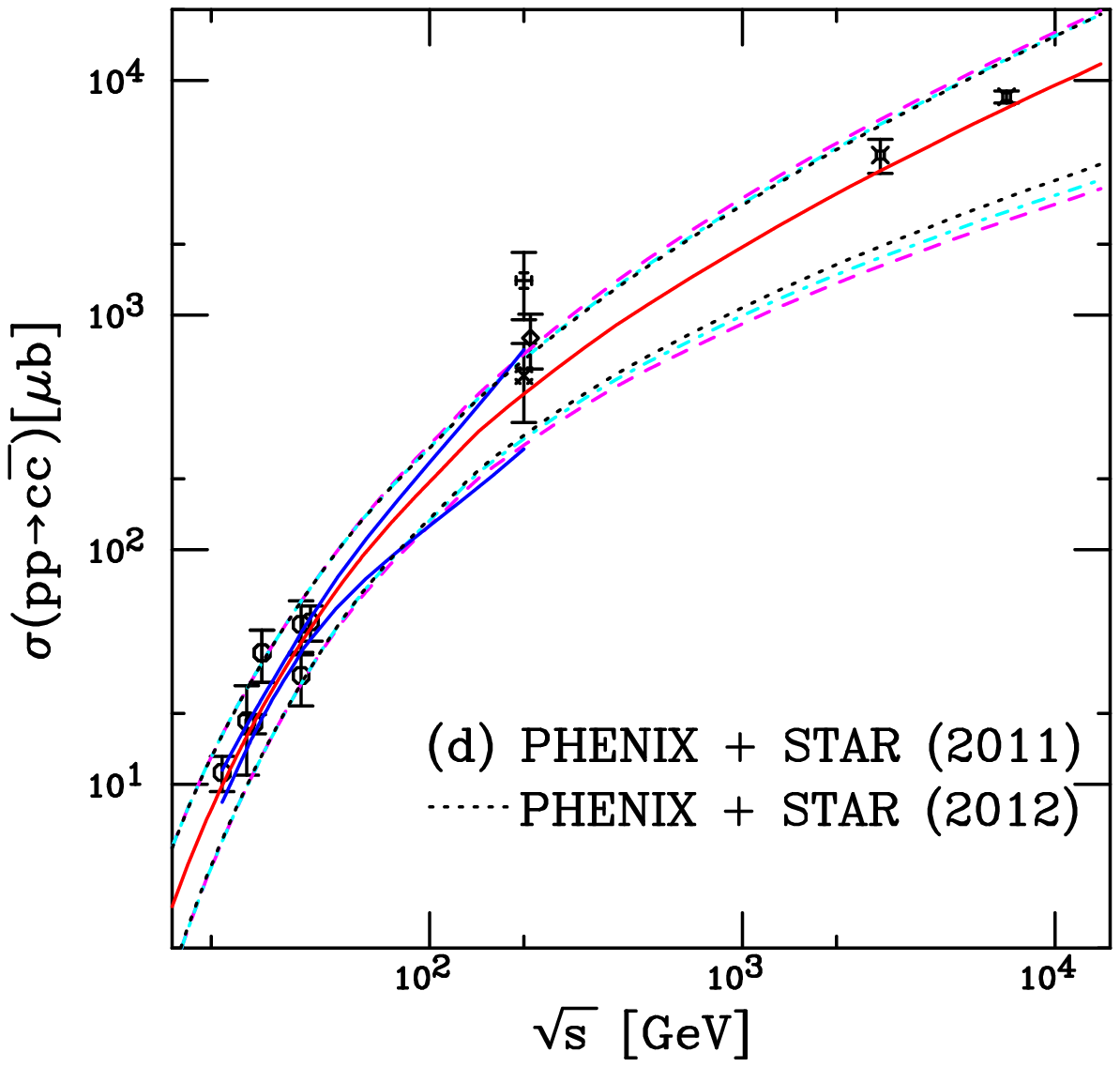}
\end{tabular}
\end{center}
\caption[]{ (Color online)
The energy dependence of the charm total cross section compared
to data for the fits to (a) fixed target data only, (b) 
including the PHENIX 200 GeV cross section, (c) including the STAR 
2004 cross section and (d) including the STAR 2011 cross section
but excluding the STAR 2004 cross section.  The best fit values are given for
the furthest extent of the $\Delta \chi^2 = 1$ contours.
The central value of the fit in each case is given by the solid red curve
while the dashed magenta curves and dot-dashed cyan curves show the extent of
the corresponding uncertainty bands.  The dashed curves outline the most extreme
limits of the band.  In (d), the dotted black curves show 
the uncertainty bands obtained with the 2012 STAR results while the solid blue 
curves in the range
$19.4 \leq \sqrt{s} \leq 200$ GeV represent the uncertainty obtained from the
extent of the $\Delta \chi^2 = 2.3$ contour in the bottom right panel of 
Fig.~\protect\ref{chi2fig}.}
\label{quadsigfig}
\end{figure}

Figure~\ref{quadsigfig} shows the resulting energy dependence of the total
charm cross section for the four different fits and the corresponding
uncertainty based on results using the one standard deviation uncertainties on
the quark mass and scale parameters.  If the central, upper and lower limits
of $\mu_{R,F}/m$ are denoted as $C$, $H$, and $L$ respectively, then the seven
sets corresponding to the `fiducial'  region are  $\{(\mu_F/m,\mu_F/m)\}$ =
\{$(C,C)$, $(H,H)$, $(L,L)$, $(C,L)$, $(L,C)$, $(C,H)$, $(H,C)$\}.  The
upper and lower limits on the PDG value of the charm quark mass are 1.36 and 
1.18 GeV.  The uncertainty band can be obtained for the best fit sets using 
Eqs.~(\ref{sigmax}) and (\ref{sigmin}).  The uncertainty bands are
shown for two cases: the regular fiducial region and including the most
extreme cases $(\mu_F/m,\mu_R/m) = (H,L)$ and $(L,H)$.  These two combinations
give the most extreme values of the cross section because the maximum
value of $\mu_F/m$ produces the fastest evolution of the parton densities while
the minimum value of $\mu_R/m$ reslts in the largest values of the strong
coupling constant with $(\mu_F/m,\mu_R/m) = (H,L)$ while the opposite is
true for  $(\mu_F/m,\mu_R/m) = (L,H)$. The difference between
the outer magenta curves, which include these extremes, and the cyan curves,
which do not, is very small.  Therefore, it is reasonable to neglect the
effect of these extremes.

Note that the fits all result in an asymmetric uncertainty band for $\sqrt{s} 
\geq 100$~GeV.  This arises because the uncertainty in the fits of $\mu_F/m$ 
is asymmetric, see Table~\ref{chi2table}, with the upper value
significantly higher than the lower.  As $\mu_F$ increases so that $\mu_F \gg
\mu_0$, the evolution of the gluon density with $\mu_F$ is reduced for
the upper limit of $\mu_F/m$.
However, the closer the lower limit of the fitted $\mu_F$ is to $\mu_0$, the
stronger the factorization scale evolution of the gluon density becomes, giving
a greater difference between the central value of $\mu_F/m$ and the lower limit
than between the central value and the upper limit.

All the fit results shown in Fig.~\ref{quadsigfig} agree equally well with the
fixed-target data.  However, the fit to the fixed-target data alone gives
the lowest cross sections at collider energies, $\sqrt{s} \geq 200$ GeV.  The 
low factorization scale values result in a slowing of the growth of the total
cross section.  The narrowest uncertainty band is obtained when the 2004 STAR
measurement is used in the fit because it requires the largest factorization
scale. Despite this, the top of the calculated uncertainty band does not even 
touch the bottom of the uncertainty on measurement.  On the other hand, the
most recent STAR measurements are compatible with the upper limits of the 
uncertainty
in the fit values.  The stronger growth in the energy dependence of the total
cross section when the RHIC data are included is due to the requirement of a
larger value of $\mu_F/m$ to fit the data, steepening the slope of the energy
dependence at large $\sqrt{s}$.  The dot-dashed and dashed curves in 
Fig.~\ref{quadsigfig}(d) were calculated with the preliminary
STAR 2011 point \cite{STAR11}.  
The black dotted curves in Fig.~\ref{quadsigfig}(d) show the
limits on the cross sections calculated using the final STAR value 
\cite{STAR11final}.  The difference in the calculated upper limits 
is 0.77\% at 200 GeV and 0.70\% at 7 TeV while the difference in the 
lower limits is $-3.36$\% at 200 GeV and $-12.65$\% at 7 TeV.  
There is a smaller difference in the
upper limits due to the relatively smaller changes in the gluon distributions
at low $x$, high $\mu_F$, compared to low $x$, low $\mu_F$.

Finally, in Fig.~\ref{quadsigfig}(d) we also show the
result for a one standard deviation uncertainty in the total cross section
obtained from the $\Delta \chi^2 = 2.3$ contour in Fig.~\ref{chi2fig}.  The 
resulting band is narrower than the uncertainty band obtained from the scale
uncertainties in the region of fixed-target data but is compatible with
the scale uncertainties at $\sqrt{s} = 200$ GeV.  Since it is based on the
energies of the data in the fits, it is not extrapolated to either higher 
or lower energies.

Lastly, we have added the 2.76 and 7 TeV total cross sections obtained by the
ALICE collaboration in $pp$ collisions \cite{ALICEpp}.  These points were not
included in our fits.  Only the fits where both PHENIX and STAR data are 
included, giving more weight to the RHIC results, have central cross section
values close to the LHC data.  While both calculations lie close to the
data, the $\chi^2$ for the LHC points is 8.67 with the 2004 STAR point and 3.9
with the latest result.

\begin{figure}[htbp]
\begin{center}
\includegraphics[width=0.475\textwidth]{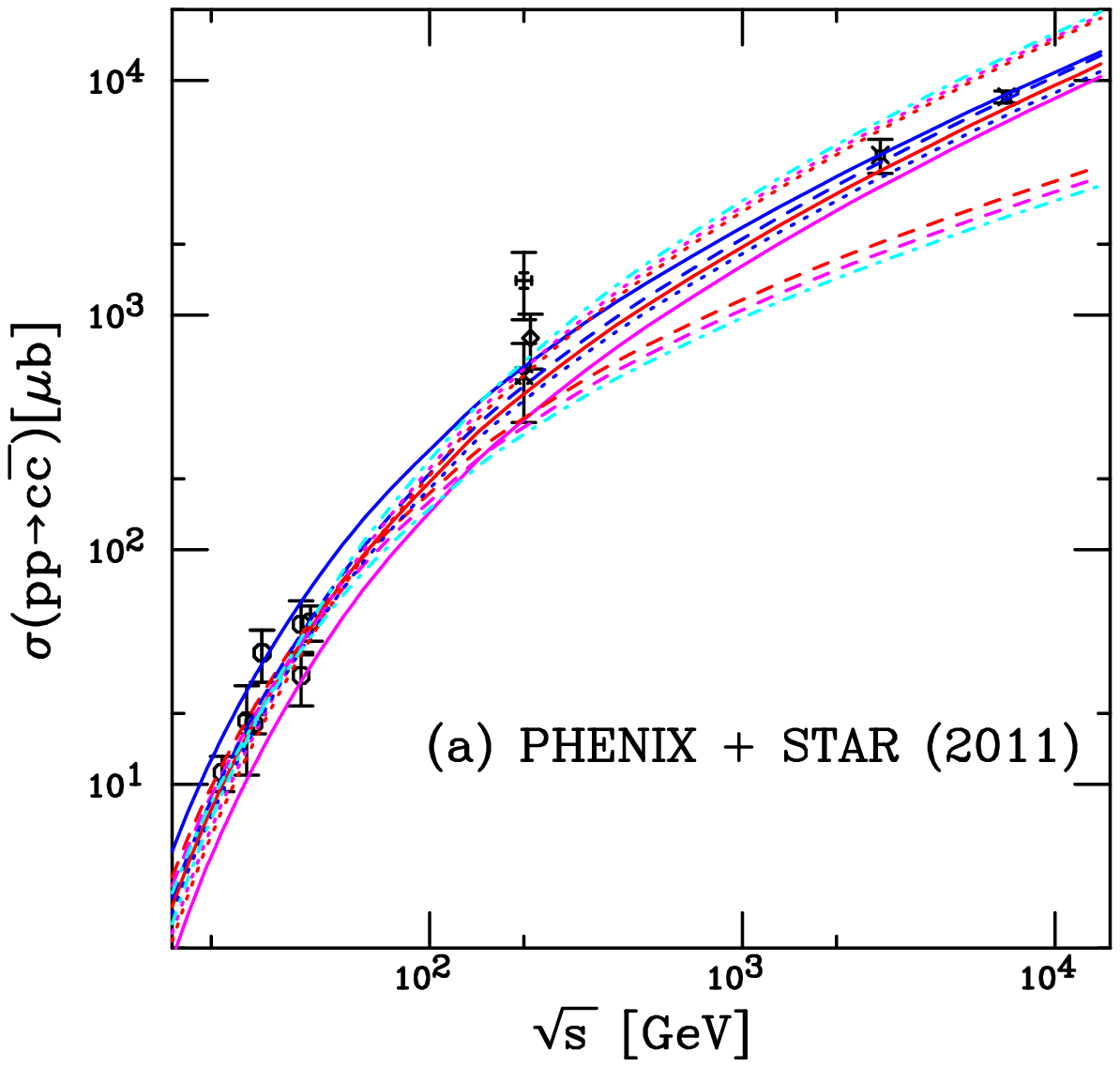}
\includegraphics[width=0.475\textwidth]{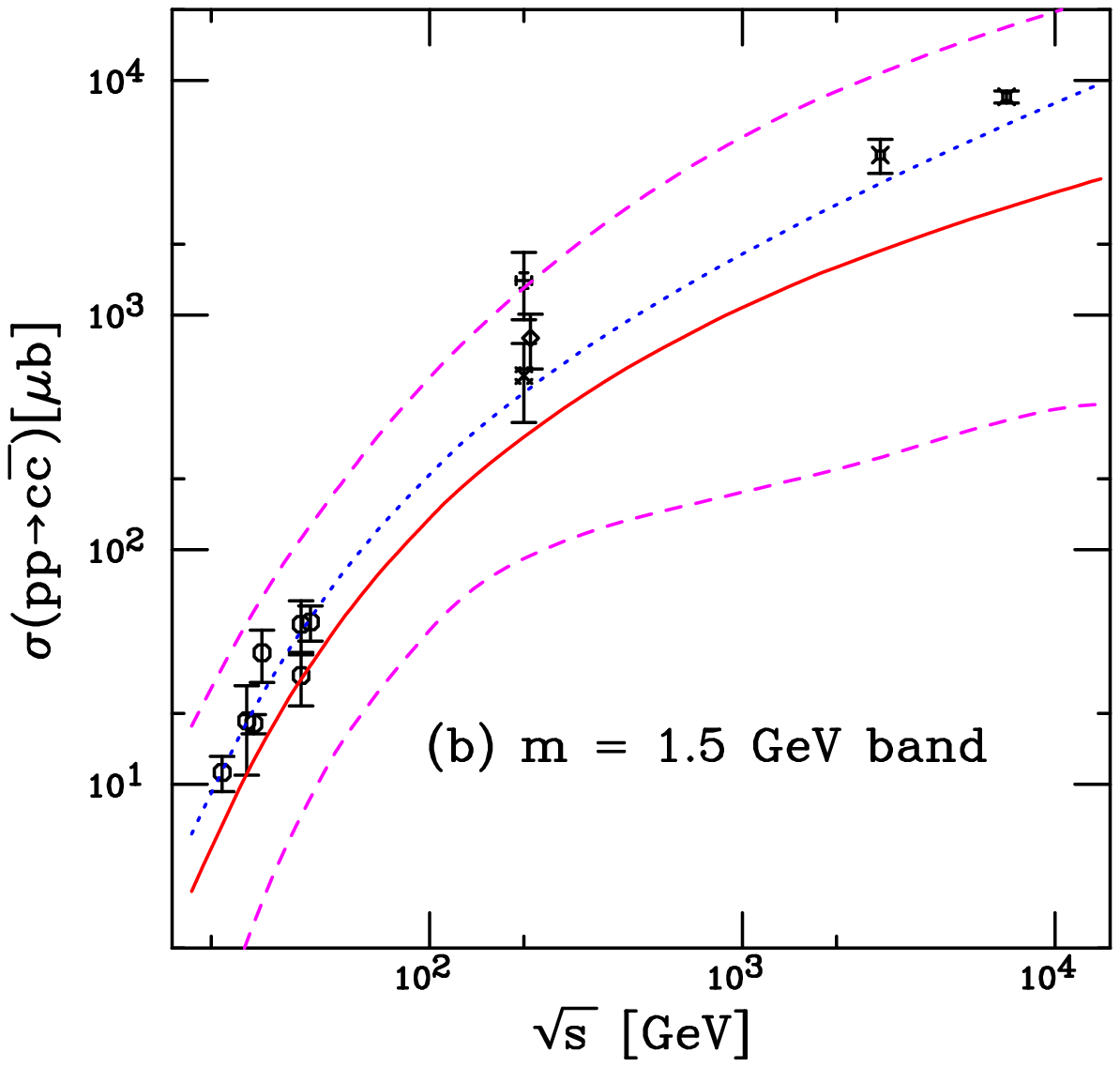}
\end{center}
\caption[]{ (Color online)
(a) The components of the uncertainty band in Fig.~\protect\ref{quadsigfig}(d). 
The central value $(m,\mu_F/m,\mu_R/m)
= (1.27 \, {\rm GeV},2.10,1.60)$ is given by the solid red curve.  The solid
blue and magenta curves outline the mass uncertainty with $(1.18 \, {\rm GeV},
2.10,1.60)$ and $(1.36\, {\rm GeV}, 2.10,1.60)$ respectively.  The dashed
curves outline the lower limits on the scale uncertainty: $(\mu_F/m,\mu_R/m) =
(2.10,1.48)$ blue; (1.25,1.60) magenta; and (1.25,1.48) red. The dotted
curves outline the upper limits on the scale uncertainty: $(\mu_F/m,\mu_R/m) =
(2.10,1.71)$ blue; (4.65,1.60) magenta; and (4.65,1.71) red.  The upper and
lower dot-dashed cyan curves correspond to $(\mu_F/m,\mu_R/m) = (4.65,1.48)$
and (1.25,1.71) respectively.  (b) The uncertainty band on the total charm
cross section obtained with the FONLL fiducial parameter set centered around
$(m,\mu_F/m,\mu_R/m) = (1.5\, {\rm GeV}, 1,1)$.  The central value is given
by the solid red curve while the limits of the uncertainty band are shown in
the dashed magenta curves.  The dotted blue curve is the result for 
$(m,\mu_F/m,\mu_R/m) = (1.2\, {\rm GeV}, 2,2)$.
}
\label{LcompRFONLLband}
\end{figure}

The individual components of the uncertainty band for the fit including the
STAR 2011 data are shown in Fig.~\ref{LcompRFONLLband}(a).
The uncertainty due to the charm quark mass (solid curves) dominates for
$\sqrt{s} < 100$ GeV where the scale uncertainty begins to become comparable.
Indeed, the scale variations at fixed-target energies are contained within 
the curves delineating the mass uncertainty.  This is very different from the
behavior of the fiducial set based on $m = 1.5$ GeV where the scale variation
dominates the uncertainty at all $\sqrt{s}$.
As the energy increases, the change in $x \sim 2m/\sqrt{s}$ due to the mass
has a much smaller effect on $xg(x,\mu_F^2)$ than the change in the evolution
of the gluon density with $\mu_F$.
At higher center of mass energies, the curves cluster according to the 
factorization scale choice.  At the top, with the largest growth as a function
of $\sqrt{s}$ are the largest values, $\mu_F/m \sim 4.65$.  The lowest value
of $\mu_F/m$, 1.25, causes the slower growth in cross section because the 
gluon distribution is increasing slowly with decreasing $x$ for this value.
The uncertainty arising from the range of $\mu_R/m$ are rather small, due to the
narrow range of fit values, and shift the
overall magnitude of the curves rather than change the slope.

The spread in the calculations can be compared to the uncertainty band obtained
using the fiducial FONLL parameter set based on $m = 1.5$ GeV 
\cite{RVjoszo,RVhp08proc}, shown in Fig.~\ref{LcompRFONLLband}(b).  
The prior by-eye fit to the
data using $m = 1.2$ GeV, $\mu_F/m = \mu_R/m = 2$ \cite{vogtHPC}
is also shown in this plot.
It gives a better representation of the data than the central FONLL parameter
set, $m = 1.5$ GeV, $\mu_F/m= \mu_R/m = 1$, and is nearly equivalent to the
best $\chi^2$/dof obtained with $m = 1.2$ GeV.  It also lies rather close
to the LHC points ($\chi^2 = 23.1$) while the central NLO cross section with 
$m = 1.5$ GeV is a factor of $\sim 3$ below these data ($\chi^2 = 142.6$).
The upper limit of the 
uncertainty band in this calculation, obtained with $\mu_R/m = 0.5$, is a factor
of $\sim 2$ larger than the fitted upper limit because $\alpha_s$ is a factor
of $\sim 1.6$ larger for $\mu_R = 0.75$ GeV than that calculated for the fit
results.  On the other hand, the lower limit is a factor of 7-8
below that calculated with the fit results at LHC energies.  
At collider energies, the slower growth in the cross section with $\sqrt{s}$ 
is due to the factorization scale $\mu_F = 0.75$ GeV ($\mu_F/m = 0.5$),
below the minimum scale of the PDFs, resulting in backward evolution of the
gluon distribution.  In the fixed-target energy range, the difference is due 
to the largest mass used, $m = 1.7$ GeV rather than 1.36 GeV.  While the 
uncertainty band obtained with the FONLL fiducial set is large enough to 
encompass all possibilities,
it is too wide to give the calculation any predictive power.  In addition,
the scale uncertainty is considerably larger than the mass
uncertainty which should not be true for the physical cross section.

\begin{figure}[htbp]
\begin{center}
\begin{tabular}{cc} 
\includegraphics[width=0.45\textwidth]{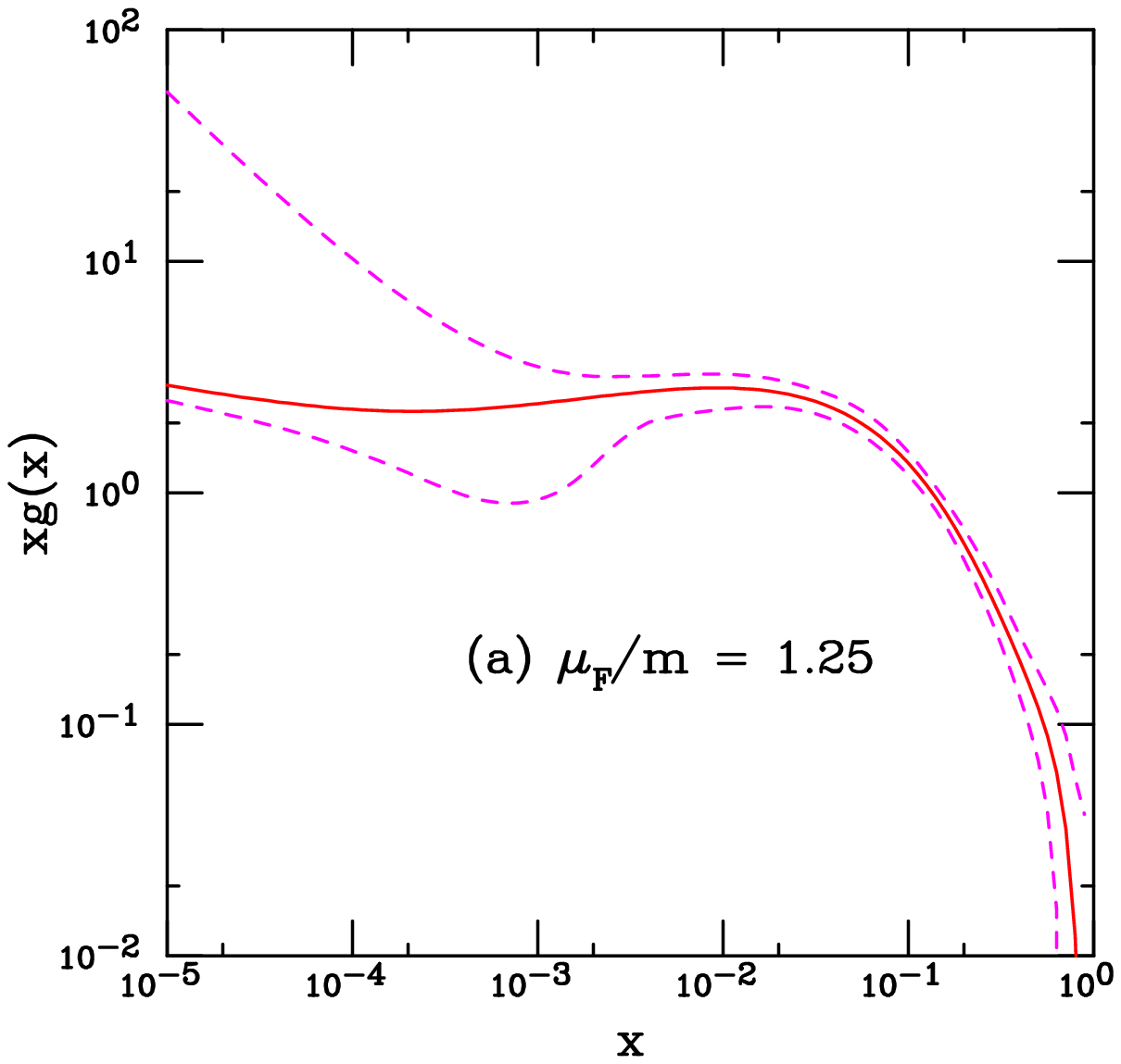} &
\includegraphics[width=0.45\textwidth]{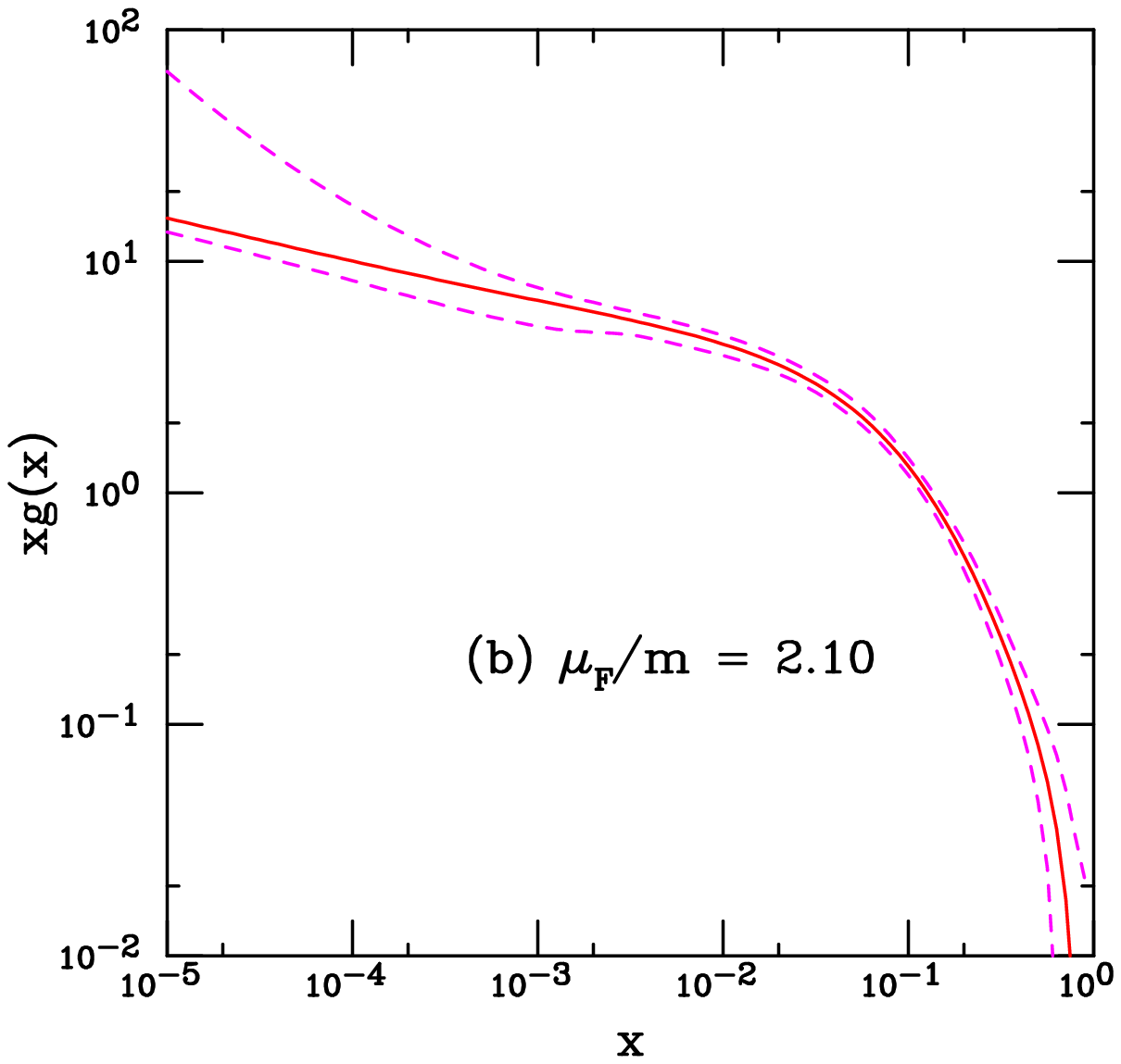} \\
\includegraphics[width=0.45\textwidth]{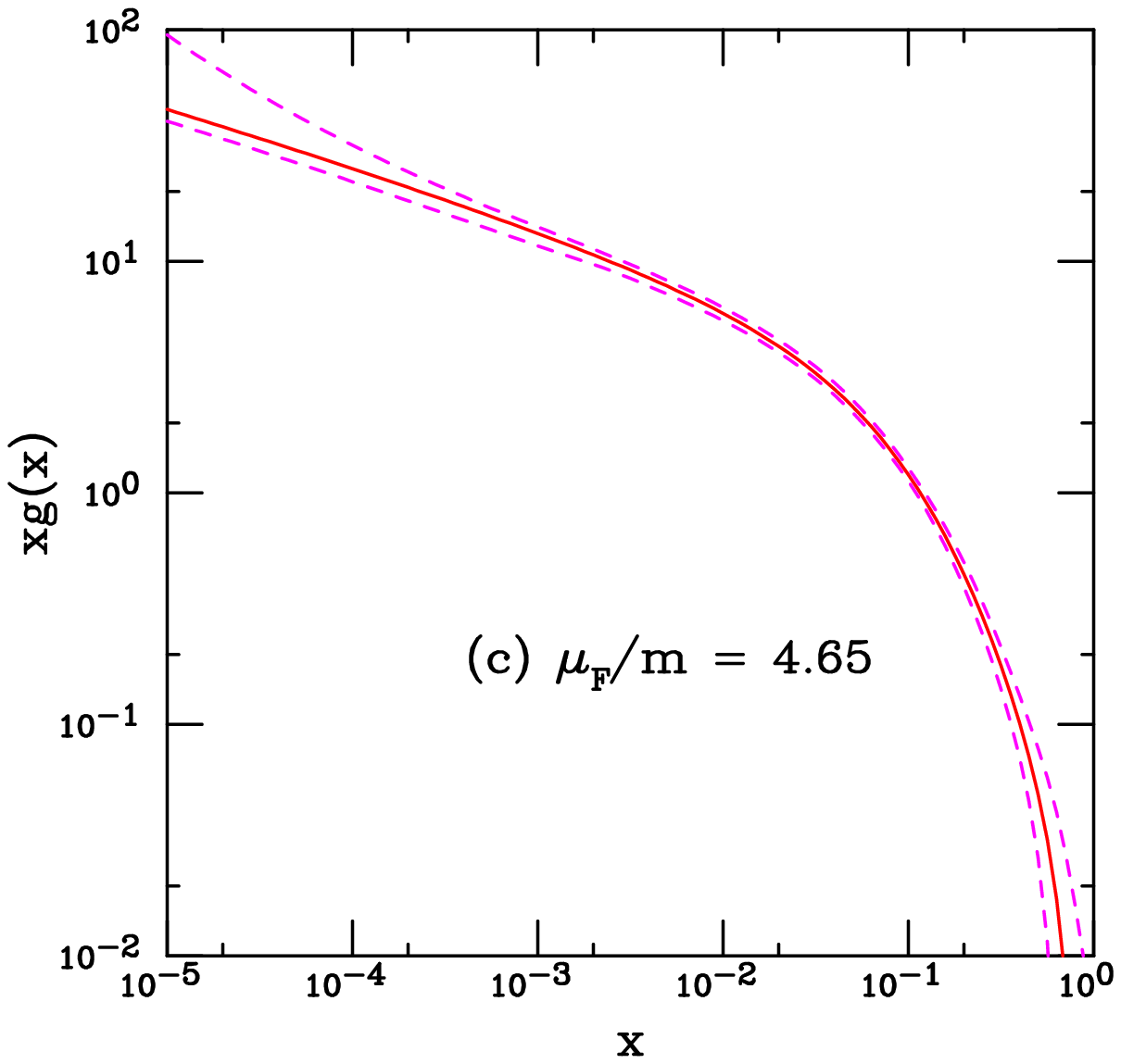} &
\includegraphics[width=0.45\textwidth]{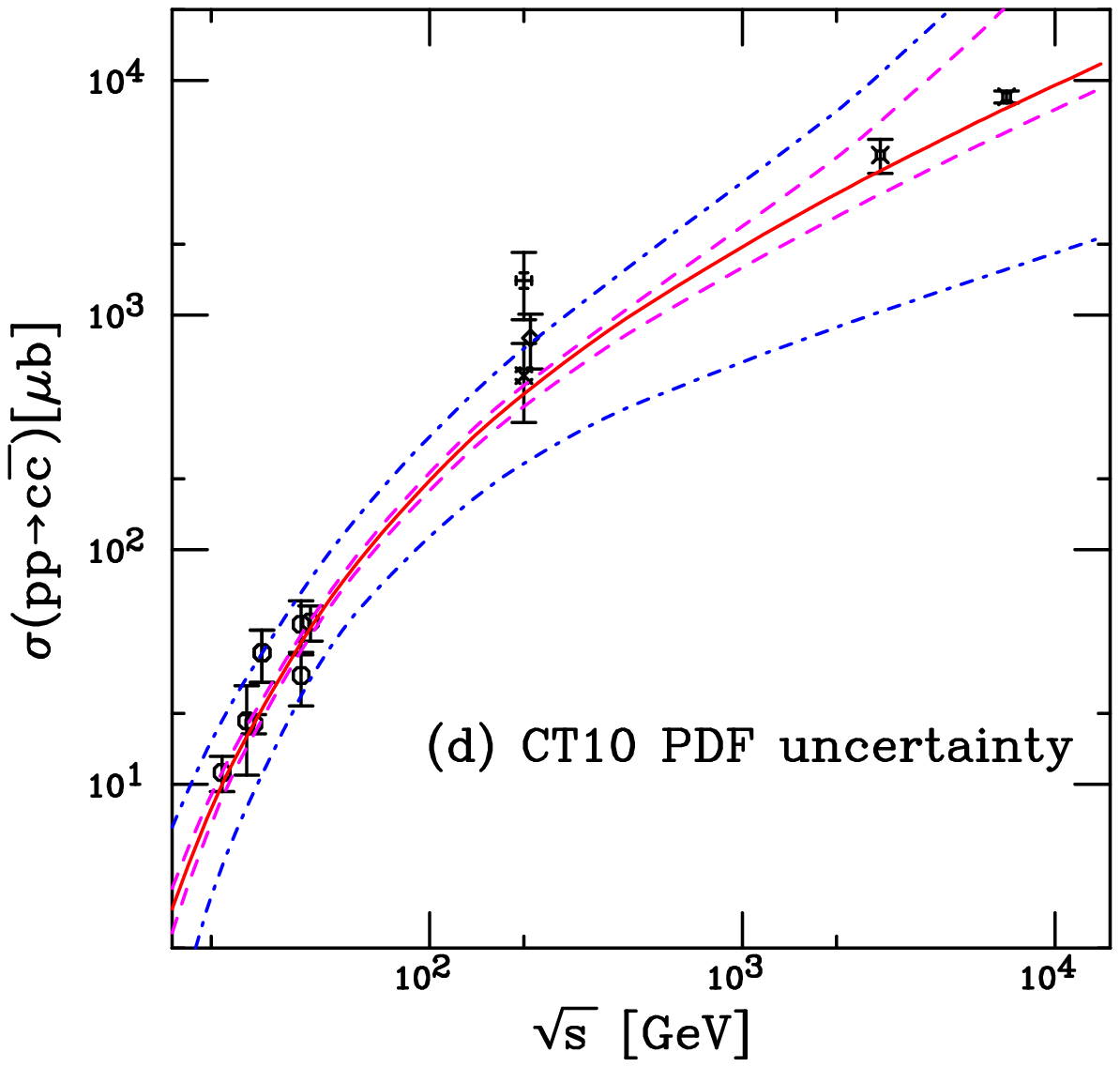} 
\end{tabular}
\end{center}
\caption[]{(Color online)
The CT10 gluon distribution, $xg(x,\mu_F)$, is shown for the
relevant values of $\mu_F/m$ for the total cross section calculation.  The
central value of the CT10 gluon distribution is given in the red solid curve
while the uncertainty band is denoted by the dashed magenta curves. The results
are shown for the lower limit of $\mu_F/m$, $\mu_F/m = 1.25$ (a); the
central value, $\mu_F/m = 2.1$ (b); and the upper limit, $\mu_F/m =
4.65$ (c).  (d) The corresponding uncertainty on the total charm cross
section due to the uncertainty in the CT10 gluon distribution is denoted by the
dashed magenta lines.  The total uncertainty due to
the mass and scale uncertainty as well as the gluon uncertainty, combined in
quadrature, is given by the dot-dashed blue curves.
}
\label{glucomp}
\end{figure}

The behavior of the gluon density corresponding to the lower, central, and
upper values of $\mu_F$ for the fits including the STAR 2011 cross section
are shown in Fig.~\ref{glucomp}.  When $\mu_F/m = 1.25$, $\mu_F$ is only $\sim
20$\% higher than $\mu_0$ so that $xg(x,\mu_F^2)$ is almost independent of $x$
for $x < 0.01$.  
As $\mu_F/m$ increases, the growth of the gluon density at
low $x$ becomes more pronounced while the uncertainty band becomes narrower
for all values of $x$.  It is clear from these results that the behavior of
$xg(x,\mu_F^2)$ determines the growth of the total cross section as a function
of center-of-mass energy.

Since the gluon density is not directly measured, the uncertainty in its 
behavior as a function of $x$ and $\mu_F$ can be important.  The largest 
uncertainty can be expected at low scales.  To quantify the uncertainty in the
gluon density that enters into our calculations, we also show the resulting
uncertainty band obtained by combining all 52 sets for the 26 eigenvectors of 
the Hessian matrix for the CT10 parton densities.  The limits on the behavior 
of $xg(x,\mu_F^2)$ show the most
variation for the lower limit of the factorization scale, see 
Fig.~\ref{glucomp}(a).  There is a sharp increase in the upper limit for
$x < 0.001$ while the lower limit on the band has a dip at the same value of
$x$.  Using the lower limit of $\mu_F/m = 1.31$ from the latest
analysis slightly reduces the variation in the band.  As $\mu_F/m$ increases, 
the growth of the gluon density at
low $x$ becomes more pronounced while the uncertainty band becomes narrower
for all values of $x$.    

The dashed curves in Fig.~\ref{glucomp}(d) show the uncertainty on the total 
charm cross section due to the variation of the proton parton density.
We have used the scale uncertainties from the fit to the 2011 STAR
result \cite{STAR11} here.  We note that while we have shown the 
uncertainty bands on the gluon density in Fig.~\ref{glucomp}(a)-(c), the cross
section uncertainty shown here includes the variations in both the quark and
gluon densities.  
In general, the uncertainty due to the parton densities
is smaller than that due to the scale choice.  
The combined effect of the mass, scale and parton density uncertainties are
given by the dot-dashed curves.  It is generally only somewhat wider than that
due to the mass and scale uncertainties alone except for the upper limit of
the band at $\sqrt{s} > 1$ TeV.

\begin{figure}[htbp]
\begin{center}
\includegraphics[width=0.475\textwidth]{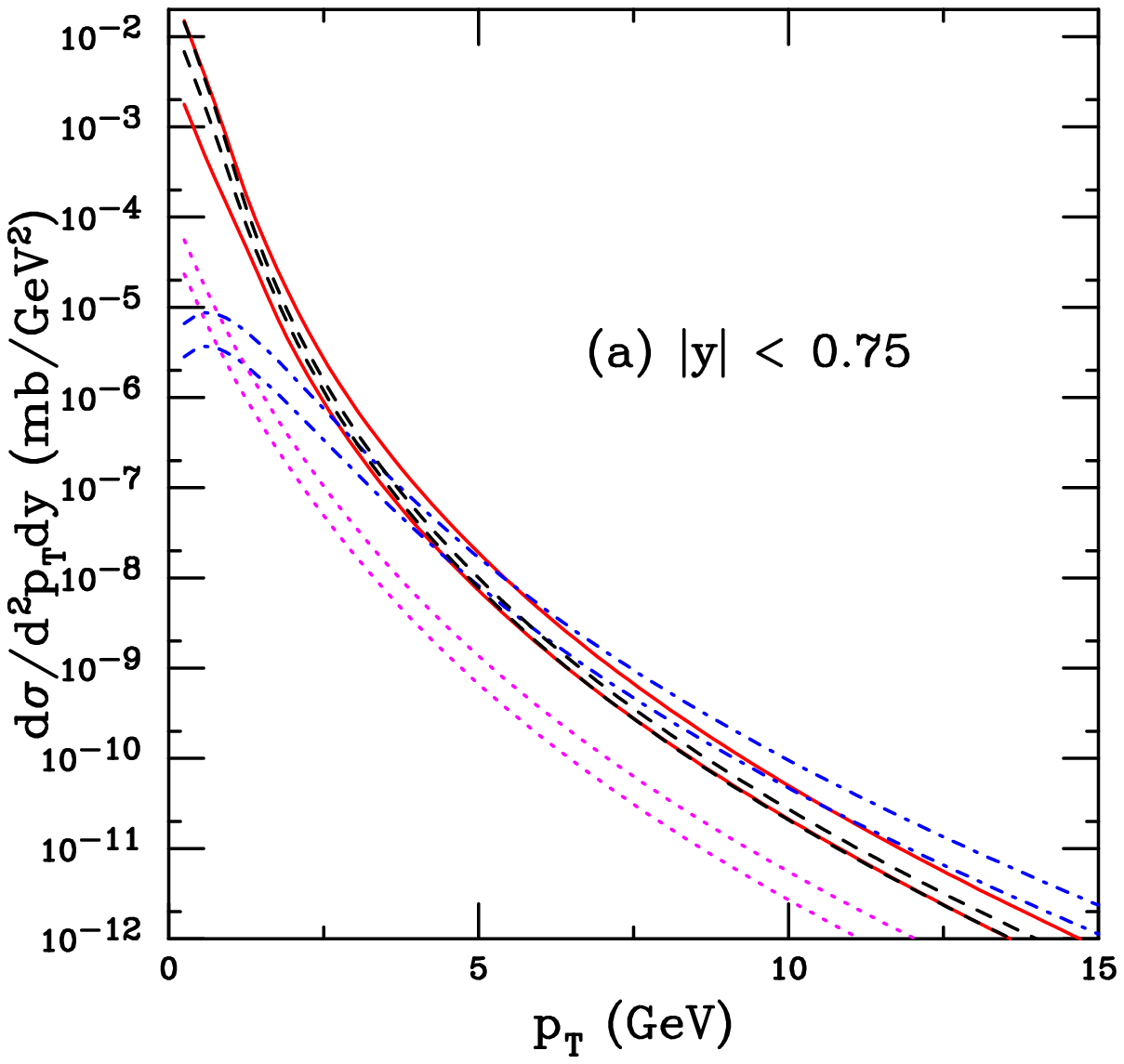}
\includegraphics[width=0.475\textwidth]{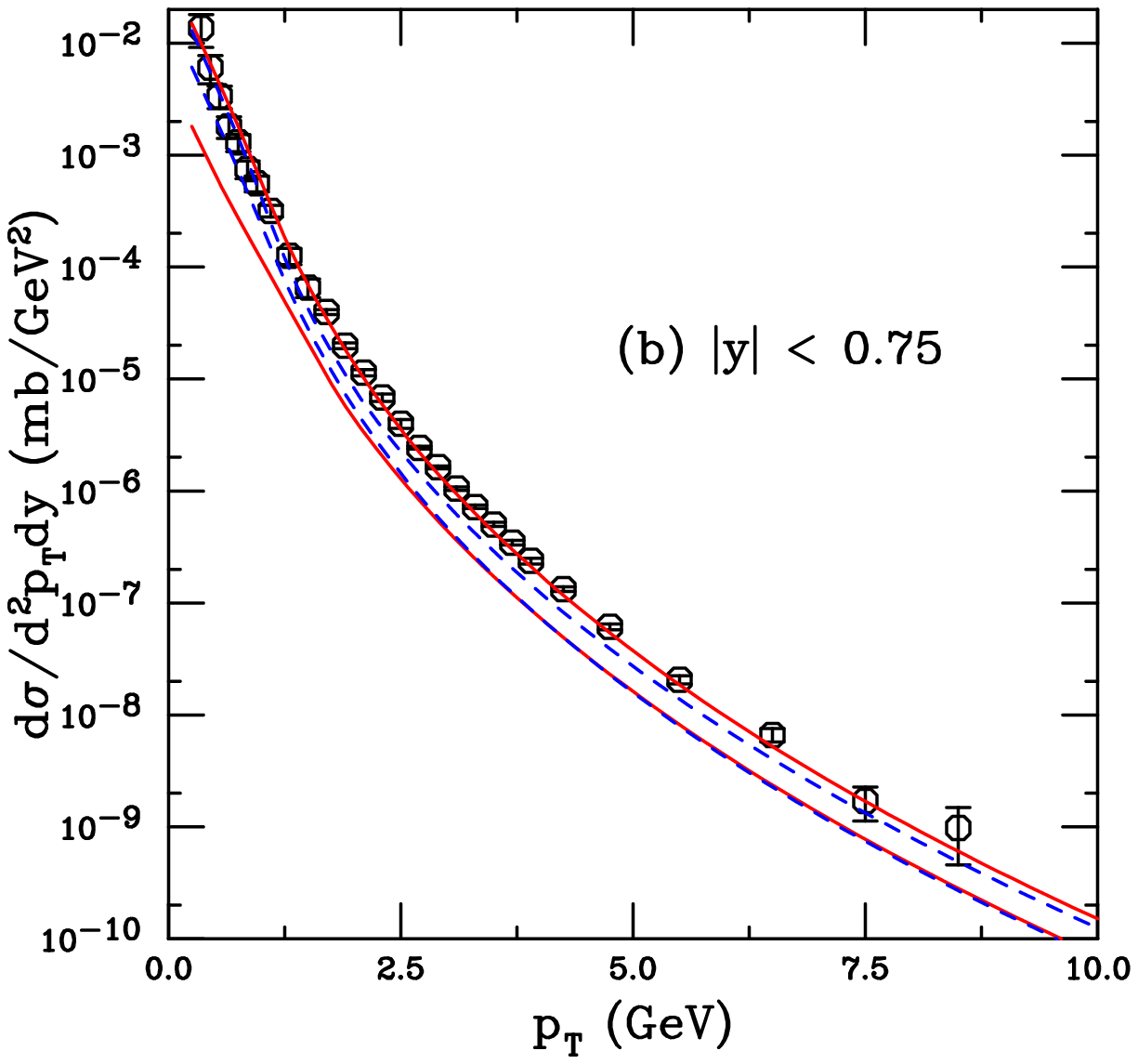}
\end{center}
\caption[]{(Color online)
(a) The components of the non-photonic 
electron spectrum: $B \rightarrow e$
(dot-dashed blue); $B \rightarrow D \rightarrow e$ (dotted magenta); 
$D \rightarrow e$ both with the FONLL parameters (solid red) and those
for $m = 1.27$ GeV in Fig.~\protect\ref{chi2fig}(d) (dashed black) at
$|y| < 0.75$ in $\sqrt{s} = 200$ GeV $pp$ collisions.    
(b) The sum of the contributions are compared with the FONLL
set for charm (solid red) and that with $m = 1.27$ GeV (dashed blue).  
The PHENIX data \protect\cite{PHENIX200} are also shown.}
\label{Dtoecomp}
\end{figure}

We now discuss how our results for the mass and scale parameters
affect the kinematic distributions of 
semileptonic decays of charm.  The state-of-the-art calculational method
for single inclusive heavy quark production and decay is the fixed-order 
next-to-leading logarithm approach (FONLL) \cite{Cacciari:1998it}.  
In addition to including the full fixed-order NLO
result~\cite{NDE,Beenakker:1990ma}, the FONLL calculation also
resums~\cite{Cacciari:1993mq} large perturbative terms proportional to
$\alpha_s^n\log^k(p_T/m)$ to all orders with next-to-leading logarithmic (NLL)
accuracy ({\it i.e.} $k=n,\,n-1$).  
The total cross sections obtained by integrating the FONLL kinematic 
distributions,
Eq.~(\ref{fonlldiff}), should be equivalent to that obtained by convoluting 
the total partonic cross sections with parton densities, Eq.~(\ref{nlo}) when
the same number of light flavors are employed.

The main difference in the two approaches that might affect the total
charm cross section is the number of active flavors.  In the
FONLL approach, the heavy quark is treated as an active light flavor at $p_T \gg
m$.  Thus the number of light flavors used to calculate $\alpha_s$ includes the
heavy quark, {\it i.e.} $n_{\rm lf} + 1$ where, for charm, $n_{\rm lf} = 3$ 
($u$, $d$ and $s$).  The same number of flavors, $n_{\rm lf} + 1$, is also 
used in the fixed-order component of the FONLL calculation for self-consistency.
Therefore, a total charm cross section calculated in the FONLL approach will
automatically be lower than the result with the same mass and scale parameters
in Eq.~(\ref{nlo}) with $n_{\rm lf} = 3$ since $\alpha_s(n_{\rm lf} = 4) < 
\alpha_s(n_{\rm lf} = 3)$.  When the renormalization scale is on the order of
the quark mass, the difference in the total cross sections at $\sqrt{s} = 
200$~GeV is less than 20\%
\cite{RVhp08proc}.  However, for $\mu_R/m < 1$, $\alpha_s(\mu_R)$ grows faster
with decreasing $\mu_R$ so that the upper limit on NLO cross section is up to
a factor of two larger than that obtained with FONLL.  On the other hand, the
lower limit, obtained with $\mu_R/m = 2$, is very similar in the two 
calculations.  Thus whether charm is treated as
a heavy ($n_{\rm lf}$) or an active ($n_{\rm lf} + 1$) flavor in the calculation
turns out to be one of the most important influences on the limits of the charm
uncertainty comparing the NLO and FONLL results.  
When the total charm cross section is
calculated with $n_{\rm lf}$ in the FONLL approach, {\it i.e.} the charm quark 
is treated as a heavy rather than an active flavor, the results 
are in agreement with the NLO calculations \cite{matteopriv}.  

The calculation of the inclusive electron spectrum
from heavy flavor decay involves three components:  the $p_T$ and rapidity
distributions of the heavy quark $Q$, calculated in perturbative QCD; 
fragmentation of the heavy quarks into heavy hadrons, $H_Q$, described by
phenomenological input extracted from $e^+e^-$ data; and the decay of
$H_Q$ into electrons according to spectra available from other
measurements, schematically written as \cite{CNV}
\begin{eqnarray}
\frac{E d^3\sigma(e)}{dp^3} &=& \frac{E_Q d^3\sigma(Q)}{dp^3_Q} \otimes
D(Q\to H_Q) \otimes f(H_Q \to e)
\label{fonlldiff}
\end{eqnarray}
where the symbol $\otimes$ denotes a generic convolution.  The fragmentation
of quarks into hadrons is denoted $D(Q \to H_Q)$.
The electron decay
spectrum, $f(H_Q \to e)$, accounts for the semileptonic branching ratios.  

Figure~\ref{Dtoecomp} shows the lepton spectra arising from semileptonic
heavy flavor decays at $\sqrt{s} = 200$ GeV, all calculated in the FONLL
approach.  The $B \rightarrow e$ and
$B \rightarrow D \rightarrow e$ bands, as well as the red $D\rightarrow e$
band, are calculated with the same fiducial 
set of parameters as in Ref.~\cite{CNV}.  The black dashed curves represent
the $D \rightarrow e$ band calculated for our best fit parameter set,
including both the PHENIX and 2011 STAR total cross sections.  The new 
$D \rightarrow e$ calculation is much
narrower.  It lies completely within the uncertainty band based on the fiducial
parameter set with the central value of the charm quark mass fixed at 
$m = 1.5$ GeV. At low $p_T$,  $p_T < 2.5$ GeV, the new $D \rightarrow e$ band is
near the top of the fiducial FONLL $D \rightarrow e$ band while for $p_T \geq
7.5$ GeV, the new set gives a result near the bottom of the fiducial $D
\rightarrow e$ band.  The transition from dominance of
the electron spectra by charm decays to bottom decays happens at lower $p_T$
with the $m = 1.27$ GeV set.

The right-hand side of Fig.~\ref{Dtoecomp} shows the sum of the $D \rightarrow
e$, $B\rightarrow e$ and $B \rightarrow D\rightarrow e$ for the two cases.
The PHENIX non-photonic electron data are compatible with the top of
the sum of the uncertainty bands with the fiducial FONLL set.  However, 
the data now lie somewhat above the band obtained when the best fit $D
\rightarrow e$ contribution to charm production
replaces the fiducial contribution.  The agreement with the data is worst
at intermediate values of $p_T$, $p_T \sim 5$~GeV, where the $c$ and $b$
contributions to the electron spectra are nearly equal.
This discrepancy is not so surprising because 
the best fit parameters were obtained using the NLO QCD calculation with 
$n_{\rm lf} = 3$ flavors while the FONLL calculation, both in the fixed-flavor
scheme and in the full FONLL result, uses $n_{\rm lf} + 1 = 4$ since
the heavy quark is treated
as an active light flavor over all $p_T$.  The value of 
$\alpha_s(\mu_R^2)$ obtained with four light flavors is smaller than that
obtained with three light flavors, even for the same value of $\mu_R$.  More
importantly, the range of factorization scales is larger for our best fit
case, $1.59 < \mu_F < 5.9$~GeV ($1.25 \leq \mu_F/m \leq 4.65$) instead of 
$0.75 < \mu_F < 3$~GeV ($0.5 \leq \mu_F/m \leq 2$) for the 
fiducial set.  The higher factorization scales cause the $p_T$ distribution to 
fall off faster with $p_T$ in the best fit case.  
The difference is apparent already
in the charm quark $p_T$ distributions and would be enhanced for the
semileptonic charm decays to electrons since the decay leptons carry only
$\sim 30$\% of the parent hadron $p_T$ \cite{PHENIXnote}

\begin{figure}[htbp]
\begin{center}
\begin{tabular}{cc}
\includegraphics[width=0.475\textwidth]{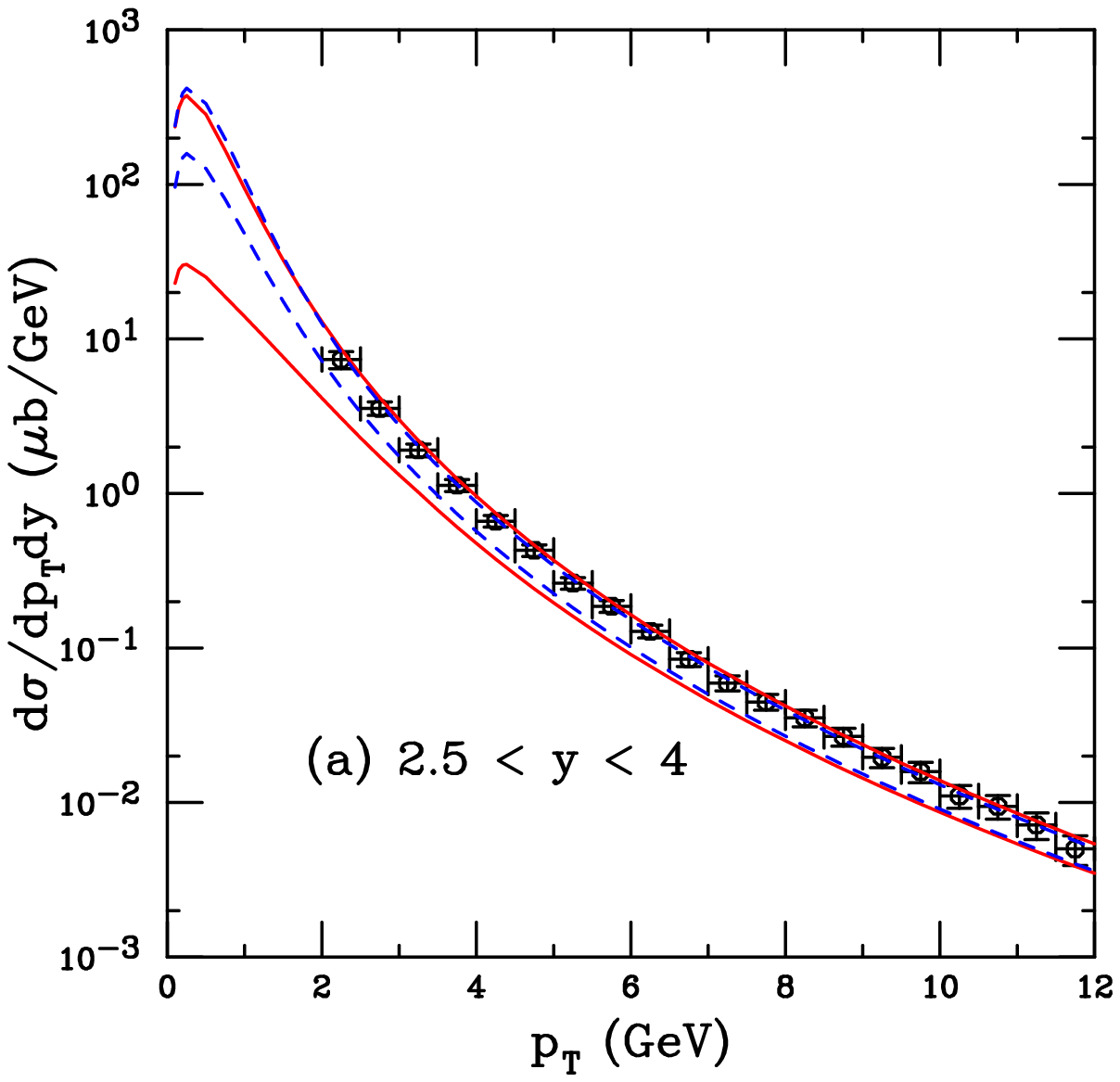} & 
\includegraphics[width=0.475\textwidth]{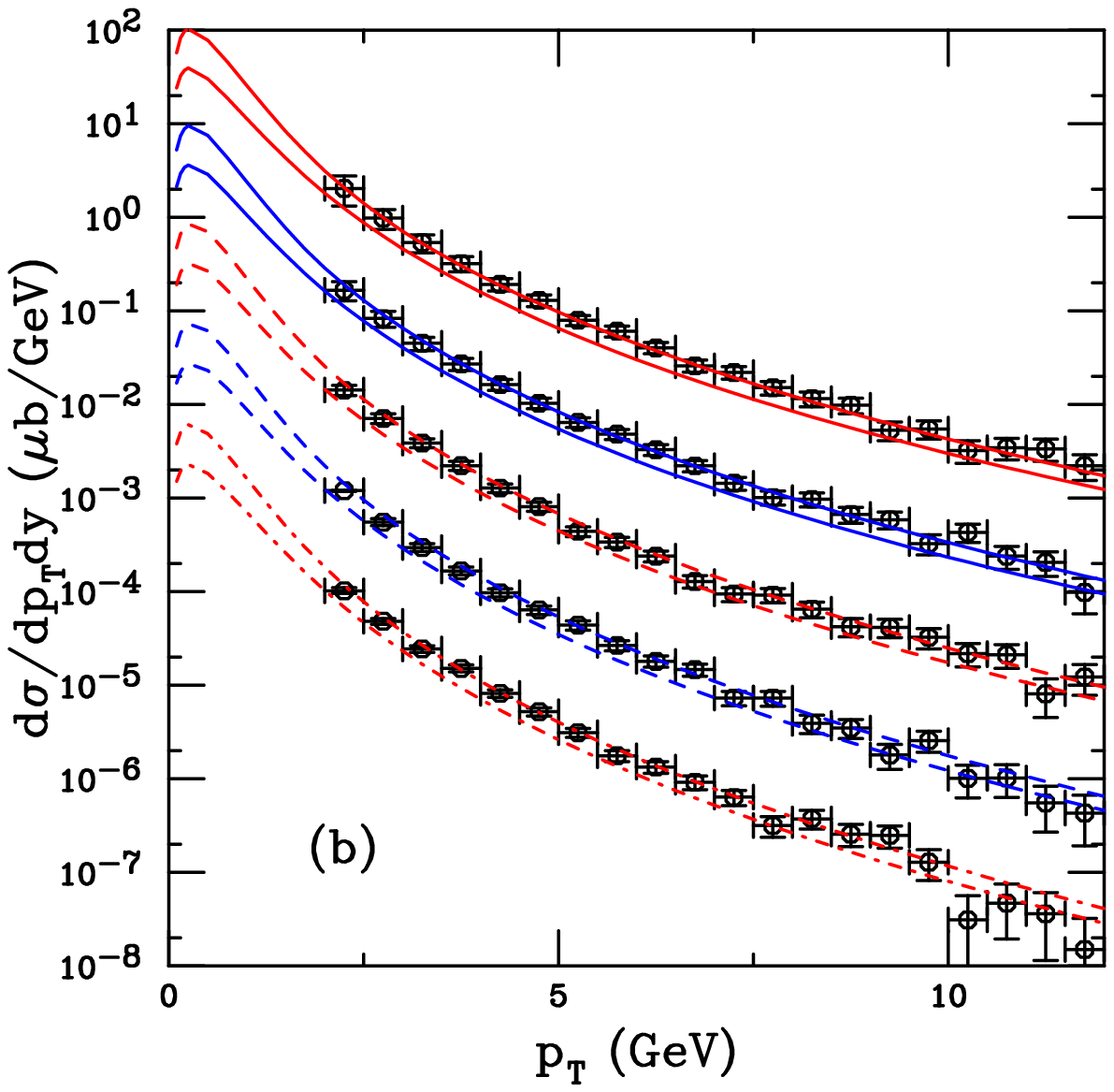} \\
\includegraphics[width=0.475\textwidth]{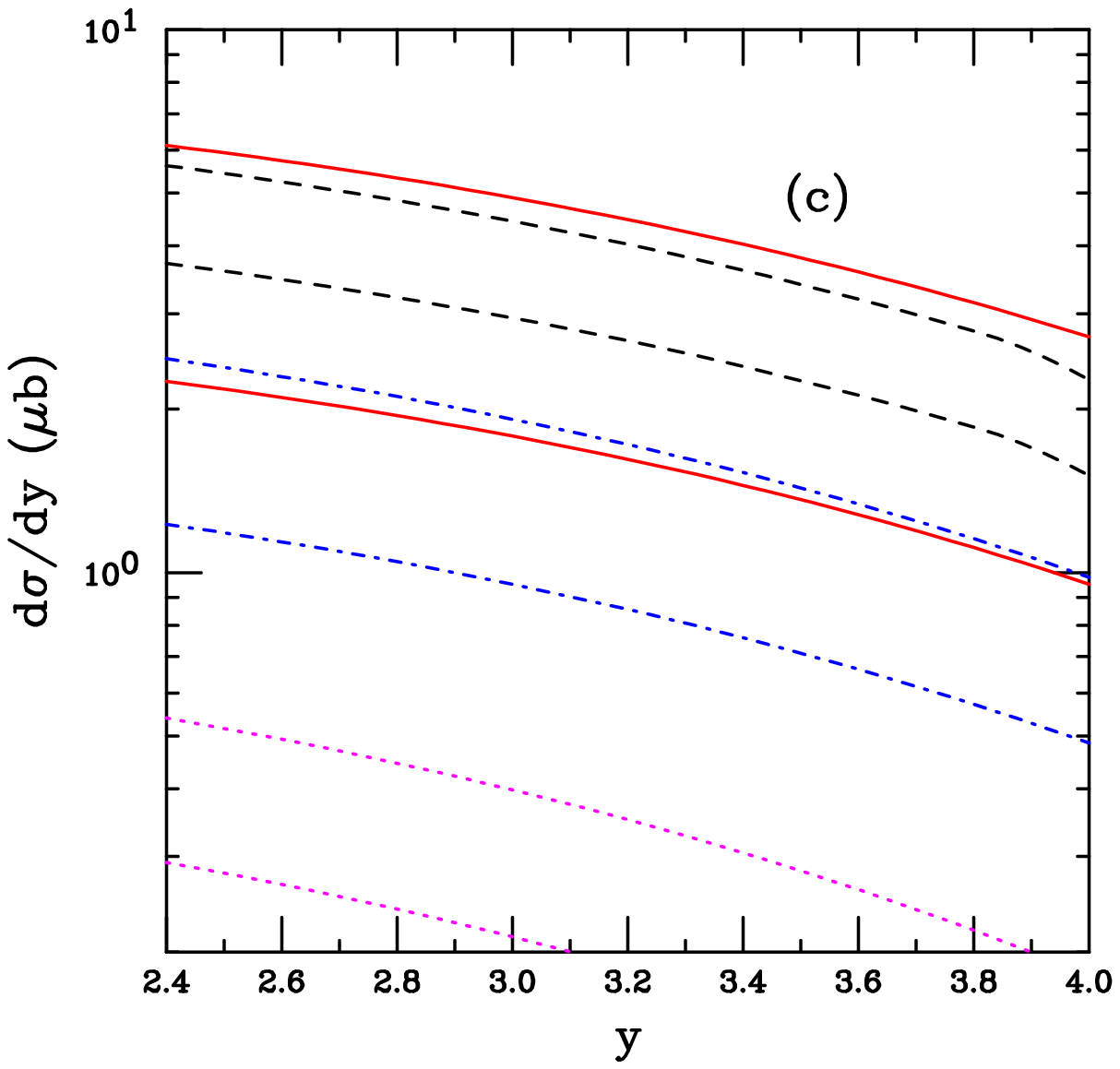} & 
\includegraphics[width=0.475\textwidth]{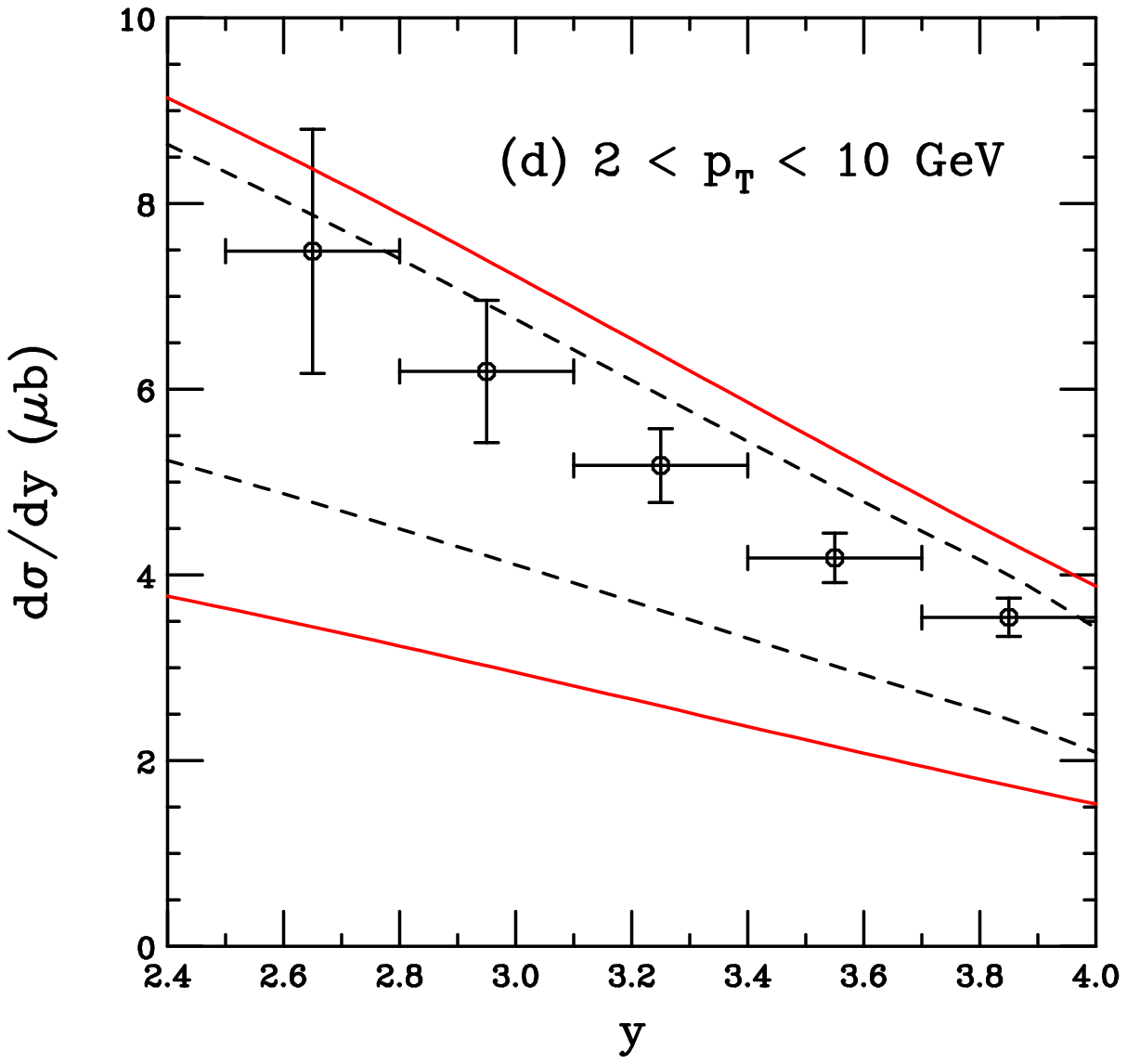} \\
\end{tabular}
\end{center}
\caption[]{(Color online) Our calculations are compared with the ALICE inclusive
single muon data from heavy flavor decays \protect\cite{ALICEmuons} at 
$\sqrt{s} = 7$ TeV.  (a) Comparison of the single lepton $p_T$ 
distributions in the rapidity interval $2.5 < y < 4$ at $\sqrt{s} = 7$ TeV 
calculated with the FONLL set for charm (solid red) and the fitted set with 
$m = 1.27$ GeV (dashed black).  (b) The contributions to the $p_T$ distributions
in (a) divided into rapidity bins, from top to bottom: $2.5 < y < 2.8$ (solid
red); $2.8 < y < 3.1$ (solid blue); $3.1 < y < 3.4$ (dashed red); $3.4 < y <
3.7$ (dashed blue); and $3.7 < y < 4$ (dot-dashed red).  The top curves are
shown at their calculated value, the others are scaled down by successive 
factors of 10 to separate them. (c)  The components of the rapidity 
distribution at $\sqrt{s} = 7$ TeV with $2 \leq p_T \leq 10$ GeV,  
$B \rightarrow \mu$
(dot-dashed blue); $B \rightarrow D \rightarrow \mu$ (dotted magenta); 
$D \rightarrow \mu$ both with the FONLL parameters (solid red) and those
for $m = 1.27$ GeV in Fig.~\protect\ref{chi2fig}(d) (dashed black).  (d)
The sum of the contributions are compared with the FONLL
set for charm (solid red) and that with $m = 1.27$ GeV (dashed black).  
}
\label{DtoecompALICE}
\end{figure}

Figure~\ref{DtoecompALICE} compares our calculations with the ALICE single
muon data in the forward rapidity region, $2.5 < y < 4$ \cite{ALICEmuons}. 
The data are given for $2 < p_T < 12$ GeV, both over the full rapidity region, 
Fig.~\ref{DtoecompALICE}(a), and separated into five rapidity bins, each 0.3
units wide, Fig.~\ref{DtoecompALICE}(b).  The calculations with both the 
fiducial charm parameter set\footnotetext{For a complete discussion of LHC
predictions using the fiducial FONLL parameter set, see Ref.~\cite{Matteonew}.}
(solid) and our charm fit (dashed) are compared to the data in 
Fig.~\ref{DtoecompALICE}(a).  The two bands are indistinguishable for $p_T > 5$
GeV.  Therefore, for clarity, we compare the muon $p_T$ 
distributions in the narrow rapidity bins to only our calculations with the
mass and scale parameters from the charm fit.  The calculations agree well with
the measurements over the entire $p_T$ range.

In Fig.~\ref{DtoecompALICE}(c) and (d) we present the results as a function
of rapidity integrated over the same $p_T$ range as the data, $2 \leq p_T \leq
10$ GeV.  Figure~\ref{DtoecompALICE}(c) shows the upper and lower limits of 
the FONLL calculations of $B \rightarrow
\mu$ and $B \rightarrow D \rightarrow \mu$ in the dot-dashed and
dotted curves respectively.  The FONLL $D \rightarrow \mu$ uncertainty
bands with the fiducial charm parameter set are shown by the solid curves
while the dashed curves are calculated with the charm fit parameters.  The
sum of the heavy flavor decay contributions to the rapidity distribution are
compared on a linear scale in Fig.~\ref{DtoecompALICE}(d).  The 
$p_T$-integrated ALICE data agree well with both calculations.  The
results with the fitted charm parameter set narrow the uncertainty band
without sacrificing consistency with the measured data.

\begin{figure}[htbp]
\begin{center}
\begin{tabular}{ccc}
\includegraphics[width=0.33\textwidth]{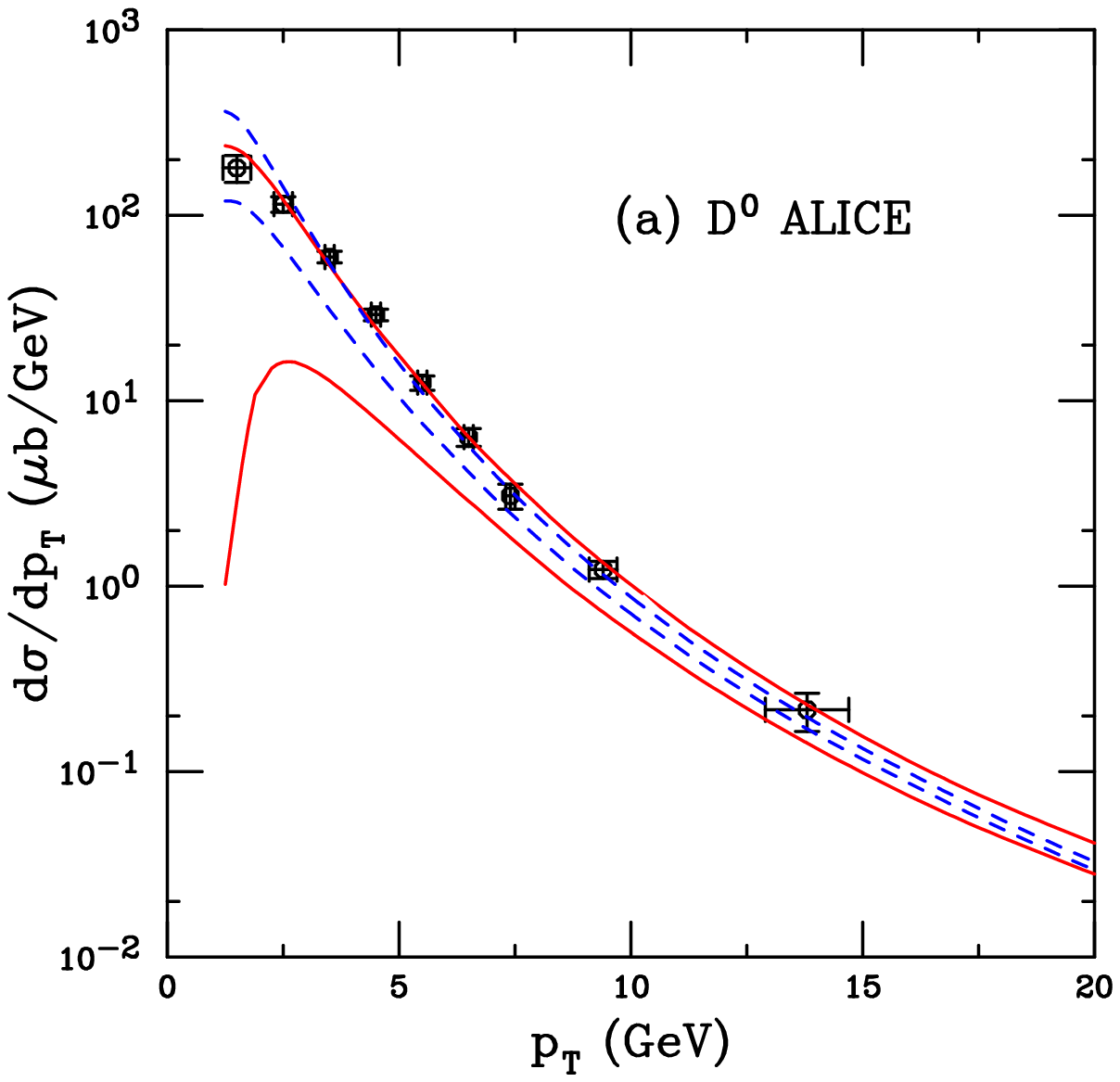} &
\includegraphics[width=0.33\textwidth]{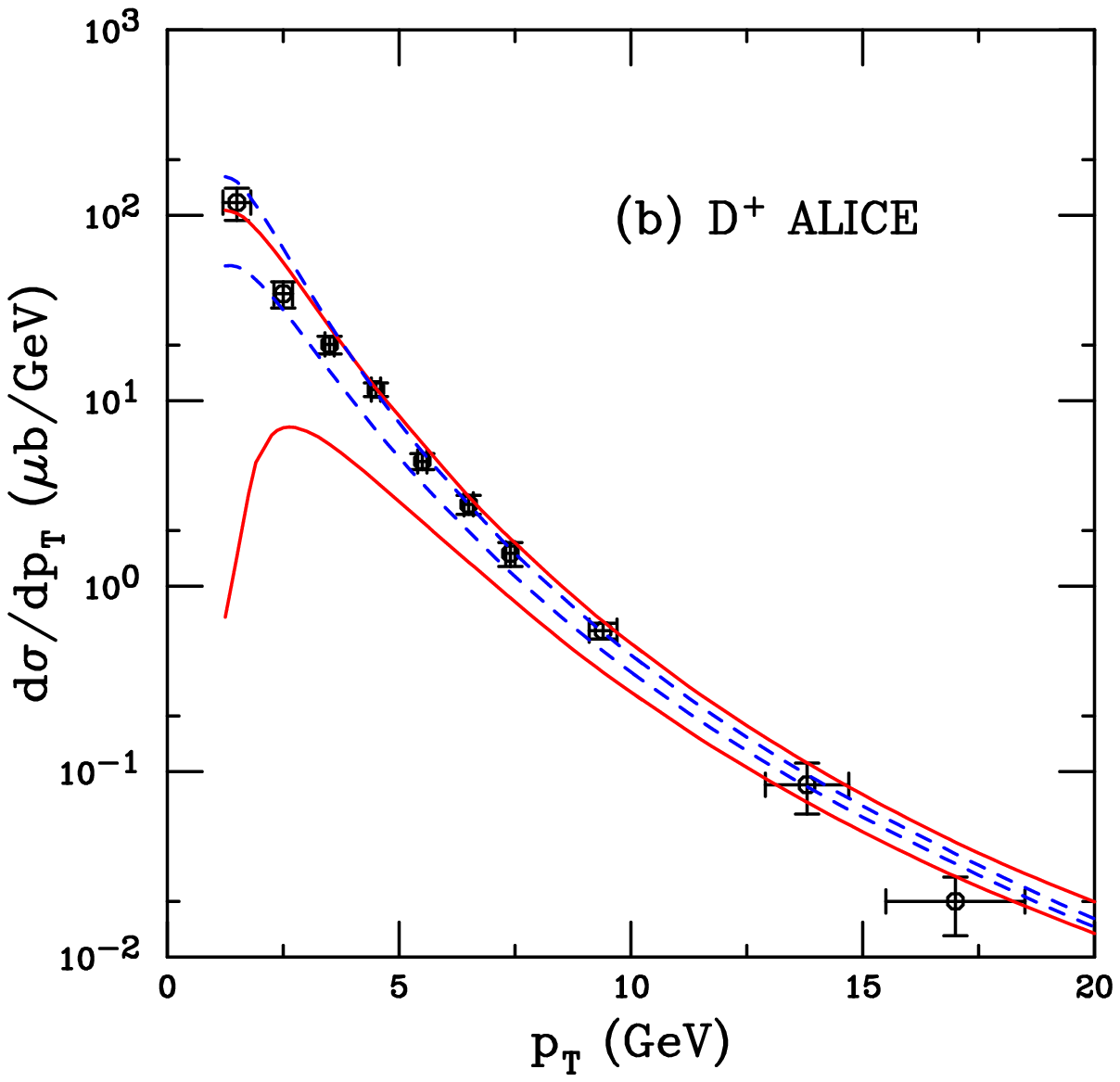} &
\includegraphics[width=0.33\textwidth]{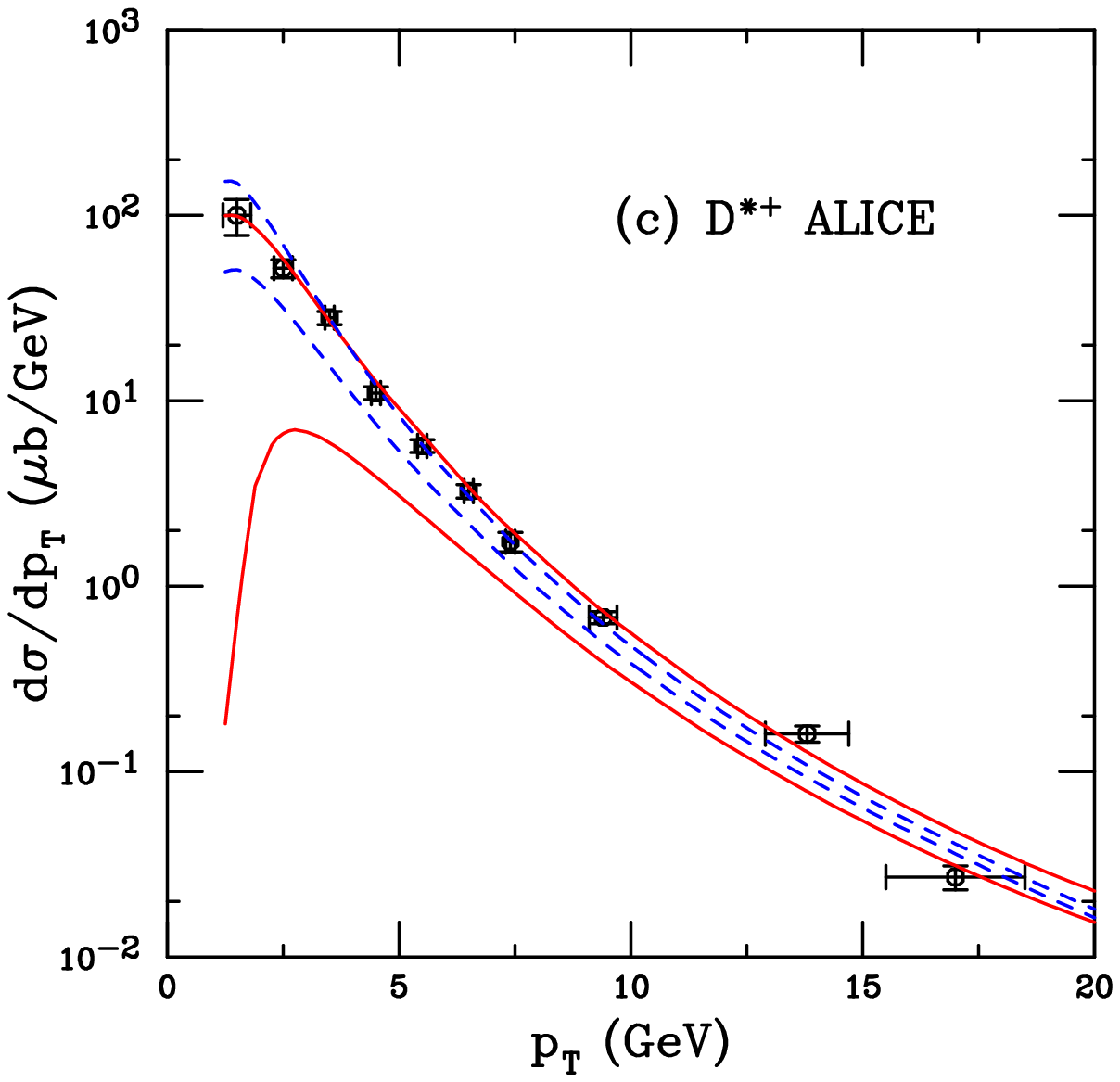} 
\end{tabular}
\end{center}
\caption[]{(Color online) Our calculations are compared with the reconstructed
ALICE (a) $D^0$, (b) $D^+$, and (c) $D^{*+}$ meson data 
\protect\cite{ALICEDmesons} at $\sqrt{s} = 7$ TeV in $|y| \leq 0.5$.
 The FONLL uncertainty
bands with the fiducial charm parameter set are shown by the red solid curves
while the blue dashed curves are calculated with the charm fit parameters.}
\label{DmesALICE}
\end{figure}

\begin{figure}[htbp]
\begin{center}
\begin{tabular}{ccc}
\includegraphics[width=0.33\textwidth]{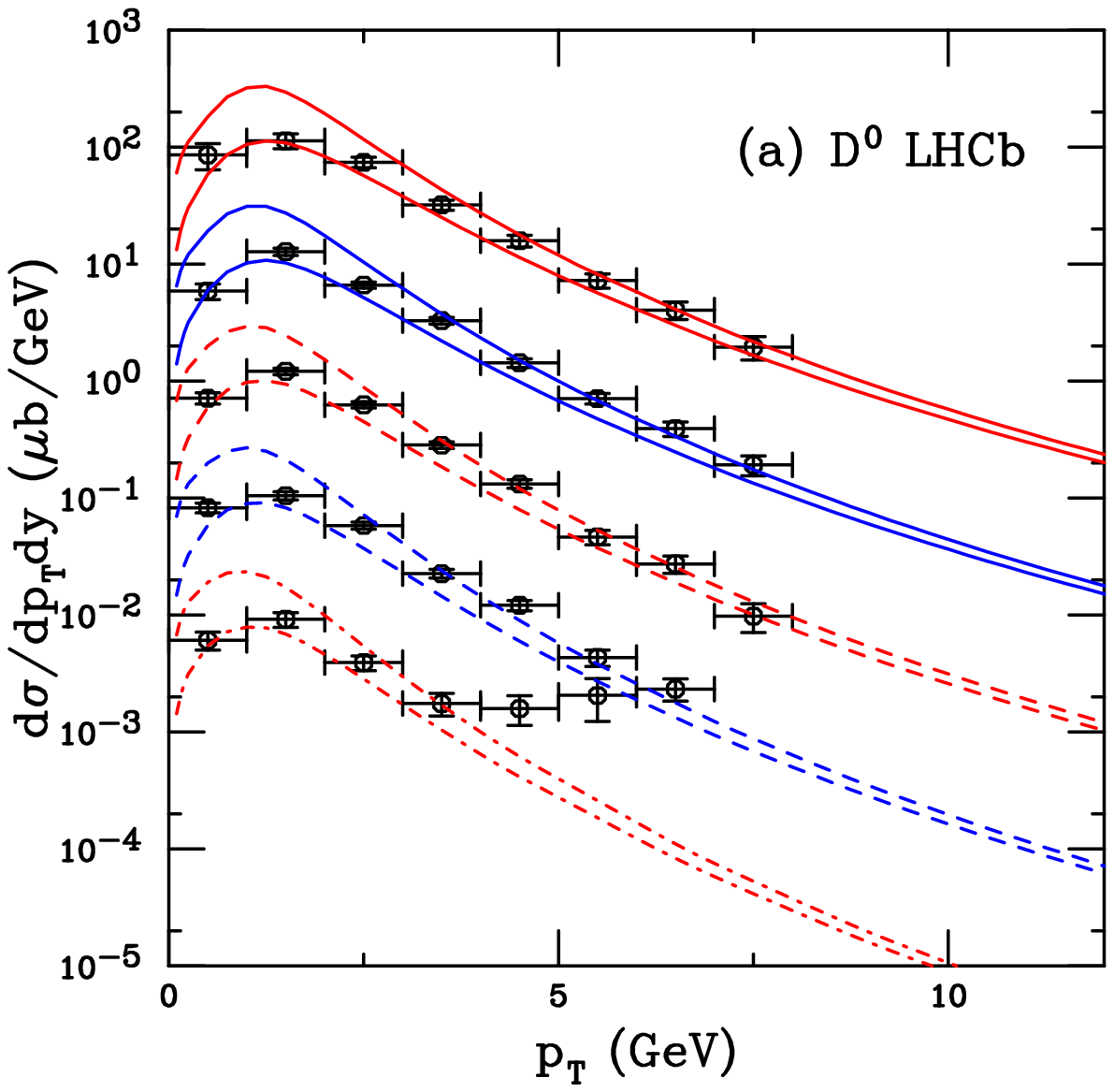} &
\includegraphics[width=0.33\textwidth]{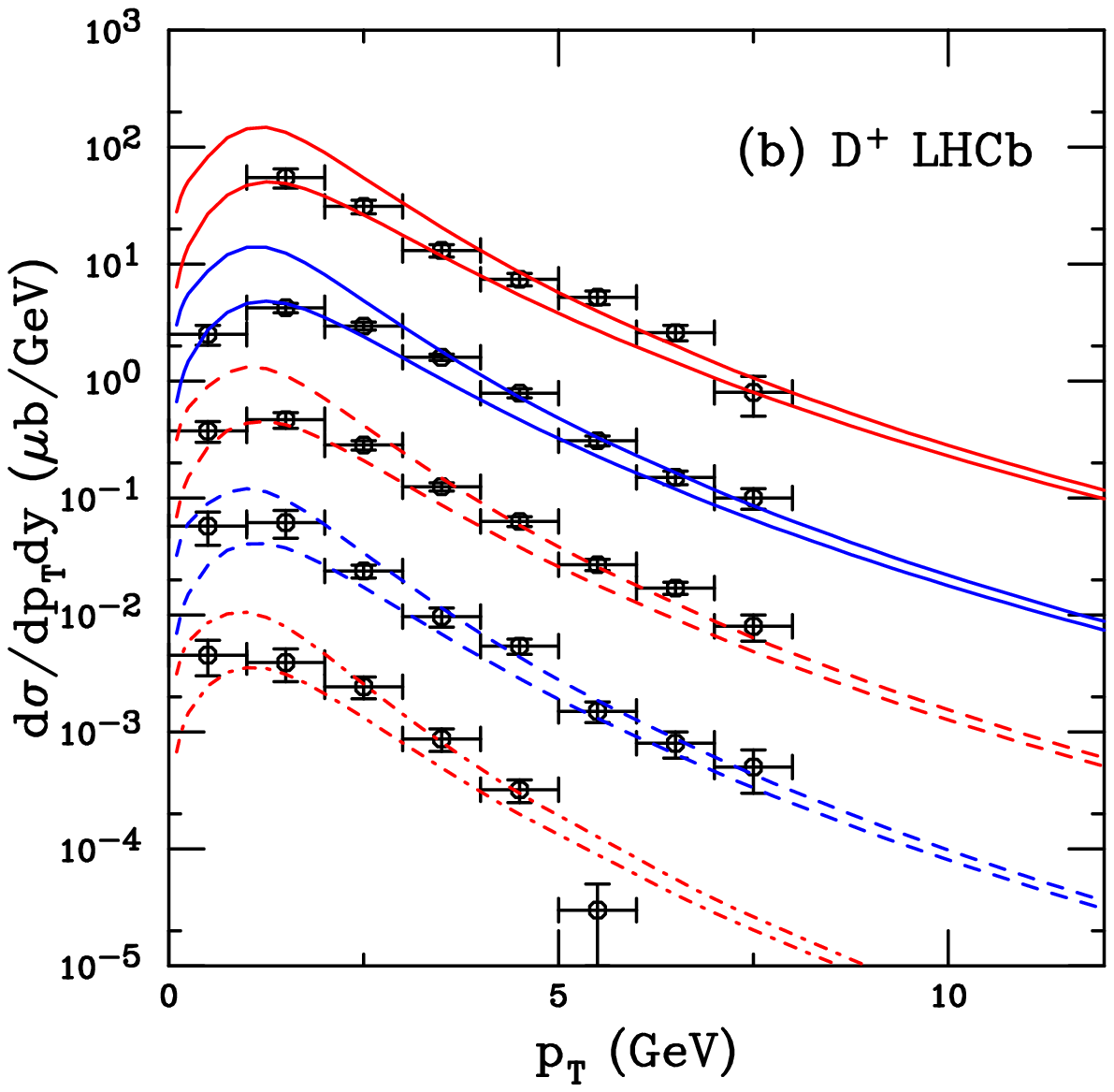} &
\includegraphics[width=0.33\textwidth]{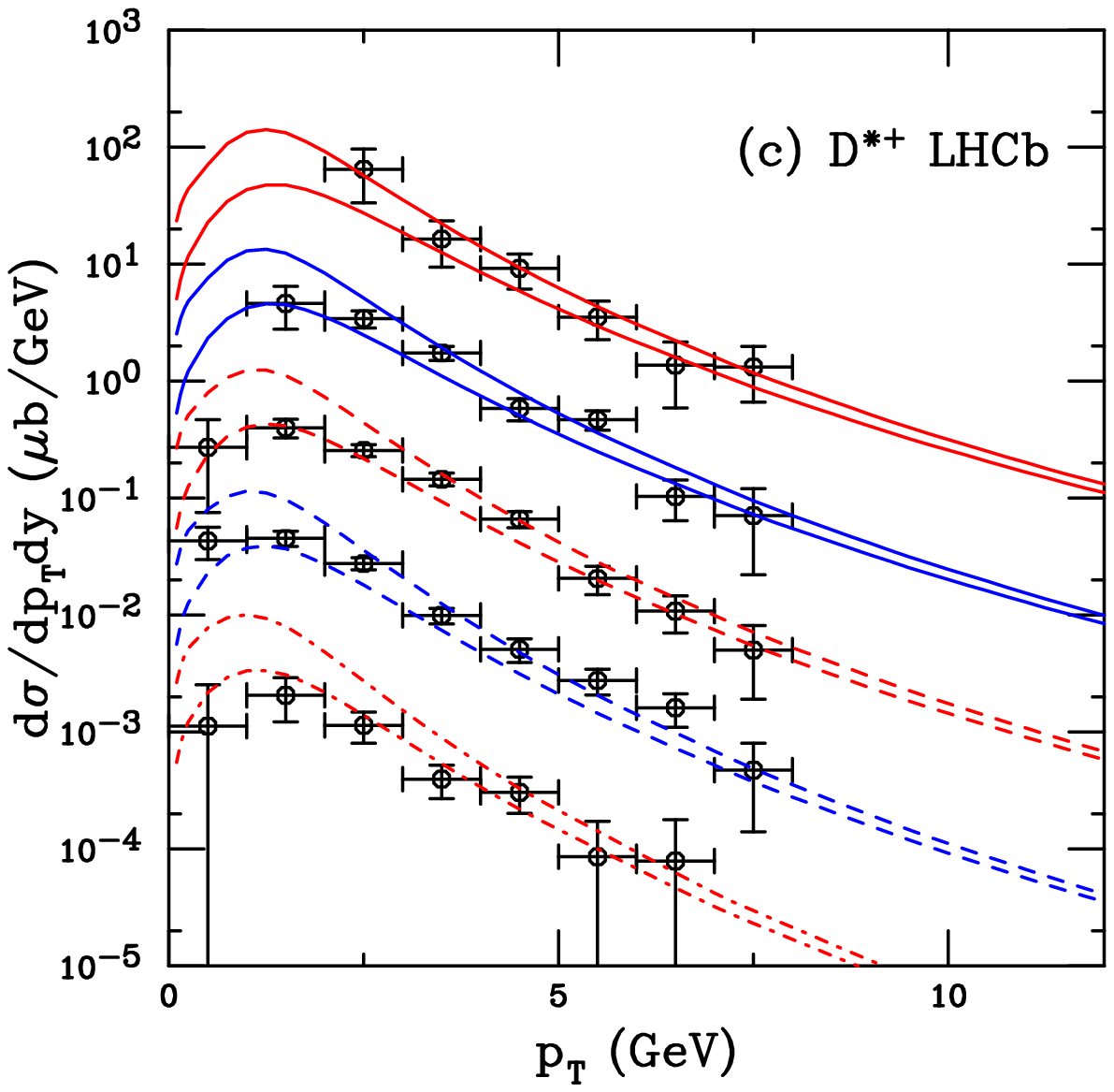} 
\end{tabular}
\end{center}
\caption[]{(Color online) Our calculations are compared with the reconstructed
LHCb (a) $D^0$, (b) $D^+$, and (c) $D^{*+}$ meson data 
\protect\cite{LHCbDmesons} at $\sqrt{s} = 7$ TeV in the rapidity intervals:
$2<y<2.5$ (solid red); $2.5 <y<3$ (solid blue); $3<y<3.5$ (dashed red);
$3.5<y<4$ (dashed blue); and $4<y<4.5$ (dot-dashed red).  The curves are
calculated with the charm fit parameters.  The sets of results are separated
by a factor of 10 between rapidity intervals to facilitate comparison. The
lowest rapidity interval, $2 < y < 2.5$, is not scaled.
}
\label{DmesLHCb}
\end{figure}

While the agreement between the lepton measurements at RHIC and the LHC and
our calculations is encouraging, as noted here and in Ref.~\cite{CNV}, there 
is significant admixture of semileptons charm and bottom decays, particularly
at lepton $p_T > 4$ GeV.  A better test of our results would be a comparison
to open charm hadron data.  Thus, 
in Figs.~\ref{DmesALICE} and \ref{DmesLHCb}, we show the  $D^0$ (a), $D^+$ (b) 
and $D^{*+}$ (c) distributions in the ALICE \cite{ALICEDmesons} and the LHCb 
\cite{LHCbDmesons} acceptances at midrapidity and forward rapidity respectively.
  
Figure~\ref{DmesALICE} compares the FONLL calculations with the fiducial 
parameter set (in red) with the fitted parameters based on $m = 1.27$ GeV
(in blue).  The upper and lower limits of both bands are shown.  While the
ALICE data are in agreement with the upper limits of both calculations, the
large $D$ meson uncertainty is reduced at low $p_T$ with the fitted parameter
set.  

Figure~\ref{DmesLHCb} shows the upper and lower limits of the FONLL
calculation based on $m = 1.27$ GeV in the five rapidity intervals of $\Delta y
= 0.5$ in the range $2 < y < 4.5$ covered by the LHCb detector.  In most cases
here also the agreement with the data is very good, the exception being the
most forward rapidity measurement of $D^{*+}$ where the calculation is above
the data.  Interestingly, while the normalization of the $D^*$ and $D^+$
calculations are rather similar over all of the rapidity intervals, compare
Figs.~\ref{DmesLHCb}(b) and (c), there is a significant drop in the measured
$D^*$ cross section at low $p_T$ between $3.5 < y < 4$ and $4 < y < 4.5$ that
is not reproduced inthe calculations.

We have shown that the calculated uncertainties on the total charm cross 
section can be considerably reduced by fitting the data with a next-to-leading
order calculation. When the same fit parameters are used to calculate the
leptons from heavy flavor decays in the FONLL approach, the results are still 
in agreement with the data.

\section{Quarkonium production in the Color Evaporation Model}
\label{closed}

We now turn to a treatment of quarkonium production within this same framework.
Perhaps the simplest approach to quarkonium production is the color 
evaporation model (CEM) which treats
heavy flavor and quarkonium production on an equal footing.  
The CEM was first discussed some time ago \cite{Barger:1979js,Barger:1980mg}
and has enjoyed considerable phenomenological success when 
applied at next-to-leading order in the total cross section and leading order
in the quarkonium $p_T$ distribution \cite{Gavai:1994in,Amundson,SchulerV}. 

In the CEM, the quarkonium 
production cross section is some fraction, $F_C$, of 
all $Q \overline Q$ pairs below the $H \overline H$ threshold where $H$ is
the lowest mass heavy-flavor hadron.  Thus the CEM cross section is
simply the $Q \overline Q$ production cross section with a cut on the pair mass
but without any constraints on the 
color or spin of the final state. The color of the
octet $Q \overline Q$ state is
`evaporated' through an unspecified process which does not change the momentum.
The additional energy needed to produce
heavy-flavored hadrons when the partonic center of mass energy, 
$\sqrt{\hat s}$, is less than $2m_H$, the $H \overline H$
threshold energy, is nonperturbatively obtained from the
color field in the interaction region.
Thus the quarkonium yield may be only a small fraction of the total $Q\overline 
Q$ cross section below $2m_H$.
At leading order, the production cross section of quarkonium state $C$ in
a $pp$ collision is
\begin{eqnarray}
\sigma_C^{\rm CEM}(s_{_{NN}})  =  F_C \sum_{i,j} 
\int_{4m^2}^{4m_H^2} ds
\int dx_1 \, dx_2~ f_i^p(x_1,\mu_F^2)~ f_j^p(x_2,\mu_F^2)~ 
\hat\sigma_{ij}(\hat{s},\mu_F^2, \mu_R^2) \, 
\, , \label{sigtil}
\end{eqnarray} 
where $ij = q \overline q$ or $gg$ and $\hat\sigma_{ij}(\hat s)$ is the
$ij\rightarrow Q\overline Q$ subprocess cross section.  

The fraction $F_C$ must be universal so that, once it is fixed by data, the
quarkonium production ratios should be constant as a function of $\sqrt{s}$,
$y$ and $p_T$.  The actual value of $F_C$ depends on the heavy quark mass, 
$m$, the scale parameters, the parton densities and 
the order of the calculation.
It was shown in Ref.~\cite{Gavai:1994in} that the quarkonium production ratios
were indeed relatively constant, as expected by the model.
In addition, Ref.~\cite{Amundson} showed that the data on 
the $J/\psi$ and open charm
cross sections as a function of $\sqrt{s}$ in hadroproduction and $W_{\gamma N}$
in photoproduction have the same energy dependence.

The data we use to obtain $F_C$ for the $J/\psi$ are from the compilation
by Maltoni {\it et al.} \cite{Maltoni:2006yp}.  The data range from fixed-target
experiments with center-of-mass energy $6.8 \leq \sqrt{s} \leq 41.6$ GeV 
\cite{maltoni23,maltoni24,maltoni25,maltoni26,maltoni27,maltoni28,maltoni29,maltoni31,maltoni32,maltoni33,maltoni34,maltoni35,maltoni36,maltoni37,maltoni40,maltoni41,maltoni42,maltoni43}  to data from the CERN ISR at $\sqrt{s} = 23$ 
\cite{maltoni30}, 30 \cite{maltoni38}, 30.6 \cite{maltoni39}, 
31 \cite{maltoni30}, 52 \cite{maltoni44,maltoni45}, 52.4 \cite{maltoni39}, 
53 \cite{maltoni30,maltoni38}, 62.7 \cite{maltoni39}, and 
63 \cite{maltoni30,maltoni38} GeV.  Data from the PHENIX experiment at RHIC 
\cite{maltoni46} are also used.  The ISR data \cite{maltoni30,maltoni38,maltoni39,maltoni44,maltoni45} are all from $pp$ measurements, as are
the data from Refs.~\cite{maltoni24,maltoni27}.  Data from single nuclear 
targets include Be \cite{maltoni25,maltoni26,maltoni33,maltoni40}, 
Li \cite{maltoni31}, C \cite{maltoni28,maltoni29}, Si \cite{maltoni41}, 
Fe \cite{maltoni34}, Au \cite{maltoni42}, and Pt \cite{maltoni27}.  Other 
experiments took data on multiple nuclear targets 
\cite{maltoni23,maltoni35,maltoni36,maltoni37,maltoni43}.  Both the total
forward cross section ($x_F > 0$) \cite{maltoni23,maltoni24,maltoni25,maltoni26,maltoni27,maltoni28,maltoni29,maltoni31,maltoni32,maltoni33,maltoni34,maltoni35,maltoni36,maltoni37,maltoni40,maltoni41,maltoni42,maltoni43,maltoni46} and
the cross section times the branching ratio to lepton pairs, $B_{ll}$, 
at $y = 0$, $B_{ll} d\sigma/dy|_{y=0}$ \cite{maltoni25,maltoni26,maltoni30,maltoni32,maltoni33,maltoni34,maltoni35,maltoni36,maltoni37,maltoni38,maltoni39,maltoni41,maltoni42,maltoni43,maltoni44,maltoni45,maltoni46} were reported.
Several of the ISR experiments \cite{maltoni30,maltoni38,maltoni39,maltoni45}
only provided the cross section at $y=0$, likely due to their limited phase
space coverage.  In cases where the total cross section was reported, the
uncertainty provided was on the level of 40\%.  We note that several detectors
have taken data at the same energy and with the same target but reported results
with different experiment numbers that diverge by more than one standard 
deviation.  For example, the $p$+C results
reported by E331 \cite{maltoni28} and E444 \cite{maltoni29} using the same
apparatus, $\sigma = 256 \pm
30$ and $166 \pm 23$ nb/nucleon respectively,
differ by more than two standard deviations.

Maltoni and collaborators corrected prior 
measurements using up-to-date values of the $J/\psi$ branching ratios to
$\mu^+ \mu^-$ and $e^+ e^-$ \cite{pdg} and, when appropriate, averaged the
results on multiple nuclear targets assuming $\sigma_{pA}/\sigma_{pp} = 
A^\alpha$ with $\alpha = 0.96\pm 0.01$  
\cite{MikeL} at $x_F \sim 0$, obtained with an 800 GeV proton 
beam \cite{Maltoni:2006yp}.  The $A$ dependence was assumed to be independent of
center-of-mass energy.  However, a recent reanalysis of these data, assuming a
combination of shadowing and absorption effects on $J/\psi$ production, found
that, at $x_F \sim 0$, the absorption cross section decreases as a function of
incident energy whether or not the data were corrected for shadowing effects
\cite{LVW}.  Later measurements with $p_{\rm lab} =158$ GeV obtained an
absorption cross section consistent with the predicted extrapolation 
\cite{NA60pA}.   In addition, effects arising from modifications of the parton
densities in the nucleus that may be present in the
data depend on the magnitude of $A$ and have not been taken into account
in the averaging.  Thus the effective $\alpha$, which includes all relevant
nuclear effects, likely depends on incident energy.

We have fit $F_C$ to both the full data set as well as to more limited sets.
Our final result is based on the total cross section data with only $p$, Be, Li,
C, and Si targets respectively.  In this way, we avoid uncertainties due to 
ignoring any cold nuclear matter effects which are on the order of a few percent
in light targets.  We also restricted ourselves to the forward cross sections
only, rather than include the $B_{ll} d\sigma/dy|_{y=0}$ data in the fits.  
The rapidity distributions calculated in the MNR code are subject to 
fluctuations about the mean, even with high statistics calculations. The
total cross sections, not subject to these fluctuations, are thus more accurate.

Our calculations use the NLO $Q \overline Q$ code of Mangano {\it et al.}~(MNR)
\cite{MNRcode} with
the $2m_H$ mass cut in Eq.~(\ref{sigtil}), as described in 
Refs.~\cite{Gavai:1994in,rhicii}.  Because the NLO $Q 
\overline Q$ code is an exclusive calculation, we take the mass cut on the
invariant average over kinematic variables of the $c$ and $\overline c$. Thus,
instead of defining $\mu_F$ and $\mu_R$ relative to the quark mass, they are
defined relative to the transverse mass, $\mu_{F,R} \propto m_T = 
\sqrt{m^2 + p_T^2}$ where 
$p_T$ is that of the $Q \overline Q$ pair, $p_T^2 = 0.5(p_{T_Q}^2 + 
p_{T_{\overline Q}}^2)$.

We use the same values of the central charm quark
mass and scale parameters as in the previous section, both for the fiducial
parameter sets and for the best fit values, to obtain the normalization $F_C$.
We fit $F_C$ to the $J/\psi$ data for both the fiducial FONLL sets 
(central value $(m,\mu_F/m, \mu_R/m) = (1.5 \, {\rm GeV}, 1,1)$) and the fit 
results obtained in the previous section (central value 
$(m,\mu_F/m, \mu_R/m) = (1.27 \, {\rm GeV}, 2.1,1.6)$).  
We determine $F_C$ only for the central parameter set in each case and scale
all the other calculations for that case by the same value of $F_C$ to
obtain the extent of the $J/\psi$ uncertainty band employing
Eqs.~(\ref{sigmax}) and (\ref{sigmin}).

\begin{figure}[htbp] 
\begin{center}
\includegraphics[width=0.475\textwidth]{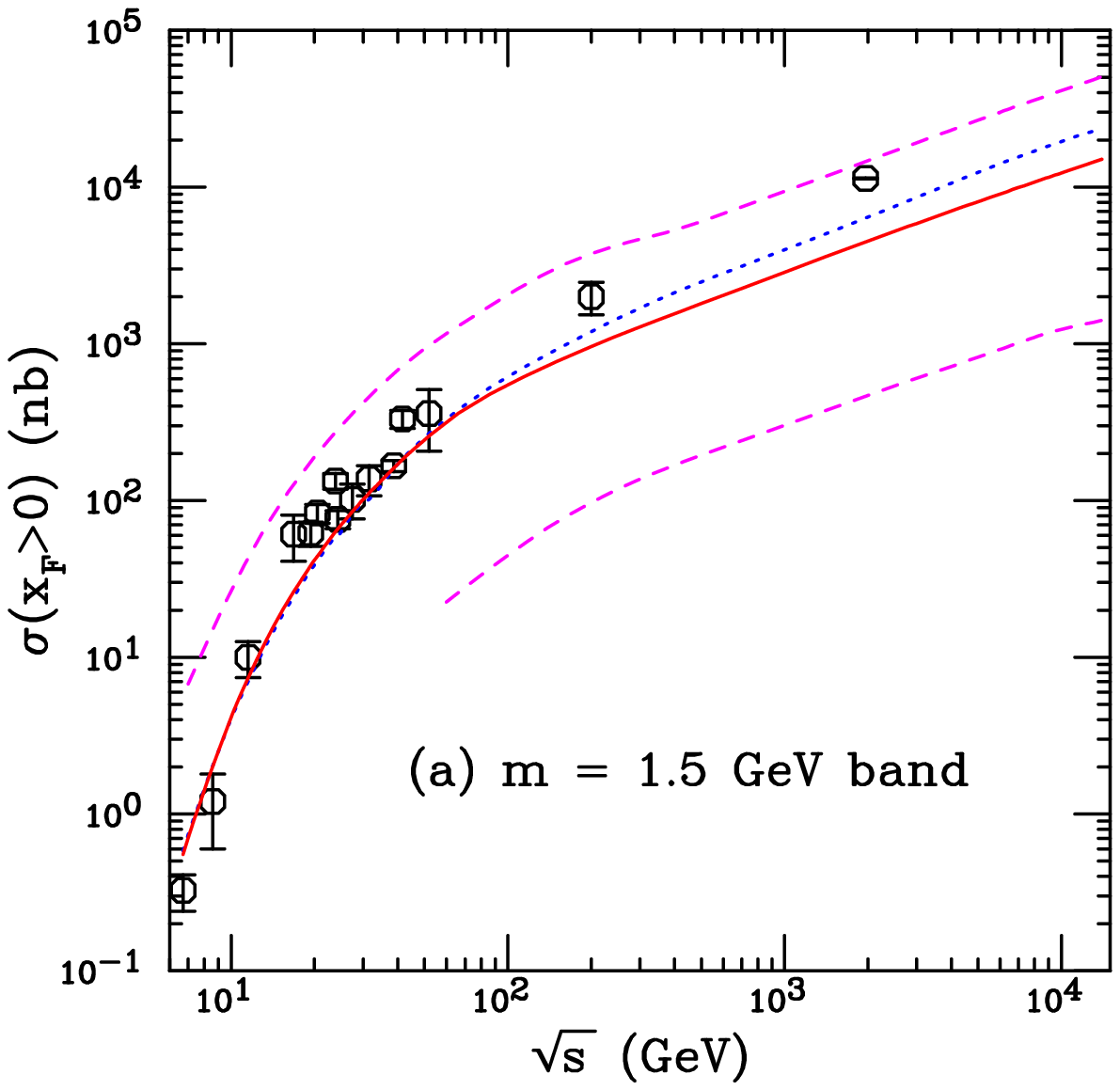}
\includegraphics[width=0.475\textwidth]{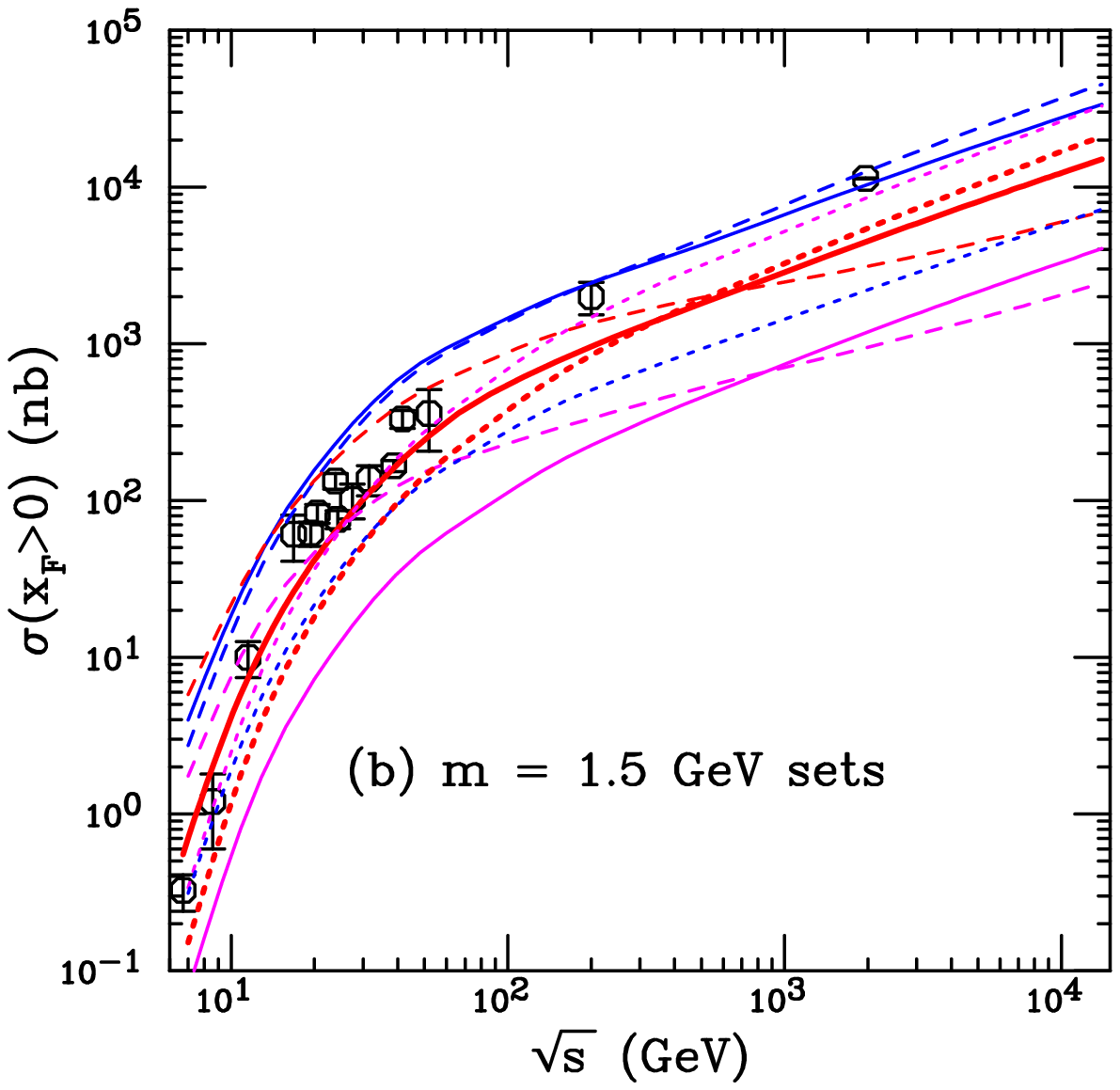}
\end{center}
\caption[]{(Color online)
The forward $J/\psi$ cross sections calculated with the FONLL fiducial
parameter set with $m=1.5$ GeV. (a) The uncertainty
band obtained employing the FONLL parameter set.  
The solid red curve is the central value while the limits of the 
uncertainty band are given by the dashed magenta curves.  The dotted blue 
curve is a result with $(m, \mu_F/m_T, \mu_R/m_T) = (1.2 \, {\rm GeV},2,2)$.
(b) The solid red curve is the central value 
$(m, \mu_F/m_T,\mu_R/m_T) = (1.5\, {\rm GeV},1,1)$.  The solid
blue and magenta curves outline the mass uncertainty with $(1.3 \, {\rm GeV},
1,1)$ and $(1.7\, {\rm GeV}, 1,1)$ respectively.  The dashed
curves are associated with $\mu/m_T = 0.5$: $(\mu_F/m_T,\mu_R/m_T) =
(1,0.5)$ blue; (0.5,1) magenta; and (0.5,0.5) red. The dotted
curves are associated with $\mu/m_T = 2$: $(\mu_F/m_T,\mu_R/m_T) =
(1,2)$ blue; (2,1) magenta; and (2,2) red.   
}
\label{psitotFONLLfig}
\end{figure}

\begin{figure}[htb] 
\begin{center}
\includegraphics[width=0.475\textwidth]{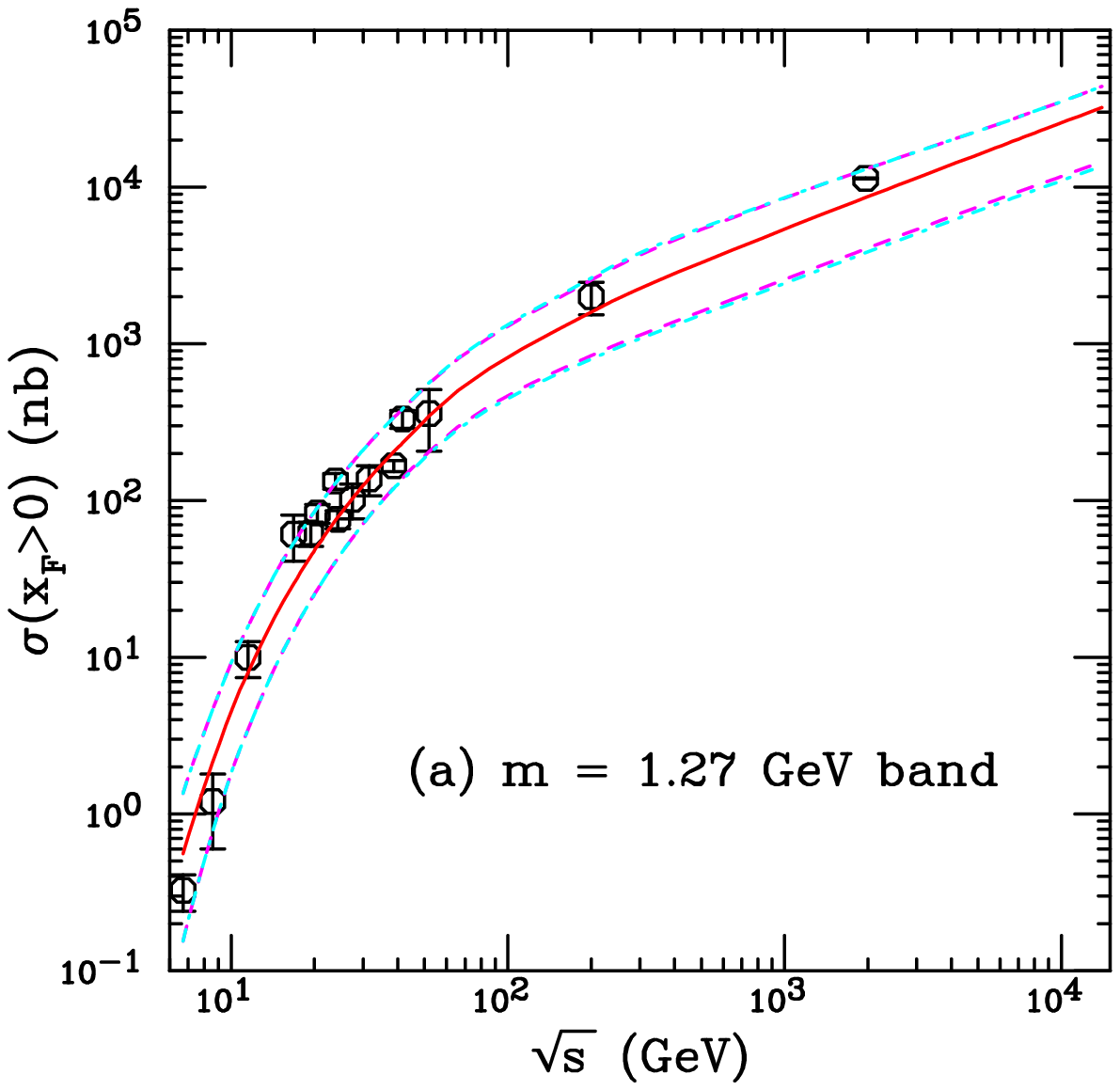}
\includegraphics[width=0.475\textwidth]{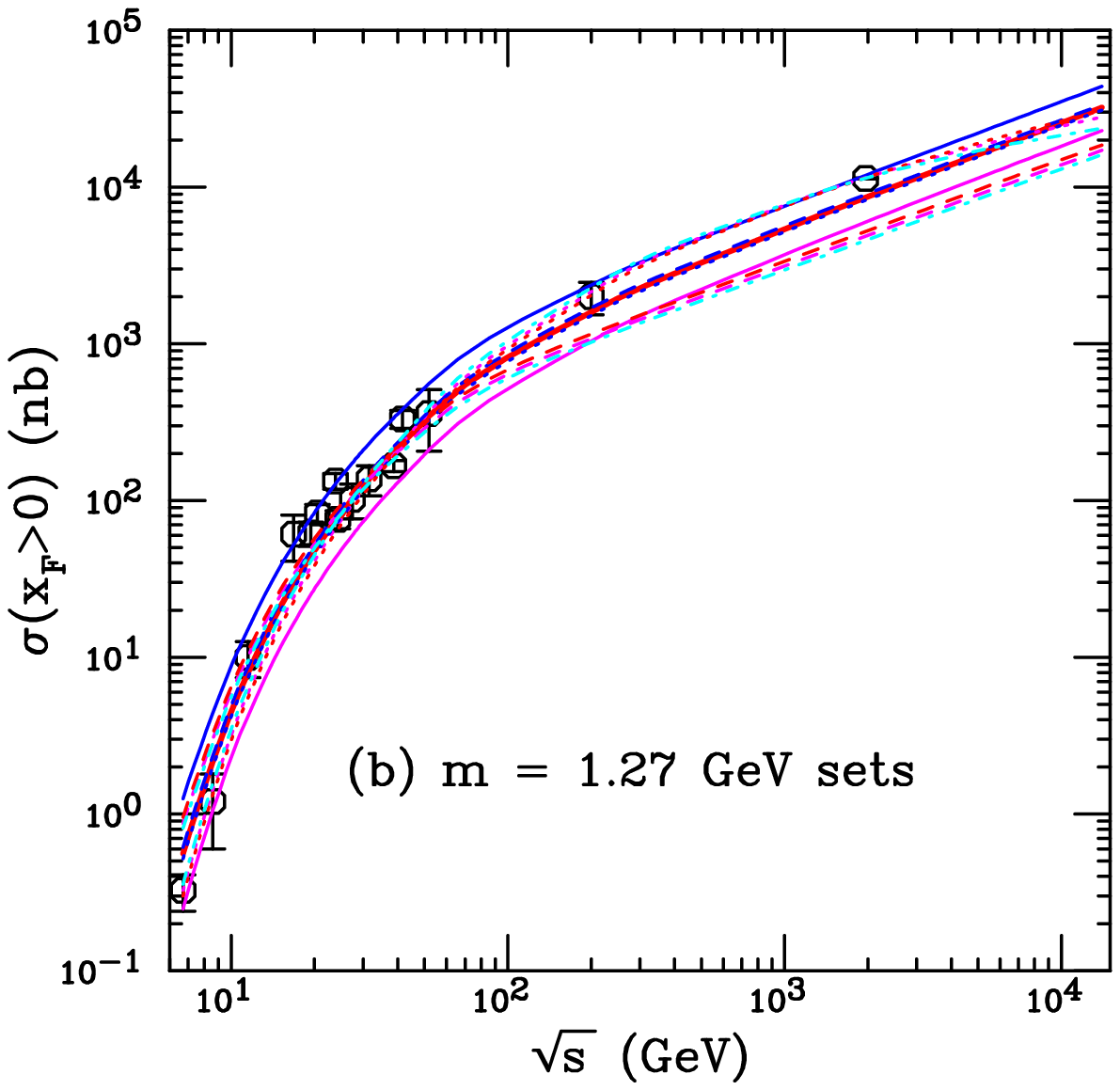}
\end{center}
\caption[]{ (Color online)
(a) The uncertainty band on the forward $J/\psi$ cross section calculated
based on the $c \overline c$ parameter fit in Fig.~\protect\ref{quadsigfig}(d).
The dashed magenta curves and dot-dashed cyan curves show the extent of
the corresponding uncertainty bands.  The dashed curves outline the most extreme
limits of the band. 
(b)  The components of the uncertainty band.  The central value 
$(m,\mu_F/m_T,\mu_R/m_T)
= (1.27 \, {\rm GeV},2.10,1.60)$ is given by the solid red curve.  The solid
blue and magenta curves outline the mass uncertainty with $(1.18 \, {\rm GeV},
2.10,1.60)$ and $(1.36\, {\rm GeV}, 2.10,1.60)$ respectively.  The dashed
curves outline the lower limits on the scale uncertainty: 
$(\mu_F/m_T,\mu_R/m_T) =
(2.10,1.48)$ blue; (1.25,1.60) magenta; and (1.25,1.48) red. The dotted
curves outline the upper limits on the scale uncertainty: $(\mu_F/m_T,\mu_R/m_T) 
= (2.10,1.71)$ blue; (4.65,1.60) magenta; and (4.65,1.71) red.  The upper and
lower dot-dashed cyan curves correspond to $(\mu_F/m_T,\mu_R/m_T) = (4.65,1.48)$
and (1.25,1.71) respectively.
}
\label{psitotfitfig}
\end{figure}

We find $F_C = 0.040377$ for the central result,
$(m,\mu_F/m, \mu_R/m) = (1.5 \, {\rm GeV}, 1,1)$,
and $F_C = 0.020393$ for the central CT10 result
with $(m,\mu_F/m, \mu_R/m) = (1.27 \, {\rm GeV}, 2.1,1.6)$.
A significantly larger value of $F_C$ is necessary for the larger quark mass
since the fraction of the total charm cross section remaining after the mass
cut is smaller.  The results for the energy dependence of the forward inclusive 
$J/\psi$ cross section are shown in Figs.~\ref{psitotFONLLfig} and 
\ref{psitotfitfig} for the central mass values of 1.5 GeV and 1.27 GeV
respectively.  The uncertainty bands are shown on the left-hand sides of
these figures while the individual parameter sets contributing to the bands are
shown on the right-hand sides.

The most obvious result in Fig.~\ref{psitotFONLLfig}(a) is that there is no
well defined lower limit on the total cross section with the fiducial parameter
set, only an upper limit.  The reason is apparent from 
Fig.~\ref{psitotFONLLfig}(b):  the combined differences in the minimum values
of the masses and scales added in quadrature are larger than the central value
for $\sqrt{s} < 63$ GeV.  When the fiducial parameter sets are applied to the
CEM calculation of $J/\psi$ production, the upper limit of the charm quark
mass, $m = 1.7$ GeV, gives a very narrow invariant 
mass interval for the CEM calculation 
in Eq.~(\ref{sigtil}), from $2m = 3.4$~GeV to $2m_D = 3.86$~GeV.  The 
difference between the results with different quark masses is more pronounced at
low center of mass energies while the energy dependence of the calculations
with different values of $m$ begins to converge at large $\sqrt{s}$.  Indeed,
the `hump' in the upper limit of the fiducial uncertainty band is due to the
slower growth of $(m,\mu_F/m_T,\mu_R/m_T) = (1.3 \, {\rm GeV},1,1)$ relative
to $(1.5 \, {\rm GeV},1,0.5)$ for $\sqrt{s} > 400$ GeV, note the crossing of
the solid and dashed blue curves in Fig.~\ref{psitotFONLLfig}(b).  
We also note that the
fiducial set does not give very good agreement with the total $J/\psi$ cross
section reported by CDF \cite{CDFpsitot} since the calculated $\sqrt{s}$
dependence is too slow to match the measured growth of the forward cross 
section.

The best fit band, shown in Fig.~\ref{psitotfitfig}(a), on the other hand,
gives very good agreement with the $J/\psi$ data over the entire energy
range, even for the CDF cross section, not included in the fit.  The data
are almost all encompassed by the width of the band.  Now, as
was the case for the total charm cross section, the uncertainty due to the 
quark mass dominates over that due to the scale choice for $\sqrt{s} < 200$ GeV.

\begin{figure}[htb]
\begin{center}
\begin{tabular}{ccc}
\includegraphics[width=0.33\textwidth]{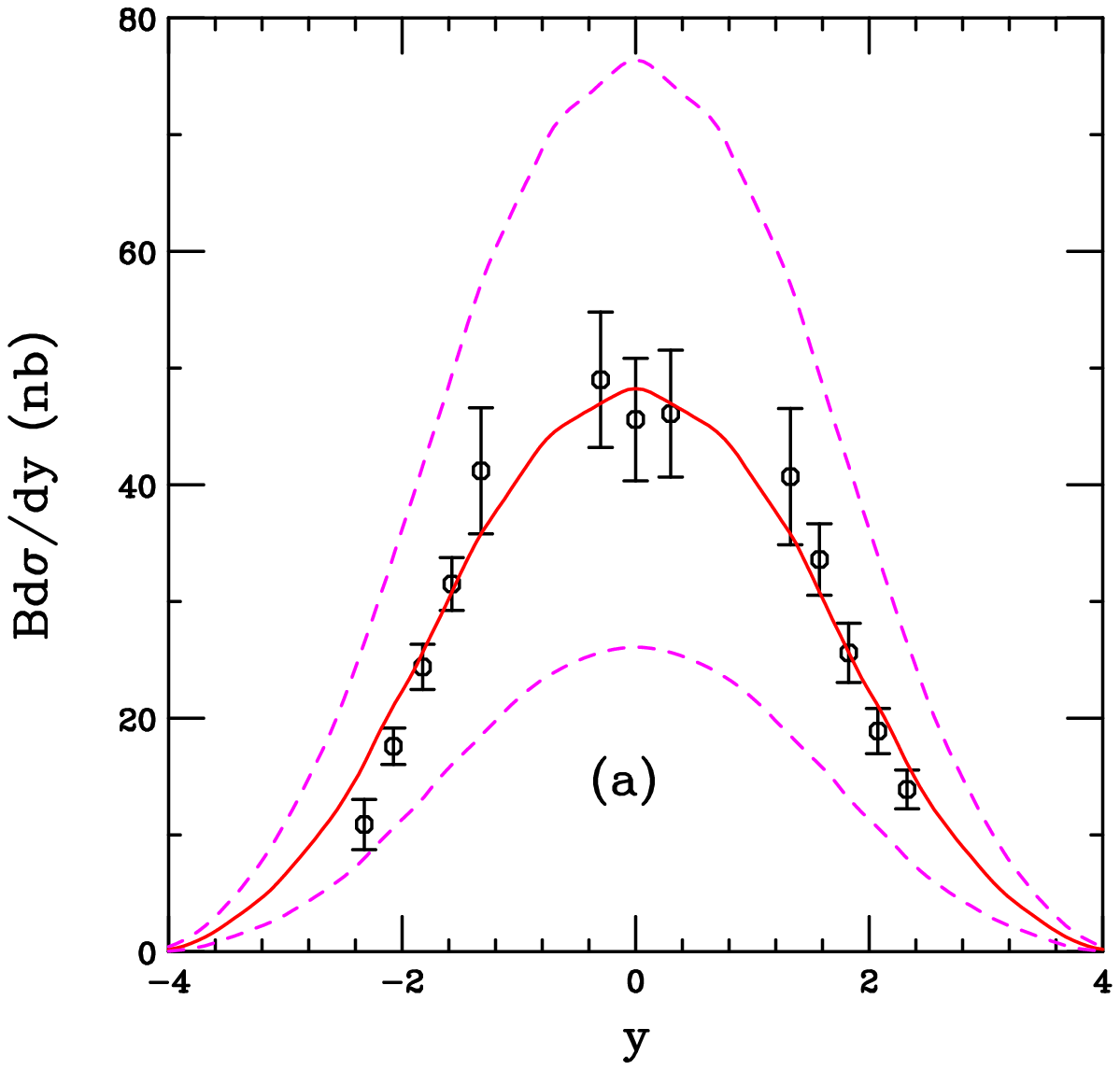}&
\includegraphics[width=0.33\textwidth]{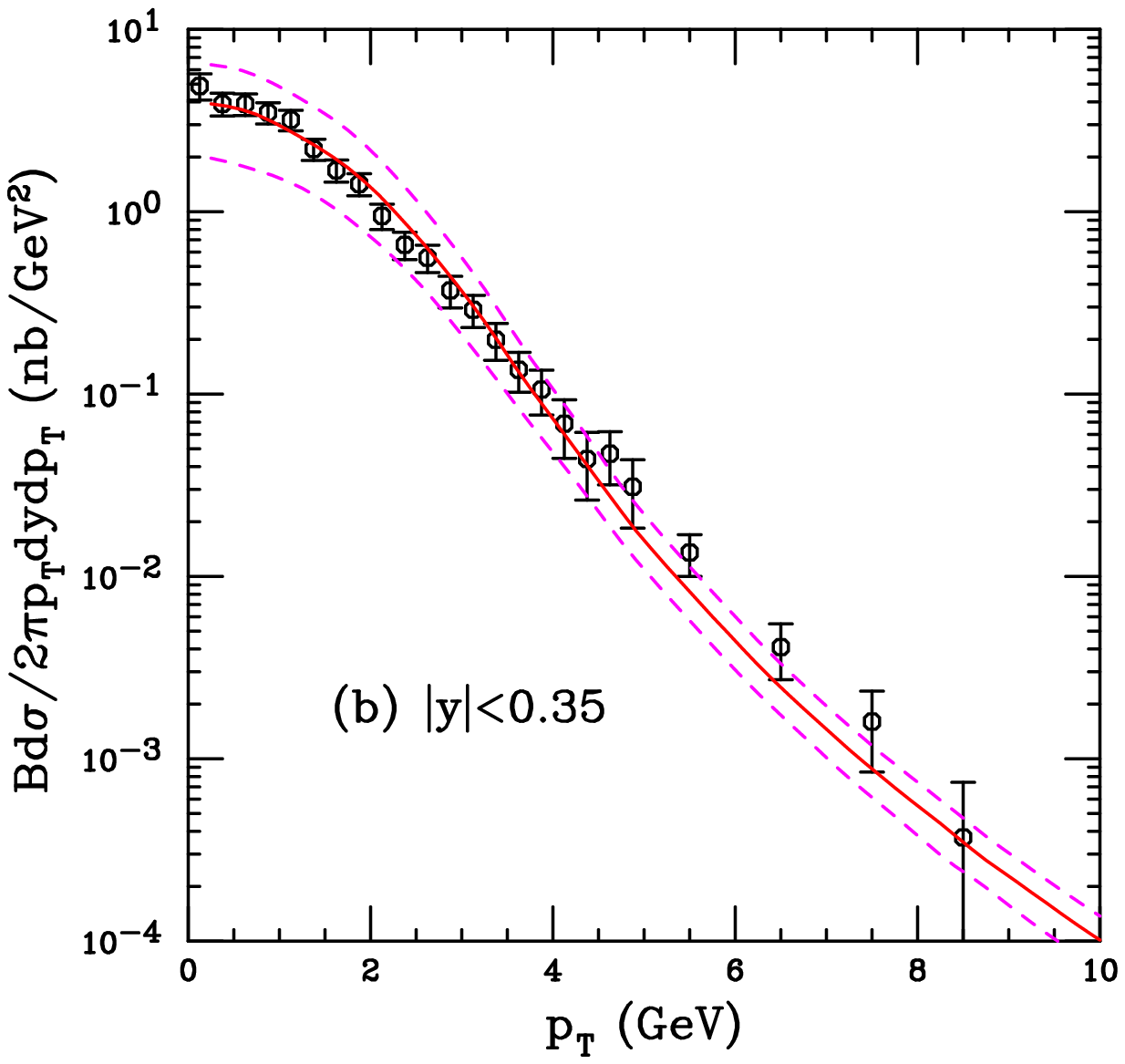}&
\includegraphics[width=0.33\textwidth]{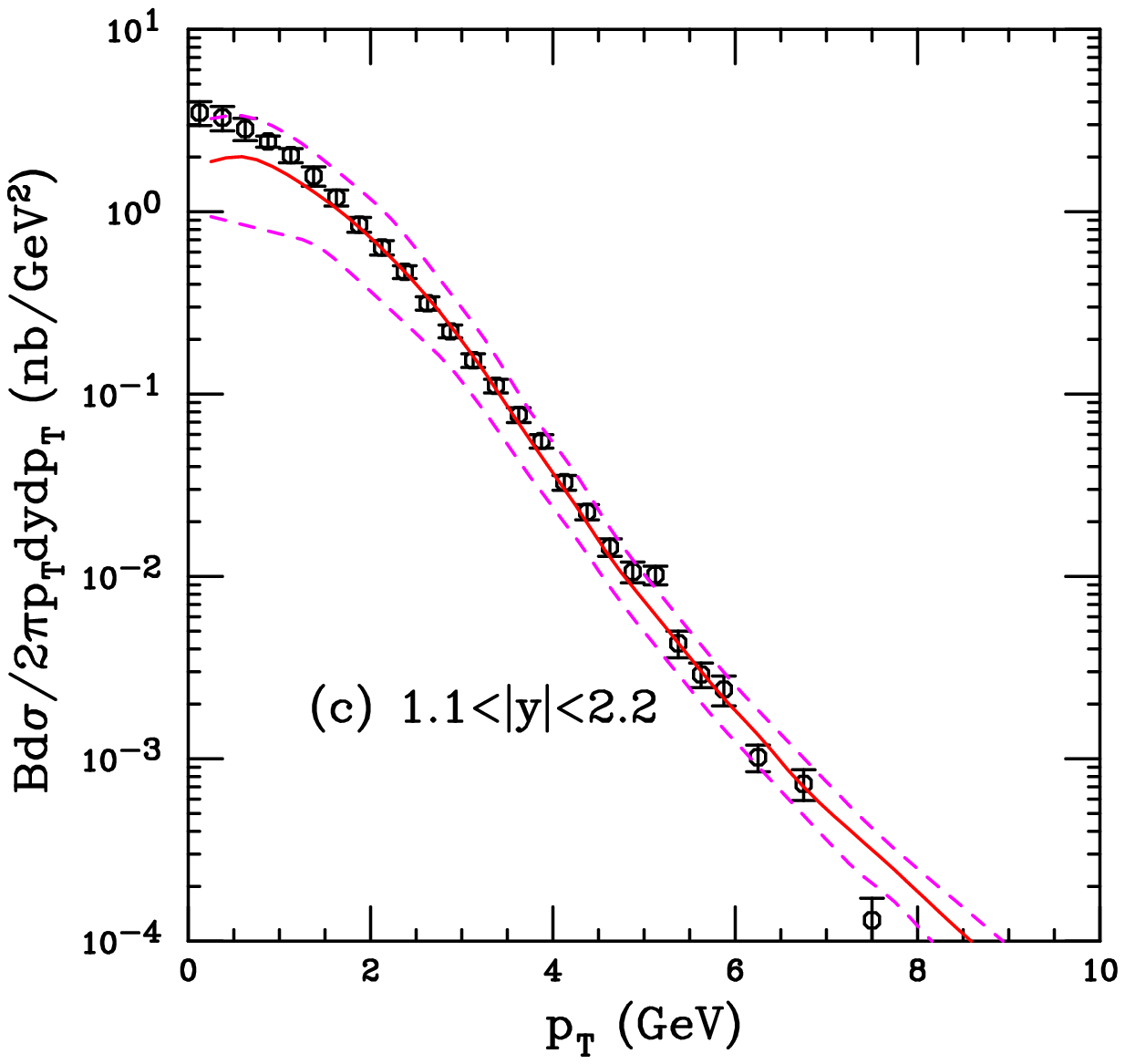}
\end{tabular}
\end{center}
\caption[]{(Color online) The $J/\psi$ rapidity distribution (a) and the 
midrapidity (b)
and forward rapidity (c) $p_T$ distributions and their uncertainties calculated 
with the same parameters as Fig.~\protect\ref{chi2fig}(d).  
The results are compared to PHENIX $pp$ measurements at
$\sqrt{s} = 200$ GeV \protect\cite{PHENIXpp}.  The correlated and uncorrelated
systematic errors in Ref.~\protect\cite{PHENIXpp} are added in quadrature.
No additional scaling factor has been applied.
The solid red curve shows the central value while the
dashed magenta curves outline the uncertainty band.
A $\langle k_T^2 \rangle$ kick of 1.19 GeV$^2$ is applied to the $p_T$ 
distributions, as discussed in the text.  
}
\label{psiRHICdistfig}
\end{figure}

\begin{figure}[htb]
\begin{center}
\begin{tabular}{ccc}
\includegraphics[width=0.33\textwidth]{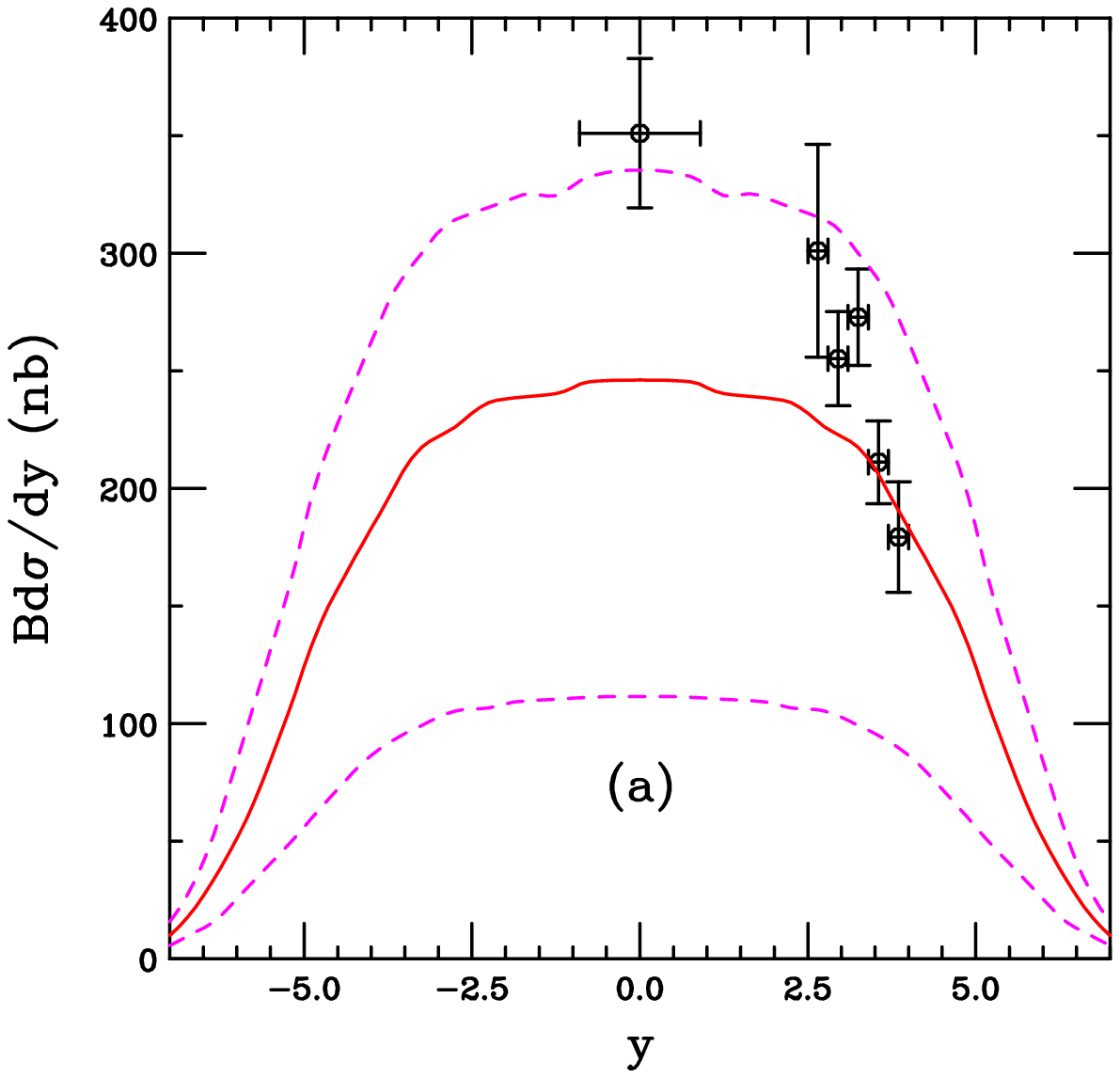}&
\includegraphics[width=0.33\textwidth]{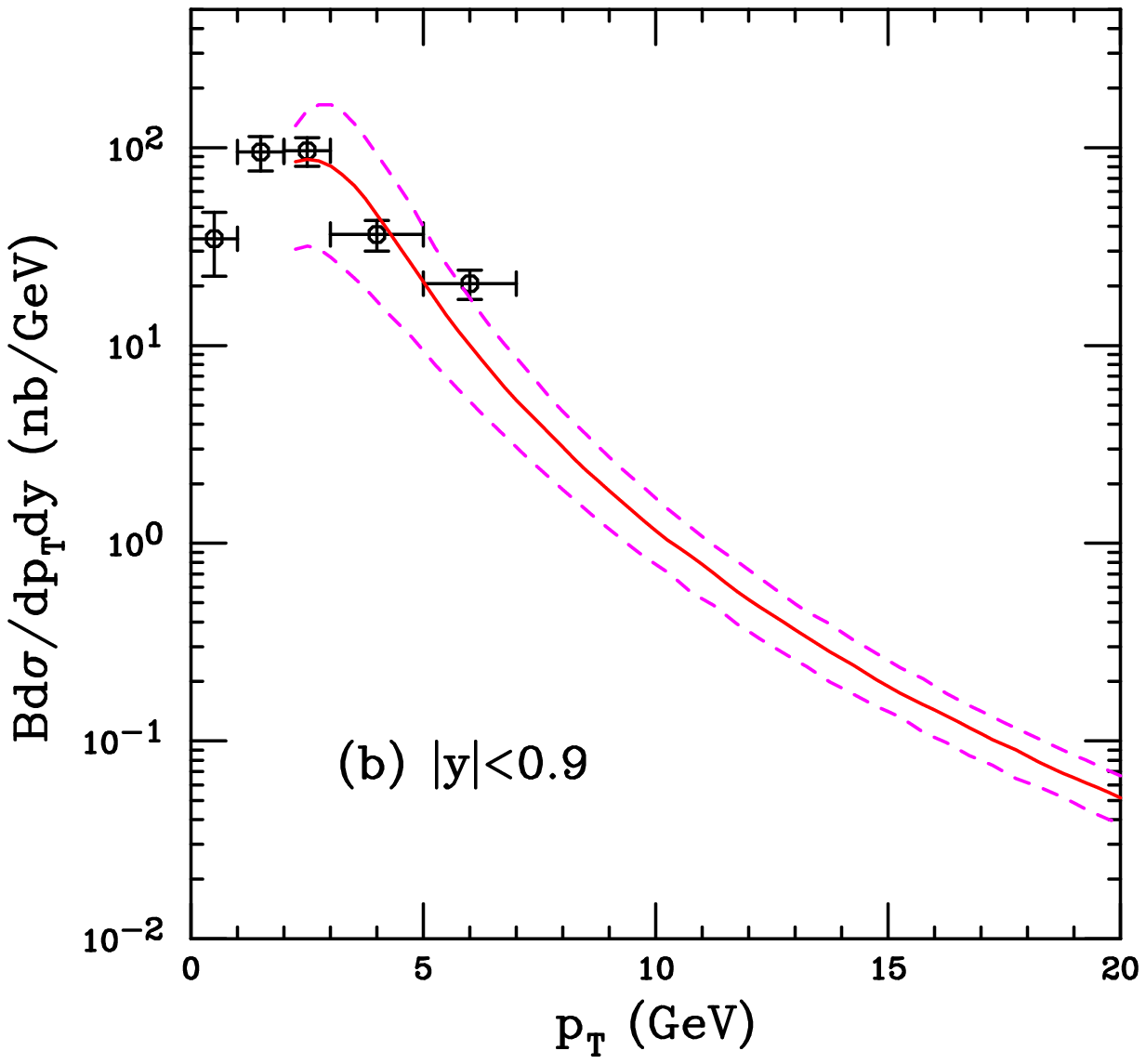}&
\includegraphics[width=0.33\textwidth]{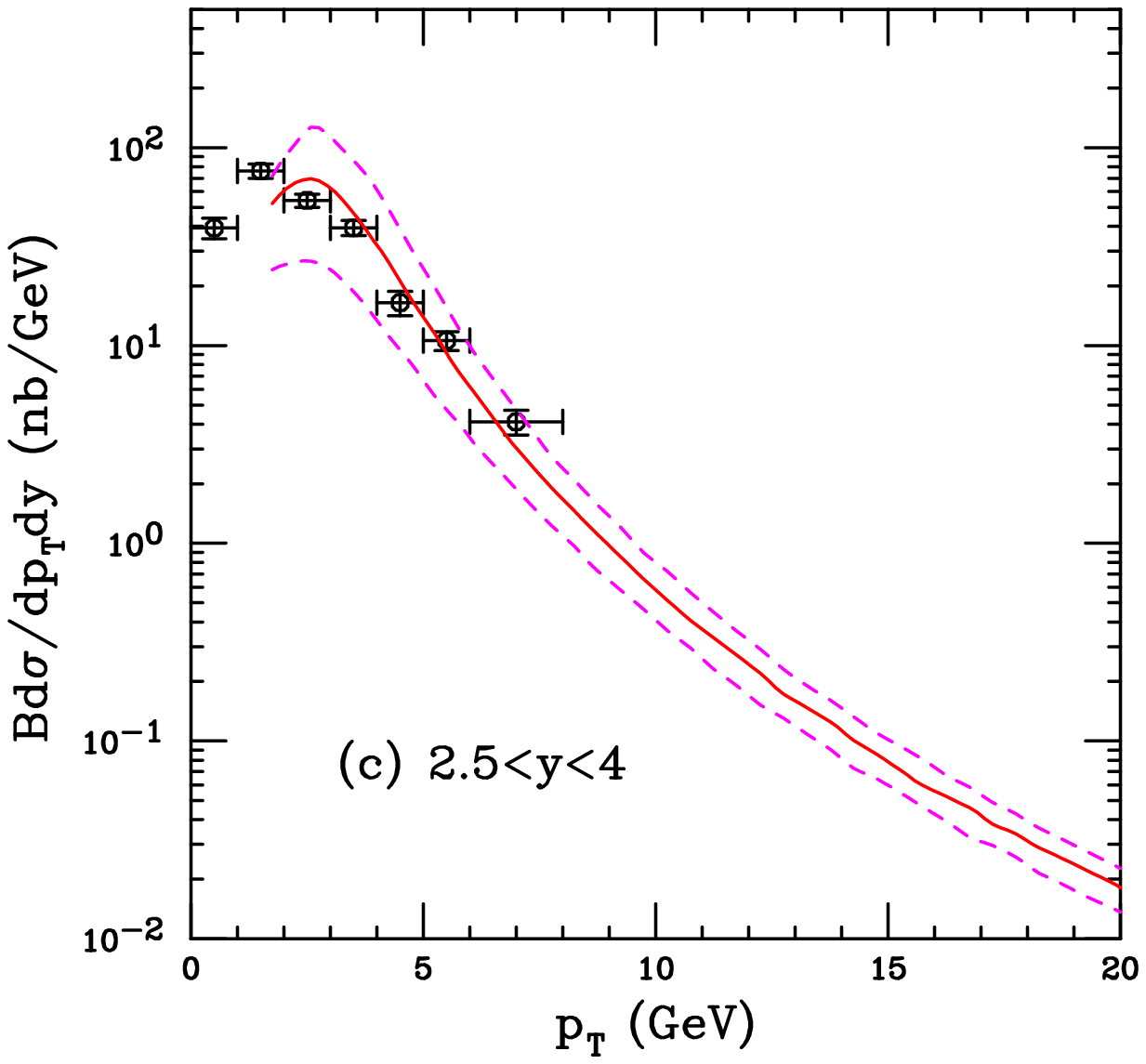}
\end{tabular}
\end{center}
\caption[]{(Color online) 
The $J/\psi$ rapidity distribution (a) and the midrapidity, $|y| < 0.9$ (b),
and forward rapidity, $2.5 < y < 4$ (c) $p_T$ distributions at 
$\sqrt{s} = 7$~TeV and their uncertainties calculated 
with the same parameters as in Fig.~\protect\ref{chi2fig}(d).  
The results are compared to the ALICE rapidity distribution as well as the
mid and forward rapidity $p_T$ distributions
\protect\cite{ALICEpp7TeV}.  
No additional scaling factor
has been applied.  The solid red curve shows the central value while the
dashed magenta curves outline the uncertainty band.
A $\langle k_T^2 \rangle$ kick of 1.49 GeV$^2$ is applied to the $p_T$ 
distributions, as discussed in the text.  
}
\label{psiALICEdistfig}
\end{figure}

We now turn to the $J/\psi$ rapidity and $p_T$ distributions, shown in
Fig.~\ref{psiRHICdistfig} for $\sqrt{s} = 200$ GeV and 
Fig.~\ref{psiALICEdistfig} for $\sqrt{s} = 7$ TeV.
At leading order in the total cross section,  the $Q \overline Q$ pair
$p_T$ is zero.  Thus, while our calculation is
next-to-leading order in the total cross section, it is leading
order in the quarkonium $p_T$ distributions. 
In the exclusive NLO calculation ~\cite{MNRcode}
both the $Q$ and $\overline Q$ variables
are integrated to obtain the pair distributions,
recall $\mu_{F,R} \propto m_T$.

Results on open heavy flavors indicate that some level of
transverse momentum broadening is needed to obtain agreement with the low $p_T$
data.  This is often done by including some intrinsic transverse momentum,
$k_T$, smearing to the initial-state parton densities.
The implementation of intrinsic $k_T$ in the MNR code is not handled in the 
same way as calculations of other hard processes due to the nature of the code.
In the MNR code, the cancellation of divergences is done numerically.  
Since adding additional numerical Monte-Carlo integrations would slow the
simulation of events, in addition to requiring multiple runs with the same
setup but different intrinsic $k_T$ kicks, the kick is added in the
final, rather than the initial, state. 
In Eq.~(\ref{sigtil}), the Gaussian function $g_p(k_T)$,
\begin{eqnarray}
g_p(k_T) = \frac{1}{\pi \langle k_T^2 \rangle_p} \exp(-k_T^2/\langle k_T^2
\rangle_p) \, \, ,
\label{intkt}
\end{eqnarray}
\cite{MLM1}, multiplies the parton
distribution functions for both hadrons, 
assuming the $x$ and $k_T$ dependencies in the initial partons completely
factorize.  If factorization applies, 
it does not matter whether the $k_T$ dependence
appears in the initial or final state if the kick is not too large, as 
described below.  In Ref.~\cite{MLM1}, $\langle k_T^2 \rangle_p = 1$ GeV$^2$
was found to best 
describe fixed-target charm production. 

In the code, the $Q \overline Q$ system
is boosted to rest from its longitudinal center-of-mass frame.  Intrinsic 
transverse
momenta of the incoming partons, $\vec k_{T 1}$ and $\vec k_{T 2}$, are chosen
at random with $k_{T 1}^2$ and $k_{T 2}^2$ distributed according to
Eq.~(\ref{intkt}).   A second transverse boost out of the pair rest frame
changes the initial transverse momentum of
the $Q \overline Q$ pair, $\vec p_T$, to $\vec p_T
+ \vec k_{T 1} + \vec k_{T 2}$.  The initial
$k_T$ of the partons could have alternatively been given to the entire
final-state system, as is essentially done if applied in the initial state,
instead of to the $Q \overline Q$ pair.  There is no difference if the
calculation is leading order only but at NLO an additional light parton can 
also appear in the final state so the correspondence is not exact.  
In Ref.~\cite{MLM1}, the difference between the two implementations is claimed 
to be small if $k_T^2 \leq 2$ GeV$^2$.  We note that the
rapidity distribution, integrated over all $p_T$, is unaffected by the
intrinsic $k_T$.

The effect of the intrinsic $k_T$ on the shape of the $J/\psi$ $p_T$ 
distribution can be expected to decrease as 
$\sqrt{s}$ increases because the average $p_T$ of the $J/\psi$ also increases 
with energy.  However, the value of $\langle k_T^2 \rangle$ may increase with
$\sqrt{s}$.  
We can check the energy dependence of $\langle k_T^2 \rangle$ by
the shape of the $J/\psi$ $p_T$ distributions at central and forward rapidity
at RHIC.  We find that $\langle k_T^2 \rangle = 1 + 
(1/12)\ln(\sqrt{s}/20) \approx 1.19$~GeV$^2$ at $\sqrt{s} = 200$ GeV
agrees well with the $J/\psi$ $p_T$ 
distributions measured by PHENIX at both midrapidity and forward rapidity,
see Fig.~\ref{psiRHICdistfig}.
The rapidity distributions, as well as the $p_T$ distributions in the two
rapidity regions, all agree well with the $J/\psi$ cross sections calculated
with the central set of parameters.  Only the low $p_T$ part of the forward
rapidity $p_T$ distribution is somewhat underestimated.  The integrated forward
cross section is about 50\% lower than the midrapidity value. In addition, the
$p_T$ distribution falls off faster at high $p_T$ in the forward rapidity 
region.

\begin{figure}[htb]
\begin{center}
\begin{tabular}{cc}
\includegraphics[width=0.33\textwidth]{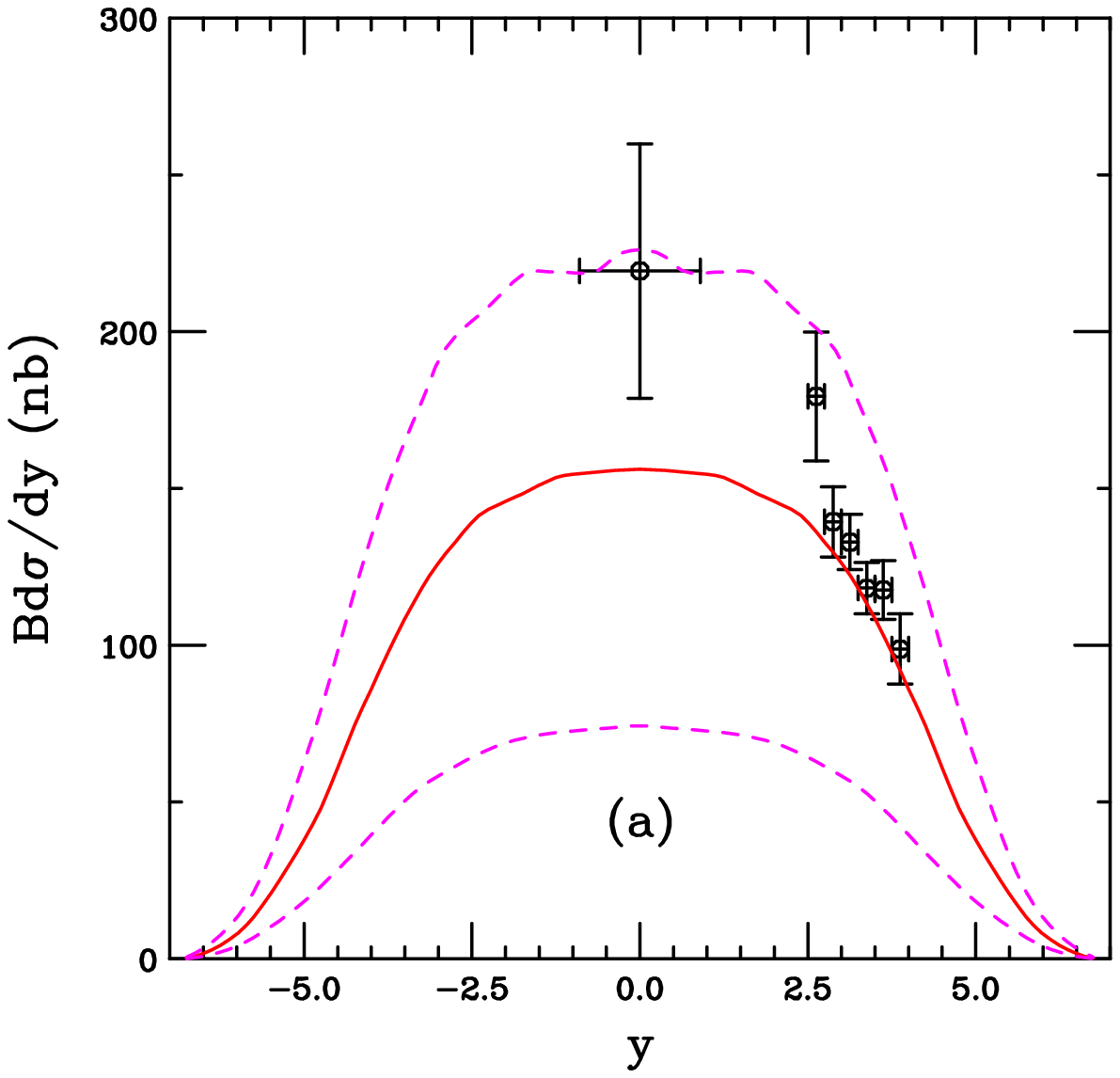}&
\includegraphics[width=0.33\textwidth]{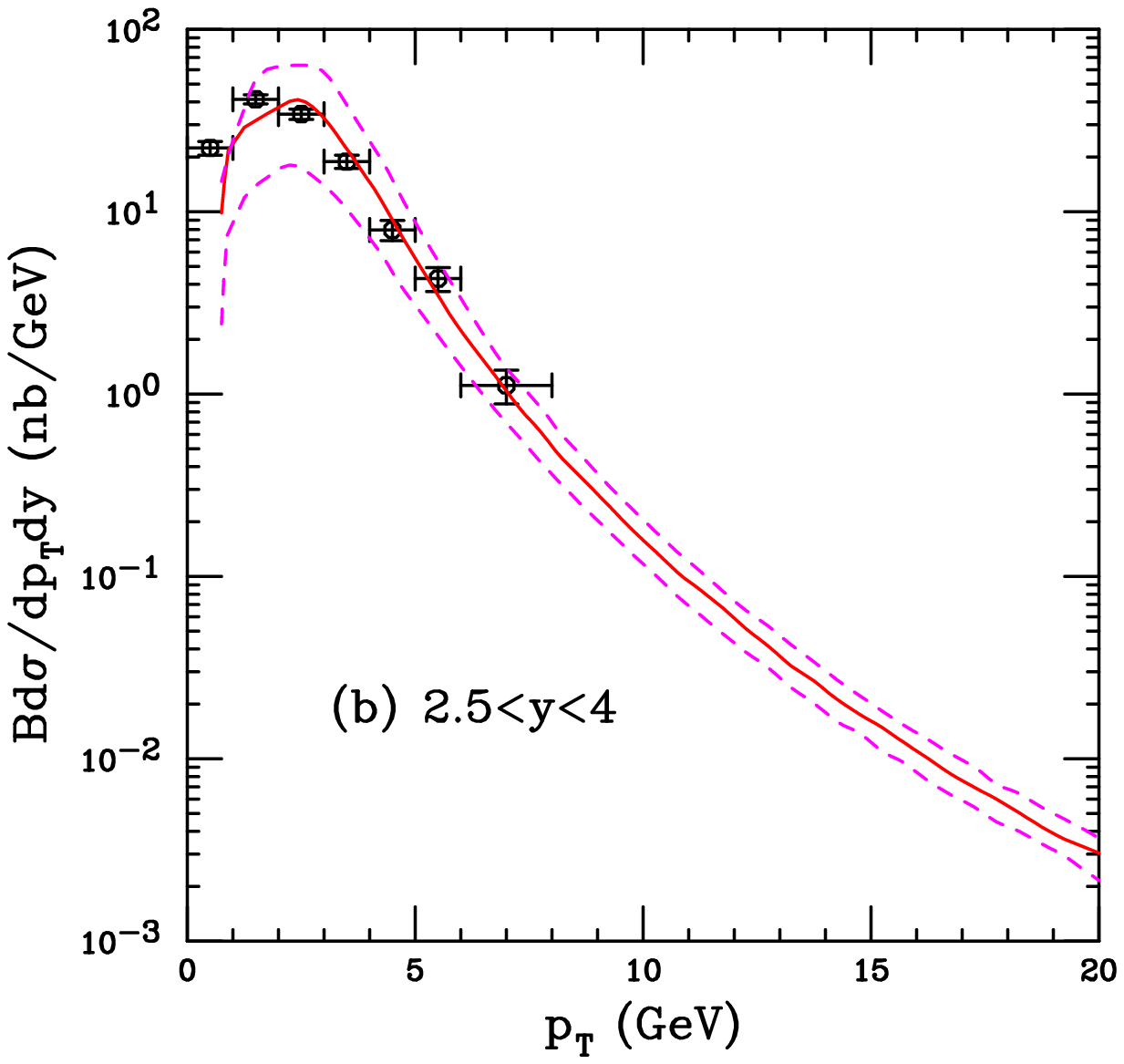}
\end{tabular}
\end{center}
\caption[]{(Color online) 
The $J/\psi$ rapidity distribution (a) and the forward rapidity, 
$2.5 < y < 4$ $p_T$ distribution (b) at 
$\sqrt{s} = 2.76$~TeV and their uncertainties calculated 
with the same parameters as in Fig.~\protect\ref{chi2fig}(d).  
The results are compared to the ALICE rapidity distribution as well as the
forward rapidity $p_T$ distribution \protect\cite{ALICEpp276TeV}.  
No additional scaling factor
has been applied.  The solid red curve shows the central value while the
dashed magenta curves outline the uncertainty band.
A $\langle k_T^2 \rangle$ kick of 1.41 GeV$^2$ is applied to the $p_T$ 
distributions, as discussed in the text.  
}
\label{psiALICE276distfig}
\end{figure}

The ALICE 7 TeV $p_T$ distributions, shown in 
Fig.~\ref{psiALICEdistfig}(b) and (c), include the ALICE rapidity cuts for
the central and forward rapidity regions, $|y| < 0.9$ and $2.5 < y < 4$, 
respectively.  
The rapidity distribution at $\sqrt{s} = 7$ TeV, Fig.~\ref{psiALICEdistfig}(a),
is flat over several units of rapidity.  Thus the integrated cross sections in
the two rapidity intervals, normalized per unit of rapidity, are very similar.
However, the forward rapidity $p_T$ distribution is still a stronger function
of $p_T$ than the midrapidity distribution.

Finally, the inclusive $J/\psi$ rapidity distribution and forward rapidity $p_T$
distribution at $\sqrt{s} = 2.76$ TeV are compared to the ALICE data in
Fig.~\ref{psiALICE276distfig}.  Here the calculated rapidity distribution is 
not as broad and the agreement with the data is rather good although the 
midrapidity point remains high relative to the central value of the calculation.
The agreement of the calculated $p_T$ distribution with the forward rapidity
data is quite good with the exception of the lowest $p_T$ point where the
calculated distribution turns over more quickly than the data.

\section{Summary}
\label{end}

We have narrowed the uncertainty band on the open heavy flavor cross section
and, in so doing, have also provided a realistic uncertainty band on $J/\psi$
production in the color evaporation model.  The central result, $m = 1.27$ GeV,
$\mu_F/m = 2.1$ and $\mu_R/m = 1.6$, is quite compatible with previous
calculations using a `by-eye' fit to the data with $m = 1.2$ GeV, $\mu_F/m =
\mu_R/m = 2$ \cite{Gavai:1994in,vogtHPC}. 

While the fits have been made by comparing the calculated NLO charm production
cross section to available data at fixed-target energies and at RHIC, they are
in good agreement with the extracted total charm cross sections at the LHC.
The same parameter set also provides good agreement with the distributions of
single leptons from semileptonic heavy flavor decays at RHIC and the LHC.
The limit on the width of the uncertainty band is now set by the uncertainty
due to bottom quark production and decay.

We have used the same fit parameters in the calculation of $J/\psi$ production
in the color evaporation model and have thus provided the first uncertainty
band on $J/\psi$ production in this approach.  The energy dependence of the
total $J/\psi$ cross section that results is a good match to the data up to
collider energies.  The $p_T$ distributions are also in good agreement with
the data from RHIC and the LHC. In future work, we will use our new parameter
set to place limits on the contribution of $B$ meson decays to $J/\psi$ 
production and will also study cold nuclear matter effects on $J/\psi$ 
production.

\section*{Acknowledgments}

We thank M. Cheng, L. Linden Levy, P. Petreczky, R. Soltz 
and P. Vranas for discussions.
This work was performed under the auspices of the U.S.\
Department of Energy by Lawrence Livermore National Laboratory under
Contract DE-AC52-07NA27344 and was also supported in part by the
National Science Foundation Grant NSF PHY-0555660.

\end{document}